\documentclass[twocolumn]{jpsj3}
\usepackage{txfonts}

\title{
Anisotropic Magnetoresistance Effects in 
Fe, Co, Ni, Fe$_4$N, and Half-Metallic Ferromagnet: 
A Systematic Analysis
}

\author{Satoshi Kokado\thanks{E-mail address: 
tskokad@ipc.shizuoka.ac.jp} 
Masakiyo Tsunoda$^1$, 
Kikuo Harigaya$^2$, and Akimasa Sakuma$^3$
}
\inst{
Faculty of Engineering, Shizuoka University, Hamamatsu 432-8561, Japan\\
$^1$Department of Electronic Engineering, 
Graduate School of Engineering, Tohoku University, Sendai 980-8579, Japan\\
$^2$Nanosystem Research Institute, AIST, Tsukuba 305-8568, Japan\\
$^3$Department of Applied Physics, Graduate School of Engineering, 
Tohoku University, Sendai 980-8579, Japan
}
\abst{
We theoretically analyze 
the anisotropic magnetoresistance (AMR) effects of 
bcc Fe ($+$), fcc Co ($+$), fcc Ni ($+$), Fe$_4$N ($-$), 
and a half-metallic ferromagnet ($-$). 
The sign in each (~~) represents the sign of the AMR ratio 
observed experimentally.
We here use the two-current model for 
a system consisting of 
a spin-polarized conduction state and localized d states 
with spin--orbit interaction. 
From the model, 
we first derive a general expression of the AMR ratio. 
The expression consists of 
a resistivity of the conduction state of 
the $\sigma$ spin ($\sigma=\uparrow$ or $\downarrow$), 
$\rho_{s \sigma}$, 
and 
resistivities due to 
s--d scattering processes 
from the conduction state 
to the localized d states. 
On the basis of this expression, 
we next find 
a relation between 
the sign of the AMR ratio and 
the s--d scattering process. 
In addition, 
we obtain expressions of the AMR ratios 
appropriate to the respective materials. 
Using the expressions, 
we evaluate their AMR ratios, 
where the expressions take into account the values of 
$\rho_{s \downarrow}/\rho_{s \uparrow}$ of the respective materials. 
The evaluated AMR ratios 
correspond well to the experimental results. 
}
\kword{anisotropic magnetoresistance effect, weak ferromagnet, 
strong ferromagnet, half-metallic ferromagnet, 
spin--orbit interaction, s--d scattering, 
spin-polarized conduction electron, two-current model}

\begin{document}
\maketitle

\section{Introduction}

\begin{figure}[ht]
\begin{center}
\includegraphics[width=0.35\linewidth]{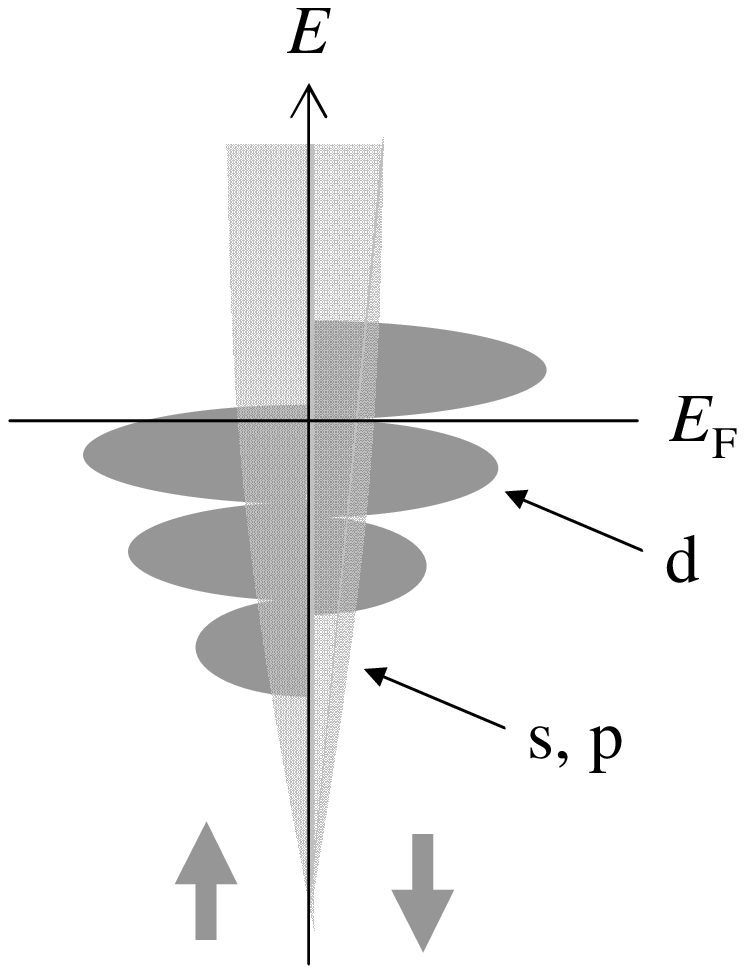} \hspace{0.5cm}
\includegraphics[width=0.35\linewidth]{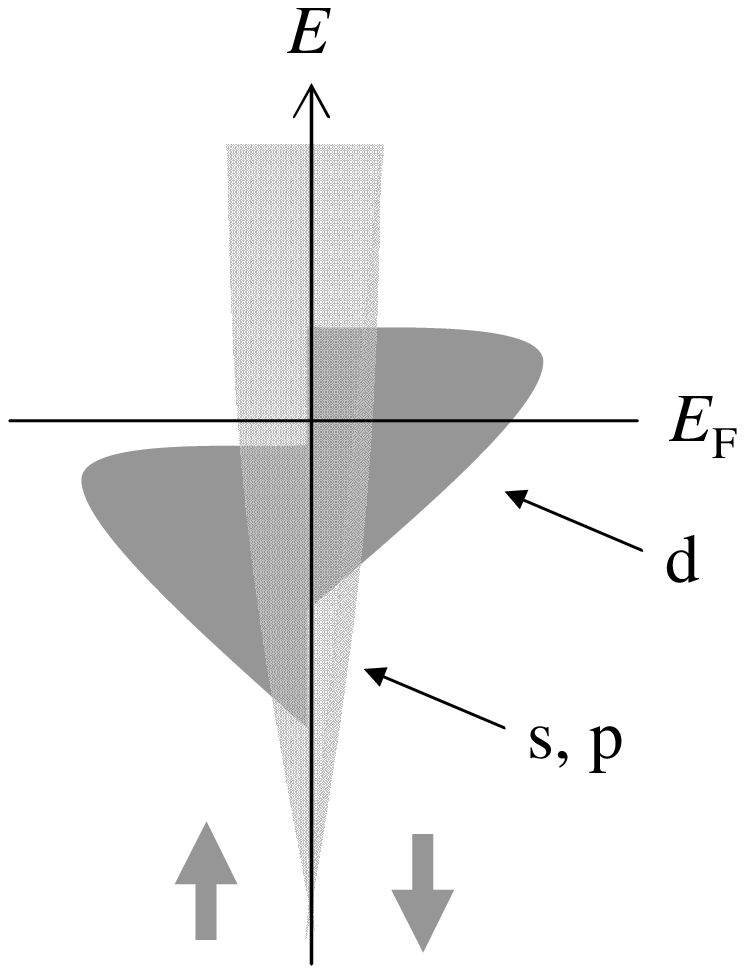} \\
\hspace{-0.5cm}
{\footnotesize (a) bcc Fe \hspace{2cm} 
(b) fcc Co and fcc Ni}\\
\vspace{0.2cm}
\includegraphics[width=0.35\linewidth]{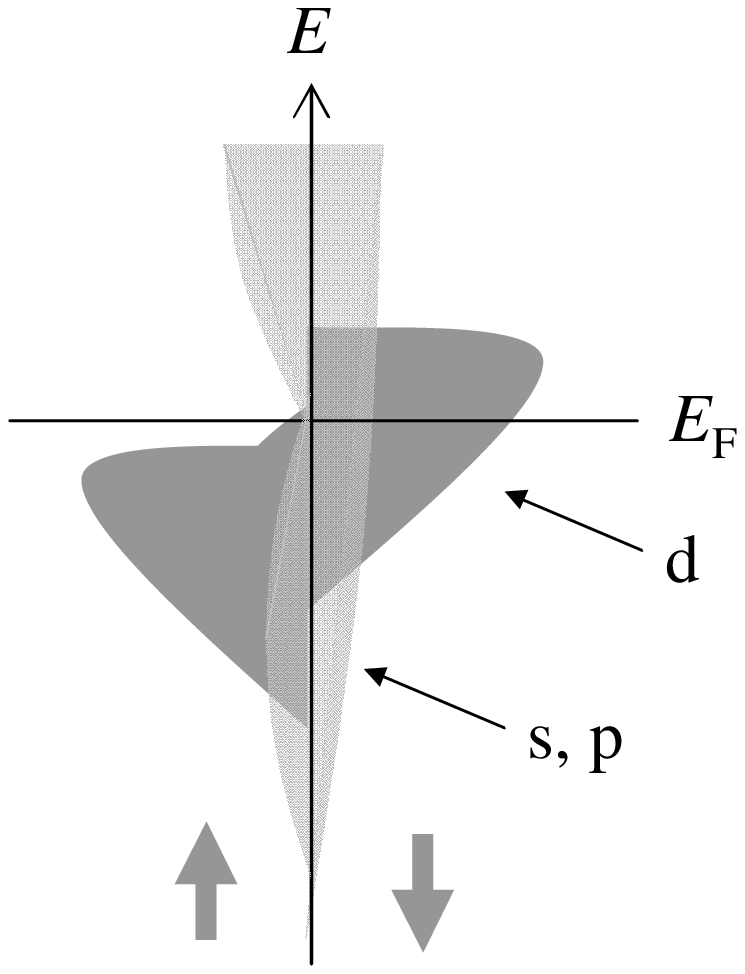} \\
{\footnotesize (c) Fe$_4$N}\\
\vspace{0.2cm}
\includegraphics[width=0.35\linewidth]{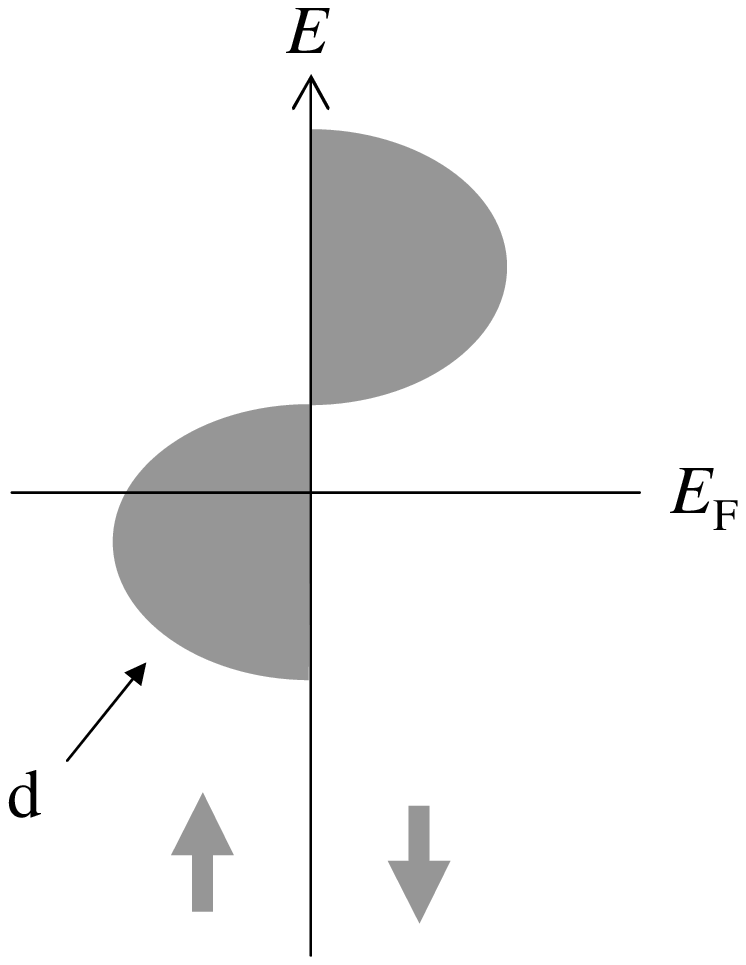} \hspace{0.5cm}
\includegraphics[width=0.4\linewidth]{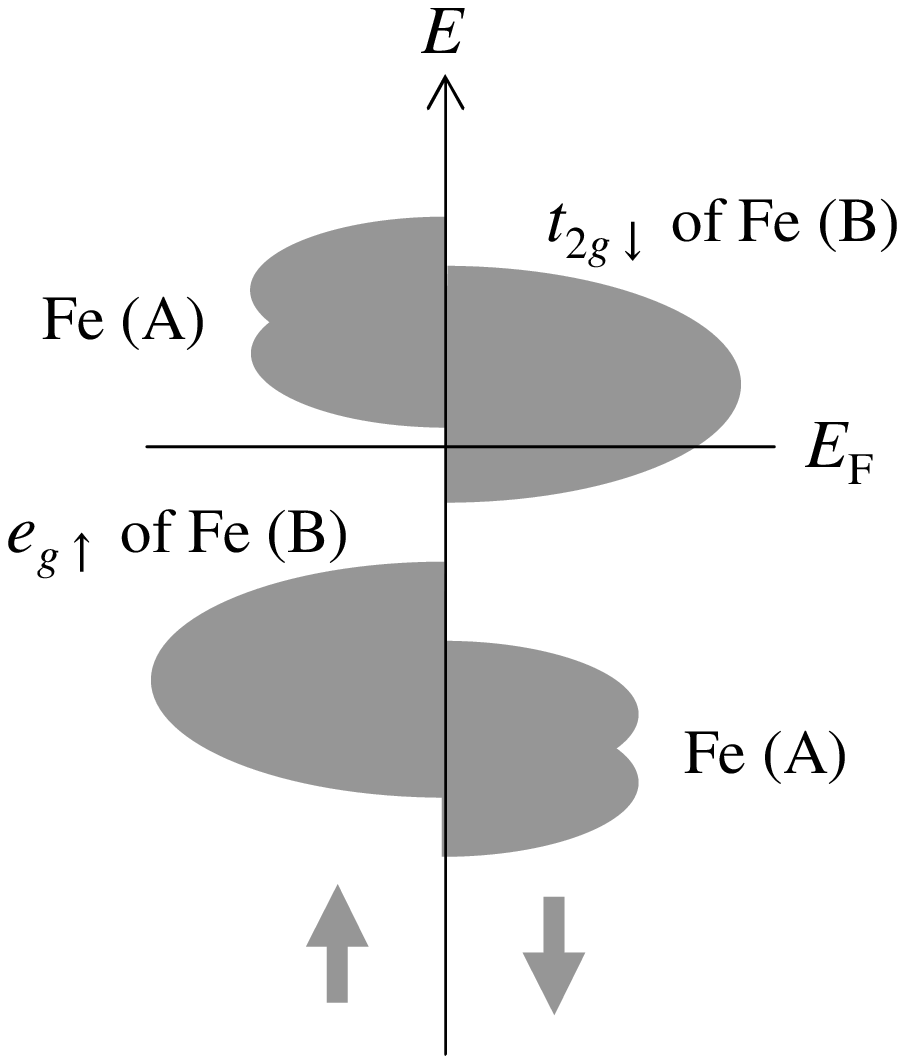} \\
\hspace{-0.5cm}
{\footnotesize 
(d) Half-metallic ferromagnet \hspace{0.3cm} 
(e) Fe$_3$O$_4$ (half-metallic ferromagnet)} \\
\caption{
Schematic illustration of 
the density of states (DOS) of the various ferromagnets. 
(a) The partial DOS of bcc Fe\cite{Wang,papa} of the weak ferromagnet. 
(b) The partial DOS of fcc Co\cite{Matar} 
and fcc Ni\cite{Vargas,papa} of the strong ferromagnet. 
(c) The partial DOS of Fe$_4$N\cite{Sakuma,Kokado} of the strong ferromagnet. 
(d) The DOS of the half-metallic ferromagnet 
such as 
Co$_2$MnAl$_{1-x}$Si$_x$,\cite{Galanaki} 
La$_{0.7}$Sr$_{0.3}$MnO$_3$,\cite{Park,Teresa} and 
La$_{0.7}$Ca$_{0.3}$MnO$_3$.\cite{Pickett} 
(e) The DOS 
of Fe$_3$O$_4$\cite{Camphausen,Zhang} of the half-metallic ferromagnet. 
In (a) - (c), 
light-gray areas (dark-gray areas) correspond to 
the sp band DOS (the d band DOS). 
The sp band is partly covered by the d band 
(see lighter areas in the d band). 
The d band consists of the conductive and localized d states, 
and the respective portions are unspecified here. 
In (d) and (e), 
only the DOS's in the vicinity of $E_{\mbox{\tiny F}}$ 
(i.e., the d band DOS) are shown. 
In (e), 
Fe (A) and Fe (B) denotes sublattices, 
and $e_{g\uparrow}$ and $t_{2g\downarrow}$ are 
3d orbitals of the Fe ion.\cite{Zhang} 
}
\label{fig_dos}
\end{center}
\end{figure}

The anisotropic magnetoresistance (AMR) effect,\cite{Smit,Gondo,Campbell,Potter,McGuire1,Dorleijn,Jaoul,McGuire,Malozemoff1,Malozemoff2,Endo,Ziese,Ziese2,Ziese1,Favre,Tsunoda,Tsunoda1,Berger,Miyazaki} 
in which the electrical resistivity depends on the relative angle between 
the magnetization direction 
and the electric current direction, 
is one of the most 
fundamental characteristics 
involving magnetic and transport properties. 
The AMR effect 
has been therefore investigated for various magnetic materials. 
In particular, the AMR ratio has been measured 
to evaluate the amplitude of the effect. 
The AMR ratio is generally defined as
\begin{eqnarray}
\label{ratio}
\frac{\Delta \rho}{\rho}=\frac{\rho_\parallel - \rho_\perp}{\rho_\perp}, 
\end{eqnarray}
where $\rho_\parallel$ ($\rho_\perp$) 
represents 
a resistivity for the case of the electrical current 
parallel to the magnetization 
(a resistivity for the case of 
the current perpendicular to the magnetization). 
Table \ref{tab1} shows the experimental values of 
the AMR ratios of typical ferromagnets, i.e., 
body-centered cubic (bcc) 
Fe\cite{McGuire} 
face-centered cubic (fcc) Co,\cite{McGuire} fcc Ni,\cite{McGuire} 
Fe$_4$N,\cite{Tsunoda,Tsunoda1} 
and the half-metallic ferromagnet.\cite{Endo,Ziese,Ziese2,Ziese1,Favre} 
Here, 
bcc Fe is categorized as a weak ferromagnet,\cite{SW_FM} 
in which 
its majority-spin d band is not filled (see Fig. \ref{fig_dos}(a)). 
In contrast, fcc Co, fcc Ni, 
and Fe$_4$N are strong ferromagnets,\cite{SW_FM} 
in which 
their majority-spin d band is filled 
(see Fig. \ref{fig_dos}(b)). 
In addition, the half-metallic ferromagnet is defined as 
having a finite density 
of states (DOS) at the Fermi energy $E_{\mbox{\tiny F}}$ 
in one spin channel 
and a zero DOS at $E_{\mbox{\tiny F}}$ in the other spin channel 
(see Figs. \ref{fig_dos}(d) and \ref{fig_dos}(e)). 
As remarkable points, Fe,\cite{McGuire} Co,\cite{McGuire} 
and Ni\cite{McGuire} exhibited positive AMR ratios, 
while 
Fe$_4$N\cite{Tsunoda,Tsunoda1} 
and the half-metallic ferromagnets\cite{Endo,Ziese,Ziese2,Ziese1,Favre} 
showed negative AMR ratios. 
Furthermore, 
in the case of Fe$_3$O$_4$\cite{Ziese,Ziese2} 
of the half-metallic ferromagnet, 
the sign of the AMR ratio changed 
from negative to positive with increasing temperature. 
For such ferromagnets, however, 
theoretical studies 
to systematically explain their AMR ratios 
have been scarce so far. 
In particular, 
a feature that strongly affects the sign of the AMR ratio 
has not yet been revealed.

\begin{table*}[ht]
\caption{
AMR ratio  
$\rho_{s\downarrow}/\rho_{s\uparrow}$ 
and $D_\downarrow^{(d)}/D_\uparrow^{(d)}$ 
of the various ferromagnets. 
The AMR ratios represent experimental values. 
Note that 
for every material except for Fe$_4$N, 
the AMR ratio defined in each paper, 
$x_{\rm AMR}=(\rho_\parallel - \rho_\perp)/[ (\rho_\parallel/3)+(2\rho_\parallel/3) ]$, 
has been transformed into $\Delta \rho/\rho$ of eq. (\ref{ratio}) 
by using 
$\Delta \rho/\rho=3 x_{\rm AMR}/(x_{\rm AMR}+3)$. 
The ratios $\rho_{s\downarrow}/\rho_{s\uparrow}$'s 
of bcc Fe, fcc Co, fcc Ni, and Fe$_4$N 
are the respective theoretical values evaluated from 
analyses using a combination of 
the first principles calculation and the Kubo formula. 
Their $D_\uparrow^{(d)}/D_\downarrow^{(d)}$'s 
are roughly estimated from 
the respective $D_{d \uparrow}^{\rm FP}/D_{d\downarrow}^{\rm FP}$'s. 
Here, 
$D_{\varsigma}^{(d)}$ is 
the DOS of each d state of the $\varsigma$ spin at $E_{\mbox{\tiny F}}$ 
(see eq. (\ref{tau_sd})), 
where 
$D_{\varsigma}^{(d)}$ is set to be 
$D_{\varsigma}^{(d)}=D_{M \varsigma}^{(d)}$ 
by ignoring $M$ for $D_{M \varsigma}^{(d)}$ of eq. (\ref{D_Ms^d}). 
In addition, 
$D_{d \varsigma}^{\rm FP}$ is the partial DOS 
of the d band at $E_{\mbox{\tiny F}}$ 
obtained by the first principles calculation. 
In a simple term, 
$D_{d \varsigma}^{\rm FP} = \sum_{M=-2}^2 D_{M\varsigma}^{(d)}$ 
is realized. 
The ratios $\rho_{s\downarrow}/\rho_{s\uparrow}$'s 
and 
$D_\uparrow^{(d)}/D_\downarrow^{(d)}$'s 
of 
the half-metallic ferromagnets are, respectively, assumed to have 
$\rho_{s\downarrow}/\rho_{s\uparrow} \to 0$ or $\infty$ 
and 
$D_\uparrow^{(d)}/D_\downarrow^{(d)} \to 0$ or $\infty$, 
judging from 
the DOS's at $E_{\mbox{\tiny F}}$ 
of Figs. \ref{fig_dos}(d) and \ref{fig_dos}(e). 
}
\begin{tabular}{lllll}
\hline \\[-0.2cm]
Category & Material & AMR ratio $\Delta \rho/\rho$ 
(experimental value) & $\rho_{s\downarrow}/\rho_{s\uparrow}$ & $D_\uparrow^{(d)}/D_\downarrow^{(d)}$ \\\\[-0.2cm]
\hline \\[-0.2cm]
Weak ferromagnet\cite{SW_FM} 
& bcc Fe & 0.0030 at 300 K (ref. \citen{McGuire}) & 
3.8 $\times$ 10$^{-1}$ (ref. \citen{Tsymbal}) & $\sim$ 2.0 (ref. 
\citen{Wang})\\ \\[-0.2cm]
Strong ferromagnet\cite{SW_FM} & fcc Co  & 0.020 at 300 K (ref. \citen{McGuire}) & 
7.3 (ref. \citen{Tsymbal}) & $\sim 0$ (ref. \citen{Matar})\\
& fcc Ni  & 0.022 at 300 K (ref. \citen{McGuire}) & 
1.0 $\times$ 10 (ref. \citen{Kokado_un1}) & $\sim 0$ 
(ref. \citen{Vargas})\\
& Fe$_4$N  & $-$0.043 - $-$0.005 for 4.2 K - 300 K (ref. \citen{Tsunoda})
& 1.6 $\times$ 10$^{-3}$ (ref. \citen{Kokado_un}) & $\sim$ 0.2 (ref. \citen{Sakuma})\\
& & $-$0.07 - $-$0.005 for 4 K - 300 K (ref. \citen{Tsunoda1}) & & \\\\[-0.2cm]
Half-metallic ferromagnet & Co$_2$MnAl$_{1-x}$Si$_x$ 
& $-$0.003 - $-$0.002 at 4.2 K (ref. \citen{Endo}) & $\to\infty$ & $\to\infty$ \\
& La$_{0.7}$Sr$_{0.3}$MnO$_3$  & $-$0.0015 at 4 K (ref. \citen{Favre}) 
& $\to\infty$ & $\to\infty$ \\ 
& La$_{0.7}$Ca$_{0.3}$MnO$_3$  & $-$0.0012 at 75 K (ref. \citen{Ziese}) & $\to\infty$ & $\to\infty$ \\
&  & $-$0.004 at 100 K (ref. \citen{Ziese1}) & &  \\
& Fe$_3$O$_4$  & $-$0.005 - 0.005 for 100 K - 300 K (refs. \citen{Ziese} and \citen{Ziese2}) & $\sim 0$ & $\sim 0$ \\
\hline 
\end{tabular}
\label{tab1}
\end{table*}

Theoretically, 
expressions of the AMR ratio 
have been derived 
by taking into account 
a resistivity due to 
the s--d scattering.\cite{Smit,Campbell,Potter,Jaoul,Malozemoff1,Malozemoff2,Ziese,Berger} 
This scattering represents that 
the conduction electron is scattered into 
the localized d states by impurities. 
The d states have 
exchange field $H_{\rm ex}$ and 
spin--orbit interaction, i.e., 
$\lambda {\mbox{\boldmath $L$}} \cdot {\mbox{\boldmath $S$}}$, 
where 
$\lambda$ is the spin--orbit coupling constant, 
${\mbox{\boldmath $L$}}$ (=$L_x$, $L_y$, $L_z$) is 
the orbital angular momentum, 
and ${\mbox{\boldmath $S$}}$ (=$S_x$, $S_y$, $S_z$) is 
the spin angular momentum. 
Here, the d states are spin-mixed owing to the spin--orbit interaction.

The applicable scope of 
the previous theories, however, 
appears to be limited to specific materials 
because 
only the partial components in the whole resistivities 
have been adopted. 
For example, 
Campbell, Fert, and Jaoul\cite{Campbell} (CFJ) derived 
an expression of the AMR ratio of 
a strong ferromagnet\cite{Malozemoff1} 
such as Ni-based alloys, i.e.,\cite{Miyazaki} 
\begin{eqnarray}
\label{CFJ}
\frac{\Delta \rho}{\rho}=\gamma ( \alpha - 1 ), 
\end{eqnarray}
with $\gamma=(3/4)(\lambda/H_{\rm ex})^2$ and 
$\alpha \approx 
\rho_{s \to d\downarrow}/\rho_{s\uparrow}$.\cite{alpha} 
Here, $\rho_{s \sigma}$ was a resistivity of 
the conduction state (named as $s$) of the $\sigma$ spin, 
with $\sigma=\uparrow$ or $\downarrow$. 
In addition, $\rho_{s \to d\varsigma}$ was 
a resistivity due to the s--d scattering, in which 
the conduction electron was scattered into 
the localized d states of the $\varsigma$ spin by impurities. 
The $\varsigma$ spin 
represented 
the spin of the dominant state in the spin-mixed state, 
where the up spin ($\varsigma=\uparrow$) 
and down spin ($\varsigma=\downarrow$) 
meant the majority spin and the minority spin, respectively. 
Note that the CFJ model adopted 
only $\rho_{s \uparrow}$ and $\rho_{s \to d\downarrow}$ 
on the basis of scattering processes 
between the dominant states at $E_{\mbox{\tiny F}}$. 
The processes were 
$s\uparrow \to s\uparrow$, $s\uparrow \to d\downarrow$, 
and $s\downarrow \to d\downarrow$,\cite{Campbell} 
where $s \sigma \to s \sigma$ represented 
the scattering process between the conduction states of the $\sigma$ spin, 
while $s \sigma \to d \varsigma$ 
was the scattering process from the conduction state of the $\sigma$ spin 
to 
the $\sigma$ spin state in the localized d states of the $\varsigma$ spin. 
On the other hand, 
Malozemoff\cite{Malozemoff1,Malozemoff2} 
extended the CFJ model to a more general model 
which 
was applicable to the weak ferromagnet 
as well as the strong ferromagnet. 
This model took into account 
$\rho_{s\uparrow}$, $\rho_{s\downarrow}$, 
$\rho_{s \to d\uparrow}$, 
and $\rho_{s \to d\downarrow}$ 
on the basis of 
the scattering processes of 
$s \uparrow \to s \uparrow$, 
$s \uparrow \to d \uparrow$, 
$s \uparrow \to d \downarrow$, 
$s \downarrow \to s \downarrow$, 
$s \downarrow \to d \downarrow$, 
and $s \downarrow\to d \uparrow$. 
In the actual application to materials, 
however, 
he often used an expression of 
the AMR ratio with 
$\rho_{s\uparrow}=\rho_{s\downarrow}=\rho_{s}$,
\cite{Malozemoff1,Malozemoff2} 
i.e., 
\begin{eqnarray}
\label{Malozemoff_AMR}
\frac{\Delta \rho}{\rho}= 
\frac{\gamma (\rho_{s \to d\downarrow} 
- \rho_{s \to d\uparrow})^2}
{(\rho_{s} + \rho_{s \to d\uparrow}) 
(\rho_{s} + \rho_{s \to d\downarrow})}, 
\end{eqnarray}
which was always positive. 
Equation (\ref{Malozemoff_AMR}) was an expression for the weak ferromagnet, 
while 
Eq. (\ref{Malozemoff_AMR}) with 
$\rho_{s \to d\uparrow}=0$ was that for the strong ferromagnet.

Furthermore, 
we point out a problem, namely, that 
the previous theories have not taken into account 
the spin dependence of 
the effective mass and the number density 
of electrons in the conduction band 
in expressions of the resistivities. 
For example, 
the half-metallic ferromagnets 
which have the DOS's of Figs. \ref{fig_dos}(d) and \ref{fig_dos}(e) 
may show significant spin dependence.

On the basis of this situation, 
we suggest improvements for 
a systematic analysis of the AMR effects of various ferromagnets. 
First, the expression of the AMR ratio should 
treat $\rho_{s\downarrow}/\rho_{s\uparrow}$ as a variable. 
The reason is that 
$\rho_{s\downarrow}/\rho_{s\uparrow}$ actually 
depends strongly on the materials (see Table \ref{tab1}). 
Namely, 
$\rho_{s\downarrow}/\rho_{s\uparrow}$ has been evaluated to be 
3.8 $\times$ 10$^{-1}$ 
for bcc Fe,\cite{Tsymbal} 
7.3 for fcc Co,\cite{Tsymbal} 
1.0 $\times$ 10 for fcc Ni,\cite{Kokado_un1,papa} 
and 1.6 $\times$ 10$^{-3}$ for Fe$_4$N,\cite{Kokado_un,Kokado} 
from 
analyses using a combination of 
the first principles calculation and the Kubo formula 
within the semiclassical approximation. 
The half-metallic ferromagnet is also assumed to have 
$\rho_{s\downarrow}/\rho_{s\uparrow} \approx 0$ or 
$\rho_{s\downarrow}/\rho_{s\uparrow} \to \infty$. 
It is noteworthy here that 
the conduction state (called $s$ in suffixes of $\rho_{s\sigma}$) 
is considered to consist of 
not only the s and p states 
but also the conductive d state. 
In addition, the exchange splitting of the s and p states 
is attributed to the fact that 
the s and p states are coupled to 
the d states with exchange splitting through the transfer integrals. 
Second, 
in the case of the half-metallic ferromagnet, 
the expressions of the resistivities 
should take into account 
the spin dependence of 
the effective mass and the number density 
of the electrons in the conduction band. 

In this paper, 
we first derived general expressions of 
the resistivities and the AMR ratio. 
We here treated 
$\rho_{s\downarrow}/\rho_{s\uparrow}$ as a variable 
and took into account 
the spin dependence 
of the effective mass and the number density 
of the electrons in the conduction band. 
Second, on the basis of the expressions, 
we roughly determined 
a relation between the sign of the AMR ratio and 
the dominant s--d scattering process. 
Namely, when the dominant s--d scattering process 
was $s\uparrow \to d \downarrow$ or $s\downarrow \to d \uparrow$, 
the AMR ratio tended to become positive. 
In contrast, 
when the dominant s--d scattering process 
was $s\uparrow \to d \uparrow$ or $s\downarrow \to d \downarrow$, 
the AMR ratio tended to be negative. 
Finally, using the expression of the AMR ratio, 
we systematically analyzed the AMR ratios of 
Fe, Co, Ni, Fe$_4$N, and the half-metallic ferromagnet. 
The evaluated AMR ratios corresponded well with 
the respective experimental results. 
In addition, 
the sign change of the AMR ratio of Fe$_3$O$_4$ 
could be explained by 
considering the increase of 
the majority spin DOS at $E_{\mbox{\tiny F}}$.

The present paper is organized as follows: 
In \S \ref{Theory}, 
we derive general expressions of the resistivities and the AMR ratio. 
We then find the relation between 
the sign of the AMR ratio and the s--d scattering process. 
In \S \ref{Applications} and \S \ref{Half-metallic ferromagnet}, 
from the general expression, 
we obtain expressions of AMR ratio appropriate to 
the respective materials. 
Using the expressions, 
we analyze their AMR ratios. 
Concluding remarks are presented in the \S \ref{Conclusion}. 
In the Appendix \ref{WF}, we obtain wave functions 
of the localized d states (i.e., the spin-mixed states) 
from a single atom model 
that involves the spin--orbit interaction. 
In Appendixes \ref{s-d} and \ref{s-s}, 
we derive expressions of s--d and s--s scattering rates, respectively. 
In the Appendix \ref{matrix}, 
we show matrix elements in the s--d scattering rate. 
Some parameters 
are formulated in the Appendix \ref{parameters}.

\section{Theory}
\label{Theory}
We 
derive general expressions of resistivities 
due to 
electron scattering by nonmagnetic impurities 
and then obtain a general expression of the AMR ratio. 
On the basis of the resistivities and the AMR ratio, 
we explain a feature of the AMR effect. 
In addition, we find a relation between the sign of the AMR ratio 
and the scattering process.

\subsection{Model}
\label{Model}
Following the Smit model\cite{Smit} and the CFJ model\cite{Campbell}, 
we use 
a simple model 
consisting of the conduction state and the localized d states. 
The conduction state is represented by a plane wave, 
while the localized d states are described by a tight-binding model, 
i.e., the linear combination of 
atomic d orbitals.\cite{Campbell} 
The d orbitals are 
obtained 
by applying a perturbation theory to 
a Hamiltonian for the d electron in a single atom, ${\cal H}$: 
\begin{eqnarray}
\label{single_H}
&&{\cal H}={\cal H}_0 + {\cal H}', \\
\label{H_0}
&&{\cal H}_0 = -\frac{\hbar^2}{2m_e} \nabla^2 + V(r) + H_{\rm ex}S_z, \\
\label{LS}
&&{\cal H}' = \lambda {\mbox{\boldmath $L$}} \cdot {\mbox{\boldmath $S$}}.
\end{eqnarray}
Here, the unperturbed term ${\cal H}_0$ is 
the Hamiltonian for the hydrogen-like atom 
with 
Zeeman interaction due to $H_{\rm ex}$, 
where $H_{\rm ex}$ is the exchange field of the ferromagnet, 
$m_e$ is the electron mass, 
and $\hbar$ is the Planck constant $h$ divided by 2$\pi$. 
The term ${V} (r)$ is 
a spherically symmetric potential energy of the d orbitals 
created by a nucleus and core electrons, 
with $r=|{\mbox{\boldmath $r$}}|$, 
where ${\mbox{\boldmath $r$}}$ is the position vector. 
The perturbed term ${\cal H}'$ is the spin--orbit interaction 
with $|\lambda/H_{\rm ex}| \ll 1$. 
Here, 
the azimuthal quantum number $L$ and 
the spin quantum number $S$ are chosen to be 
$L$=2 and $S$=1/2, respectively. 
From this model, 
we obtain the spin-mixed states 
within the second-order perturbation (see Appendix \ref{WF}).

\subsection{Resistivity}
\label{Resistivity}
Using the localized d states and the conduction state, 
we can obtain the resistivity 
for the case of a parallel ($\parallel$) or perpendicular ($\perp$) 
configuration. 
As a starting point, we consider the two-current model\cite{Mott} 
composed of the up spin and down spin current components. 
In addition, 
this model is improved by including 
the spin-flip scattering, 
which is due to, for example, 
spin-dependent disorder\cite{FeNi,CoPd} and magnon\cite{Fert1,Ren}. 
The resistivity of $\ell$ configuration 
$\rho_\ell$ ($\ell=\parallel$ or $\perp$) is then written 
as\cite{link}
\begin{eqnarray}
\label{rho_j_ini}
\rho_\ell = \frac{\rho_{\ell,\uparrow} \rho_{\ell,\downarrow} 
+ \rho_{\ell,\uparrow} \rho_{\ell,\downarrow \uparrow} 
+\rho_{\ell,\downarrow} \rho_{\ell,\uparrow \downarrow}}
{\rho_{\ell,\uparrow} 
+ \rho_{\ell,\downarrow} 
+ (1 +a)\rho_{\ell,\uparrow \downarrow}
+ (1 + a^{-1})
\rho_{\ell,\downarrow \uparrow}},
\end{eqnarray}
with 
\begin{eqnarray}
\label{rho_js}
&&\rho_{\ell,\sigma} = \frac{m_\sigma^*}{n_\sigma e^2 \tau_{\ell,\sigma}}, \\
\label{rho_sf0}
&&\rho_{\ell,\sigma \sigma'} = \frac{m_\sigma^*}{n_\sigma e^2 
\tau_{\ell,\sigma  \sigma'}}, \\
\label{a}
&&a = \frac{m_\downarrow^* n_\uparrow}{m_\uparrow^* n_\downarrow},
\end{eqnarray}
where 
$\rho_{\ell,\sigma}$ is a resistivity of the $\sigma$ spin state 
for the $\ell$ configuration,\cite{Fert,Ibach,Grosso,Drude,Kittel1,Ren,Berger} 
while 
$\rho_{\ell,\sigma \sigma'}$ ($\sigma \ne \sigma'$) is a resistivity 
due to the spin-flip scattering process 
from the $\sigma$ spin state to the $\sigma'$ spin state 
for the $\ell$ configuration. 
It is noted that 
eq. (\ref{rho_j_ini}) with $\rho_{\ell,\sigma \sigma'}$=0 
corresponds to the resistivity of the two-current model. 
The constant $e$ is the electronic charge, 
and $n_\sigma$ ($m_\sigma^*$) 
is the number density\cite{Ibach,Grosso} (the effective mass\cite{Mathon}) 
of the electrons in the conduction band of the $\sigma$ spin, 
where the conduction band consists of the s, p, and conductive d states. 
The quantity $\tau_{\ell,\sigma}$ is a relaxation time of 
the conduction electron of the $\sigma$ spin 
for the $\ell$ configuration, 
and 
$\tau_{\ell,\sigma \sigma'}$ is a relaxation time of 
the spin-flip scattering process 
from the $\sigma$ spin state to the $\sigma'$ spin state 
for the $\ell$ configuration. 
The scattering rate $1/\tau_{\ell,\sigma}$ 
is expressed as\cite{McGuire1,Potter}
\begin{eqnarray}
\label{tau_inv}
\frac{1}{\tau_{\ell,\sigma}} = 
\frac{1}{\tau_{s \sigma}}+ 
\sum_{M=-2}^2 \sum_{\varsigma=\uparrow, \downarrow}
\frac{1}{\tau_{s \sigma \to d M \varsigma}^{(\ell)}}. 
\end{eqnarray}
Here, 
$\tau_{s \sigma}$ is 
a relaxation time of 
the conduction state 
of the $\sigma$ spin, 
where 
this state 
consists of the s, p, and conductive d states. 
In addition, 
$\tau_{s \sigma \to d M\varsigma}^{(\ell)}$ is 
a relaxation time of the s--d scattering for the $\ell$ configuration. 
This s--d scattering means that 
the conduction electron of the $\sigma$ spin 
is scattered into 
``the $\sigma$ spin state in the localized d state of $M$ and $\varsigma$'' 
by nonmagnetic impurities. 
The quantities 
$M$ ($M=-2$, $-$1, 0, 1, 2) 
and $\varsigma$ ($\varsigma=\uparrow$ or $\downarrow$) 
are, respectively, the magnetic quantum number and the spin 
of the dominant state in the spin-mixed state 
(see Appendix \ref{WF}). 
The expressions of 
$1/\tau_{s \sigma \to d M \varsigma}^{(\ell)}$ 
and $1/\tau_{s \sigma}$ 
are derived in Appendixes \ref{s-d} and \ref{s-s}, respectively.

Using eqs. (\ref{tau_sd^-1}), (\ref{22down}) - (\ref{2-2up}), 
and (\ref{mat_0_para}) - (\ref{mat_2_perp}), 
we obtain $\rho_{\ell,\sigma}$ 
of eq. (\ref{rho_js}) 
as
\begin{eqnarray}
\label{rho_j}
&&\hspace*{-0.8cm}\rho_{\parallel, \uparrow} = \rho_{s\uparrow} 
+ 2 \gamma \rho_{s\uparrow \to d1\downarrow} 
+ (1 - 2 \gamma) \rho_{s\uparrow \to d0\uparrow},  \\
\label{rho_j2}
&&\hspace*{-0.8cm}\rho_{\parallel, \downarrow} = \rho_{s\downarrow} 
+(1 - 2 \gamma) \rho_{s\downarrow \to d0\downarrow} 
+ 2 \gamma \rho_{s\downarrow \to d-1\uparrow},  \\
\label{rho_j3}
&&\hspace*{-0.8cm}\rho_{\perp, \uparrow} = \rho_{s\uparrow} 
+ \frac{\gamma}{2} \rho_{s\uparrow \to d1\downarrow}
+ \frac{\gamma}{2} \rho_{s\uparrow \to d-1\downarrow} 
+\frac{3}{8} 
\rho_{s\uparrow \to d2\uparrow} \nonumber \\
&& \hspace*{0.1cm}
+\frac{3}{8}\left( 1 - \frac{4}{3} \gamma \right) 
\rho_{s\uparrow \to d-2\uparrow}
+\frac{1}{4} \left( 1 - 2\gamma \right) 
\rho_{s\uparrow \to d0\uparrow}, \\
\label{rho_j1}
&&\hspace*{-0.8cm}\rho_{\perp, \downarrow} = \rho_{s\downarrow} 
+\frac{3}{8} 
\rho_{s\downarrow \to d-2\downarrow}
+ \frac{3}{8} \left(1- \frac{4}{3} \gamma \right) 
\rho_{s\downarrow \to d2\downarrow} \nonumber \\
&& \hspace*{0.1cm}+\frac{1}{4}\left( 1 - 2\gamma \right) 
\rho_{s\downarrow \to d0\downarrow}
+ \frac{\gamma}{2} \rho_{s\downarrow \to d1\uparrow}
+ \frac{\gamma}{2} \rho_{s\downarrow \to d-1\uparrow}, 
\end{eqnarray}
with 
\begin{eqnarray}
\label{gamma}
&&\hspace*{-0.8cm}\gamma=
\frac{3}{4} \left( \frac{\lambda}{H_{\rm ex}} \right)^2, \\
\label{r_s_s}
&&\hspace*{-0.8cm}\rho_{s\sigma}=
\frac{m_\sigma^*}{n_\sigma e^2 \tau_{s\sigma}}, \\
\label{r_ss-ds'}
&&\hspace*{-0.8cm}\rho_{s\sigma \to dM\varsigma}
=\frac{m_\sigma^*}{n_\sigma e^2 \tau_{s\sigma \to dM\varsigma}}, \\
&&\hspace*{-0.8cm}
\label{1/tau_s_s}
\frac{1}{\tau_{s\sigma}}
=\frac{2\pi}{\hbar}n_{\rm imp} 
\label{V_s^2}
|V_{s}|^2 D_{\sigma}^{(s)}, \\
\label{1/tau_s_dM}
&&\hspace*{-0.8cm}
\frac{1}{\tau_{s \sigma \to d M \varsigma}}
=\frac{2\pi}{\hbar} n_{\rm imp} N_{\rm n}
|V_{s\sigma \to d\sigma}|^2 
D_{M \varsigma}^{(d)}, \\
\label{V_ss-ds^2}
&&\hspace*{-0.8cm}|V_{s\sigma \to d\sigma}|^2= \nonumber \\
&&\frac{1}{3}
\left| v_{\rm imp}(R_{\rm n})  \int \int \int R(r)(z^2 - x^2)
\exp \left({\rm i} k_{\mbox{\tiny F}, \sigma}z \right) 
{\rm d}x {\rm d}y {\rm d}z \right|^2. \nonumber \\
\end{eqnarray}
Here, 
terms higher than the second order of $\lambda/H_{\rm ex}$ 
have been ignored. 
Accordingly, 
terms with $\gamma \rho_{s \sigma \to d \varsigma}$ 
in eqs. (\ref{rho_j}) - (\ref{rho_j1}) 
correspond to terms obtained from 
only the Smit\cite{Smit} spin-mixing mechanism\cite{Jaoul,Malozemoff2} 
with $(\lambda/2)(L_+ S_- + L_- S_+)$ (see Appendix \ref{WF}). 
In contrast, 
terms related to the $\lambda L_zS_z$ operator have been eliminated. 
A resistivity of the conduction state of the $\sigma$ spin, 
$\rho_{s \sigma}$, is 
due to the s--s scattering, 
in which 
the conduction electron of the $\sigma$ spin 
is scattered into the conduction state of the $\sigma$ spin 
by nonmagnetic impurities (see Appendix \ref{s-s}). 
In addition, 
$\rho_{s \sigma \to d M \varsigma}$ 
is a resistivity due to the s--d scattering. 
The s--d scattering means that 
the conduction electron of the $\sigma$ spin is scattered into 
``the $\sigma$ spin state in the localized d state of $M$ and $\varsigma$'' 
by the impurities, 
where 
$M$ and $\varsigma$ are as explained above 
(see Appendixes \ref{WF} and \ref{s-d}). 
The quantities 
$\tau_{s \sigma}$ and $\tau_{s \sigma \to d M \varsigma}$ are 
the relaxation times of the s--s and s--d scatterings, respectively. 
The quantity 
$V_{s}$ is 
the matrix element of the impurity potential 
for the s--s scattering (see eq. (\ref{1/tau_ssigma})), 
while $V_{s\sigma \to d\sigma}$ is 
that for the s--d scattering 
(see eqs. (\ref{v_Ms_k}), (\ref{tau_sd^-1}), and (\ref{d_orbital}), 
and Appendix \ref{matrix}), 
where 
$k_{{\mbox{\tiny F}},\sigma}$ 
is the Fermi wavevector of the $\sigma$ spin 
in the current direction. 
Here, each impurity 
is assumed to have a spherically symmetric scattering potential 
which acts only over a short range. 
The quantity 
$D_{\sigma}^{(s)}$ is 
the DOS of the conduction state of the $\sigma$ spin at $E_{\mbox{\tiny F}}$ 
(see eq. (\ref{D_s^s})), 
and $D_{M \varsigma}^{(d)}$ 
is that of the d state of $M$ and $\varsigma$ at $E_{\mbox{\tiny F}}$ 
(see eq. (\ref{D_Ms^d})). 
Furthermore, $n_{\rm imp}$ is the impurity density, 
and 
$N_{\rm n}$ is the number of the nearest-neighbor host atoms 
around the impurity 
(see eq. (\ref{N_nn})).

\begin{table*}[ht]
\caption{
s--d scattering terms 
in $\rho_{\ell,\sigma}$ of eqs. (\ref{rho_j}) - (\ref{rho_j1}) 
or 
eqs. (\ref{rho_j_s}) - (\ref{rho_j1_s}). 
The configuration $\ell$ is $\ell=\parallel$ or $\perp$, 
and $\sigma$ is $\sigma=\uparrow$ or $\downarrow$. 
The terms with $\rho_{s\sigma \to d M \varsigma}$ 
are listed for each $m$. 
Here, $m$ is the magnetic quantum number of the d orbital 
$\phi_{m,\sigma}({\mbox{\boldmath $r$}})$, 
where $\phi_{m,\sigma}({\mbox{\boldmath $r$}})$ corresponds to 
the final state in the s--d scattering process 
(see eqs. (\ref{22down}) - (\ref{2-2up})). 
For each $\rho_{\ell,\sigma}$, 
terms with $\rho_{s\sigma \to dM\downarrow}$ are written in the upper line, 
while those with $\rho_{s\sigma \to dM\uparrow}$ 
are given in the lower line. 
For each line, 
the summation of the s--d scattering terms is written 
in the right-hand column, 
where $\rho_{s\sigma \to dM\varsigma}$ is put to be 
$\rho_{s\sigma \to dM\varsigma}=\rho_{s\sigma \to d\varsigma}$. 
}
\begin{tabular}{lllll}
\hline \\[-0.2cm]
& $m=-2$ & $m=0$ & $m=2$ &  Summation \\\\[-0.2cm] \hline \\[-0.2cm]
$\rho_{\parallel,\uparrow}$ & 
& $2 \gamma \rho_{s\uparrow \to d1\downarrow}$ & & $2 \gamma \rho_{s\uparrow \to d\downarrow}$\\\\[-0.2cm]
& & $(1 - 2 \gamma) \rho_{s\uparrow \to d0\uparrow}$ & & $(1 - 2 \gamma) \rho_{s\uparrow \to d\uparrow}$\\\\
$\rho_{\parallel,\downarrow}$ & &  
$(1 - 2 \gamma) \rho_{s\downarrow \to d0\downarrow}$ & & $(1 - 2 \gamma) \rho_{s\downarrow \to d\downarrow}$\\\\[-0.2cm]
& & $2 \gamma \rho_{s\downarrow \to d-1\uparrow}$ & & $2 \gamma \rho_{s\downarrow \to d\uparrow}$\\\\
$\rho_{\perp,\uparrow}$ & 
$\frac{\gamma}{2} \rho_{s\uparrow \to d-1\downarrow}$ 
& $\frac{\gamma}{2} \rho_{s\uparrow \to d1\downarrow}$ & & $\gamma \rho_{s\uparrow \to d\downarrow}$ \\\\[-0.2cm]
& $\frac{3}{8}\left( 1 - \frac{4}{3}\gamma \right) 
\rho_{s\uparrow \to d-2\uparrow}$ & $\frac{1}{4}\left( 1 - 2\gamma \right) 
\rho_{s\uparrow \to d0\uparrow}$ & $\frac{3}{8} 
\rho_{s\uparrow \to d2\uparrow}$ & $(1 -\gamma) \rho_{s\uparrow \to d \uparrow}$ \\\\
$\rho_{\perp,\downarrow}$ & $\frac{3}{8} 
\rho_{s\downarrow \to d-2\downarrow}$ & $\frac{1}{4}\left( 1 - 2\gamma \right) 
\rho_{s\downarrow \to d0\downarrow}$ & $\frac{3}{8}\left( 1 - \frac{4}{3}\gamma \right) 
\rho_{s\downarrow \to d2\downarrow}$ & $(1 -\gamma) \rho_{s\downarrow \to d \downarrow}$\\\\[-0.2cm]
& & $\frac{\gamma}{2} \rho_{s\downarrow \to d-1\uparrow}$ & $\frac{\gamma}{2} \rho_{s\downarrow \to d1\uparrow}$ & $\gamma \rho_{s\downarrow \to d\uparrow}$\\\\[-0.2cm]
\hline 
\end{tabular}
\label{tab3}
\end{table*}

When the $M$ dependence of 
$D_{M\varsigma}^{(d)}$ in eq. (\ref{1/tau_s_dM}) is ignored 
in a conventional manner,\cite{Campbell} 
eqs. (\ref{rho_j}) - (\ref{rho_j1}) become
\begin{eqnarray}
\label{rho_j_s}
&&\rho_{\parallel, \uparrow} = \rho_{s\uparrow} 
+ 2 \gamma \rho_{s\uparrow \to d \downarrow} 
+ (1 - 2 \gamma) \rho_{s\uparrow \to d\uparrow},  \\
\label{rho_j2}
&&\rho_{\parallel, \downarrow} = \rho_{s\downarrow} 
+(1 - 2 \gamma) \rho_{s\downarrow \to d\downarrow} 
+ 2 \gamma \rho_{s\downarrow \to d\uparrow},  \\
\label{rho_j3}
&&\rho_{\perp, \uparrow} = \rho_{s\uparrow} 
+ \gamma \rho_{s\uparrow \to d\downarrow} 
+ (1 - \gamma) \rho_{s\uparrow \to d\uparrow}, \\
\label{rho_j1_s}
&&\rho_{\perp, \downarrow} = \rho_{s\downarrow} 
+(1 - \gamma) \rho_{s\downarrow \to d\downarrow} 
+ \gamma \rho_{s\downarrow \to d\uparrow}, 
\end{eqnarray}
respectively, 
with
\begin{eqnarray}
\label{rho_sd}
&&\hspace*{-0.8cm}\rho_{s\sigma \to d\varsigma} 
= \frac{m_\sigma^*}{n_\sigma e^2 \tau_{s\sigma \to d\varsigma}}, \\
\label{tau_sd}
&&\hspace*{-0.8cm}
\frac{1}{\tau_{s \sigma \to d \varsigma}}
=\frac{2\pi}{\hbar} n_{\rm imp} N_{\rm n}
|V_{s\sigma \to d\sigma}|^2 D_{\varsigma}^{(d)}, 
\end{eqnarray}
where $\gamma$, $\rho_{s\sigma}$, 
and $|V_{s\sigma \to d\sigma}|^2$ 
are given by eqs. (\ref{gamma}), (\ref{r_s_s}), 
and (\ref{V_ss-ds^2}), respectively. 
Here, $D_{\varsigma}^{(d)}$ is 
the DOS of each d state of the $\varsigma$ spin at $E_{\mbox{\tiny F}}$, 
where 
$D_{\varsigma}^{(d)}$ is set to be 
$D_{\varsigma}^{(d)}=D_{M \varsigma}^{(d)}$ 
by ignoring $M$ for $D_{M \varsigma}^{(d)}$ of eq. (\ref{D_Ms^d}).

\subsection{AMR ratio}
\label{AMR ratio}
Using eqs. 
(\ref{ratio}), (\ref{rho_j_ini}), and (\ref{rho_j_s}) - (\ref{rho_j1_s}), 
we obtain 
the general expression of the AMR ratio 
as
\begin{eqnarray}
\label{general AMR}
\frac{\Delta \rho}{\rho}
=\gamma \frac{A+B}{CD},
\end{eqnarray}
with
\begin{eqnarray}
&&\hspace*{-0.8cm}A=(\rho_{s\uparrow \to d\downarrow} 
- \rho_{s\uparrow \to d\uparrow}) \times \nonumber \\
&&\hspace*{-0.2cm}
\Bigg\{(\rho_{s\downarrow} + \rho_{s\downarrow \to d\downarrow})
(\rho_{s\downarrow} + \rho_{s\downarrow \to d\downarrow}
+\rho_{\downarrow \uparrow}
-\rho_{\uparrow \downarrow}) \nonumber \\
&&\hspace*{-0.2cm}
+ \left[
(1+ a )\rho_{\uparrow \downarrow}
+
(1+ a^{-1} )\rho_{\downarrow \uparrow}
\right](\rho_{s\downarrow}+ \rho_{s\downarrow \to d\downarrow}
+\rho_{\downarrow \uparrow})
\Bigg\},  \\
&&\hspace*{-0.8cm}B=(\rho_{s\downarrow \to d\uparrow} 
- \rho_{s\downarrow \to d\downarrow}) \times \nonumber \\
&&\hspace*{-0.2cm}
\Bigg\{(\rho_{s\uparrow} + \rho_{s\uparrow \to d\uparrow})
(\rho_{s\uparrow} + \rho_{s\uparrow \to d\uparrow}
+\rho_{\uparrow \downarrow}
-\rho_{\downarrow \uparrow}
) \nonumber \\
&&\hspace*{-0.2cm}
+ \left[
(1+ a )\rho_{\uparrow \downarrow}
+(1+ a^{-1})\rho_{\downarrow \uparrow}
\right](\rho_{s\uparrow}+ \rho_{s\uparrow \to d\uparrow}
+\rho_{\uparrow \downarrow})
\Bigg\},  \\
&&\hspace*{-0.8cm}C=
(\rho_{s\uparrow} + \rho_{s\uparrow \to d\uparrow})
(\rho_{s\downarrow} + \rho_{s\downarrow \to d\downarrow} 
+ \rho_{\downarrow \uparrow})
+ (\rho_{s\downarrow} + \rho_{s\downarrow \to d\downarrow})
\rho_{\uparrow \downarrow},\nonumber \\\\
&&\hspace*{-0.8cm}D=\rho_{s\uparrow}+\rho_{s\uparrow \to d\uparrow}
+\rho_{s\downarrow}+\rho_{s\downarrow \to d\downarrow}+
( 1+ a ) \rho_{\uparrow \downarrow}
+( 1+a^{-1}) \rho_{\downarrow \uparrow},
\nonumber \\\\
\label{rho_sf}
&&\hspace*{-0.8cm}\rho_{\sigma \sigma'}=
\frac{m_\sigma^*}{n_\sigma e^2 \tau_{\sigma \sigma'}}, 
\end{eqnarray}
where 
$\rho_{\sigma \sigma'}$ ($\sigma \ne \sigma'$) is a resistivity 
due to the spin-flip scattering process 
from the $\sigma$ spin state to the $\sigma'$ spin state, 
and $\tau_{\sigma \sigma'}$ 
is a relaxation time of this scattering. 
Here, 
$\tau_{\sigma \sigma'}$ has been assumed to be independent of 
the configuration 
(see $\tau_{\ell,\sigma \sigma'}$ of eq. (\ref{rho_sf0})).

\subsection{Feature of the AMR effect}
\label{Origin of AMR}

On the basis of the above results, 
we introduce a certain quantity based on the AMR ratio 
and then 
reveal a feature of the AMR effect. 
In particular, we find 
that the sign of the AMR ratio is determined by 
the increase or decrease 
of ``existence probabilities of the specific d orbitals'' 
due to the spin--orbit interaction. 
In addition, we roughly determine 
a relation between the sign of the AMR ratio and 
the scattering process.

\subsubsection{$Z_{\sigma;\varsigma}$}
Taking into account 
the after-mentioned (i) - (iii), 
we introduce the quantity based on the AMR ratio. 
Here, the AMR ratio 
reflects 
the difference of ``changes of the d orbitals 
due to the spin--orbit interaction'' 
between different $m$'s, 
where $m$ is the magnetic quantum number of the d orbital 
$\phi_{m,\sigma}({\mbox{\boldmath $r$}})$ of eq. (\ref{d_orbital}). 
Such a quantity $Z_{\sigma;\varsigma}$ is written as
\begin{eqnarray}
\label{AAA}
&&\hspace*{-1.2cm}Z_{\sigma;\varsigma}=
X(0,\sigma;\varsigma) - Y_{\sigma;\varsigma}, \\
\label{YYY}
&&\hspace*{-1.2cm}Y_{\sigma;\varsigma}= 
\frac{1}{4}X(0,\sigma;\varsigma)+\frac{3}{8}X(2,\sigma;\varsigma)
+\frac{3}{8}X(-2,\sigma;\varsigma), \\
\label{XXX}
&&\hspace*{-1.2cm}X(m,\sigma;\varsigma)=\sum_{M=-2}^{2} 
\left( \left| \int \phi_{m,\sigma}^*({\mbox{\boldmath $r$}}) 
\Phi_{M,\varsigma}^{(d)}({\mbox{\boldmath $r$}})
{\rm d}{\mbox{\boldmath $r$}}\right|^2 
- \delta_{m,M}\delta_{\sigma,\varsigma} \right), 
\end{eqnarray}
where 
$\Phi_{M,\varsigma}^{(d)}({\mbox{\boldmath $r$}})$ 
is given by eqs. (\ref{22down}) - (\ref{2-2up}). 
Roughly speaking, 
$Z_{\sigma;\varsigma}$ may correspond to 
the numerator of the AMR ratio of eq. (\ref{ratio}), 
$\rho_{\parallel}-\rho_\perp$. 
In particular, 
$\sum_{M=-2}^{2} 
\left| \int \phi_{0,\sigma}^*({\mbox{\boldmath $r$}}) 
\Phi_{M,\varsigma}^{(d)}({\mbox{\boldmath $r$}})
{\rm d}{\mbox{\boldmath $r$}}\right|^2$ 
in $X(0,\sigma;\varsigma)$ and 
$\sum_{M=-2}^{2} \left[
\frac{1}{4} \left| \int \phi_{0,\sigma}^*({\mbox{\boldmath $r$}}) 
\Phi_{M,\varsigma}^{(d)}({\mbox{\boldmath $r$}})
{\rm d}{\mbox{\boldmath $r$}}\right|^2 +
\frac{3}{8}\sum_{m=\pm 2}
\left| \int \phi_{m,\sigma}^*({\mbox{\boldmath $r$}}) 
\Phi_{M,\varsigma}^{(d)}({\mbox{\boldmath $r$}})
{\rm d}{\mbox{\boldmath $r$}}\right|^2 \right]$ 
in $Y_{\sigma;\varsigma}$ 
may be related to $\rho_{\parallel}$ and $\rho_\perp$, respectively. 
This $X(m,\sigma;\varsigma)$ represents 
the change 
of ``the existence probability of the d orbital 
of $m$ and $\sigma$'' 
due to the spin--orbit interaction. 
Here, 
$\left| \int \phi_{m,\sigma}^*({\mbox{\boldmath $r$}}) 
\Phi_{M,\varsigma}^{(d)}({\mbox{\boldmath $r$}})
{\rm d}{\mbox{\boldmath $r$}}\right|^2$ 
is adopted on the basis of 
the scattering rate in $\rho_{s \sigma \to d \varsigma}$ 
(see Appendix \ref{s-d}), 
and $\sum_{M=-2}^2$ comes from 
that in the right-hand side of eq. (\ref{tau_inv}). 
In addition, 
1/4, 3/8, and 3/8 in $Y_{\sigma;\varsigma}$ 
correspond to 
the coefficients of $|V_{s\sigma \to d\sigma}|^2$ of eq. (\ref{V_ss-ds^2}) 
in the scattering rates of $m$=0, 2, and $-$2, respectively 
(see Appendix \ref{matrix}). 
Such $Z_{\sigma;\varsigma}$ and $X(m,\sigma;\varsigma)$ 
have been based on 
the following (i) - (iii): 
\begin{itemize}
\item[(i)] 
By comparing eqs. (\ref{rho_j_s}) and (\ref{rho_j3}) 
or eqs. (\ref{rho_j2}) and (\ref{rho_j1_s}), 
we find that the AMR effect 
arises from 
the difference of s--d scattering terms 
between $\parallel$ and $\perp$ configurations. 
All the s--d scattering terms 
with $\rho_{s \sigma \to d \varsigma}$ 
in eqs. (\ref{rho_j_s}) - (\ref{rho_j1_s}) 
are listed in Table \ref{tab3}, 
where terms with $\rho_{s\sigma \to d M \varsigma}$ 
in eqs. (\ref{rho_j}) - (\ref{rho_j1}) 
are also listed. 
The s--d scattering terms 
in $\rho_{\parallel,\sigma}$ 
originate from 
a transition from the plane wave to the d orbital of $m$=0, 
$\phi_{0,\sigma}({\mbox{\boldmath $r$}})$ 
(see Appendix \ref{matrix}).\cite{Campbell} 
In contrast, 
the s--d scattering terms 
in $\rho_{\perp,\sigma}$ are due to 
transitions from the plane wave to the d orbitals of $m=\pm 2$ and 0, 
$\phi_{\pm 2,\sigma}({\mbox{\boldmath $r$}})$ and 
$\phi_{0,\sigma}({\mbox{\boldmath $r$}})$. 
The d orbitals of $m=\pm 1$, 
$\phi_{\pm 1,\sigma}({\mbox{\boldmath $r$}})$, 
give no contribution to 
$\rho_{\parallel,\sigma}$ and $\rho_{\perp,\sigma}$. 
\item[(ii)] In such s--d scattering terms, 
only terms with $\gamma \rho_{s \to d \varsigma}$ 
actually contribute to the AMR effect. 
The $\gamma \rho_{s \to d \varsigma}$ terms are 
induced by the spin--orbit interaction. 
As found from eqs. (\ref{rho_j_s}) - (\ref{rho_j1_s}) or 
the summation in Table \ref{tab3}, 
the case of $\gamma \ne 0$ leads to 
$\rho_{\parallel,\uparrow} \ne \rho_{\perp,\uparrow}$ and 
$\rho_{\parallel,\downarrow} \ne \rho_{\perp,\downarrow}$, 
while 
the case of $\gamma$=0 leads to 
$\rho_{\parallel,\uparrow}=\rho_{\perp,\uparrow}$ and 
$\rho_{\parallel,\downarrow}=\rho_{\perp,\downarrow}$. 
\item[(iii)] The $\gamma \rho_{s \to d \varsigma}$ terms 
stem from the change of the d orbitals 
due to the spin--orbit interaction. 
The d orbital 
is slightly changed 
by the spin-mixing term $(\lambda/2)(L_+ S_- + L_- S_+)$ 
in the spin--orbit interaction. 
It is noteworthy that 
the contributions due to 
the $\lambda L_zS_z$ term are eliminated by ignoring terms 
higher than the second order of $\lambda/H_{\rm ex}$ (see Appendix \ref{WF}). 
\end{itemize}

\begin{table}[ht]
\caption{
Change of the d orbital due to the spin--orbit interaction 
$X(m,\sigma;\varsigma)$ of eq. (\ref{XXX}) ($m$=0, $\pm 2$), 
$Z_{\sigma;\varsigma}$ of eq. (\ref{AAA}), 
and $s \sigma \to d \varsigma$. 
Here, terms higher than the second order of $\epsilon$ (=$\lambda/H_{\rm ex}$) 
have been ignored. 
In addition, $\sigma$ and $\varsigma$ of $s \sigma \to d \varsigma$ 
are extracted from $X(m,\sigma;\varsigma)$. 
Since $Z_{\sigma;\varsigma}$ may correspond approximately to 
$\rho_{\parallel}-\rho_\perp$ of the AMR ratio, 
we can roughly determine a relation between 
the sign of the AMR ratio and the s--d scattering process. 
}
\begin{tabular}{lcccc}
\hline \\[-0.2cm]
($\sigma$, $\varsigma$) & 
($\downarrow$, $\downarrow$) & ($\uparrow$, $\downarrow$) 
& ($\uparrow$, $\uparrow$) & ($\downarrow$, $\uparrow$) \\
$s \sigma \to d \varsigma$ & $s \downarrow \to d \downarrow$ & $s \uparrow \to d \downarrow$ 
& $s \uparrow \to d \uparrow$ & $s \downarrow \to d \uparrow$ \\\\[-0.2cm]
\hline \\[-0.2cm]
$X(2,\sigma;\varsigma)$ & $-\epsilon^2$ & 0 & 0 & $\epsilon^2$ \\\\[-0.2cm]
$X(0,\sigma;\varsigma)$ & $-\frac{3 \epsilon^2}{2}$ & 
$\frac{3\epsilon^2}{2}$ & $- \frac{3 \epsilon^2}{2}$ & $\frac{3 \epsilon^2}{2}$ \\\\[-0.2cm]
$X(-2,\sigma;\varsigma)$ & 0 & $\epsilon^2$ & $-\epsilon^2$ & 0 \\\\[-0.2cm]
$Z_{\sigma;\varsigma}$ 
& $-\frac{3\epsilon^2}{4}~(<0)$ & $\frac{3\epsilon^2}{4}~(>0)$ 
& $-\frac{3\epsilon^2}{4}~(<0)$ & $\frac{3\epsilon^2}{4}~(>0)$ \\\\[-0.2cm]
\hline
\end{tabular}
\label{tab_wf_diff}
\end{table}

\subsubsection{Sign of $Z_{\sigma;\varsigma}$ and s--d scattering}
\label{sign_Y}
\begin{figure}[ht]
\begin{center}
\includegraphics[width=0.45\linewidth]{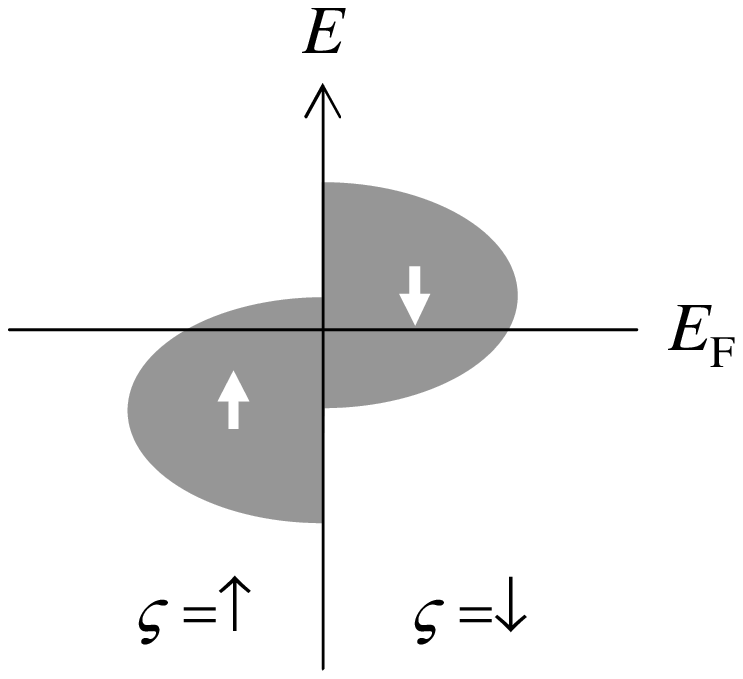} \hspace{0.5cm}
\includegraphics[width=0.45\linewidth]{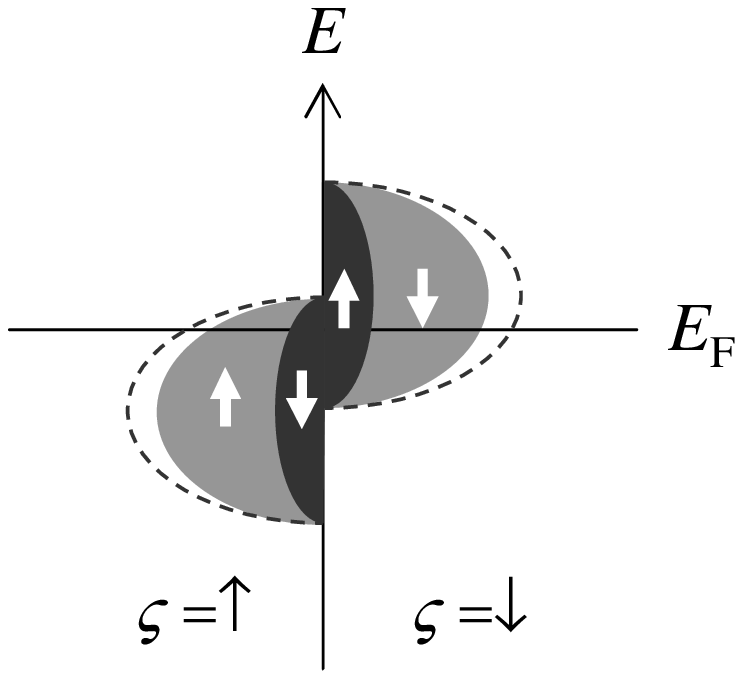}  \\
\hspace{-0.5cm}
{\footnotesize (a) $\lambda=0$ \hspace{3.5cm} (b) $\lambda \ne 0$}\\
\caption{
Effect of the spin--orbit interaction 
on the DOS of a typical d band. 
(a) The case of $\lambda=0$. 
Here, $\lambda$ is the spin--orbit coupling constant (see eq. (\ref{LS})). 
(b) The case of $\lambda \ne 0$. 
In (b), the partial DOS of 
the pure d orbital with $\phi_{m,\varsigma}$ 
is indicated by the gray areas, 
while 
that of the other d orbital with $\phi_{m,\sigma}$ 
is shown by the black areas, 
where $\sigma \ne \varsigma$. 
The orbital $\phi_{m,\sigma}$ or $\phi_{m,\varsigma}$ 
is given by eq. (\ref{d_orbital}), 
where $\varsigma$ denotes 
the spin of the dominant state in the spin-mixed state. 
In (b), 
a slight amount of $\phi_{m,\sigma}$ is mixed with $\phi_{m,\varsigma}$. 
This mixing reduces the existence probability of $\phi_{m,\varsigma}$ 
(see Appendix \ref{WF}). 
The dashed curves in (b) represent the shape of the DOS of (a). 
}
\label{dos_sign}
\end{center}
\end{figure}

In order to obtain $Z_{\sigma;\varsigma}$, 
we first investigate $X(m,\sigma;\varsigma)$ of eq. (\ref{XXX}). 
As seen from 
Table \ref{tab_wf_diff}, 
$X(2,\downarrow;\downarrow)$, 
$X(0,\downarrow;\downarrow)$, 
$X(0,\uparrow;\uparrow)$, and 
$X(-2,\uparrow;\uparrow)$ become negative, 
while 
$X(0,\uparrow;\downarrow)$, 
$X(-2,\uparrow;\downarrow)$, 
$X(2,\downarrow;\uparrow)$, and 
$X(0,\downarrow;\uparrow)$ are positive. 
Here, 
the former $X(m,\sigma;\varsigma)$'s 
are obtained from 
the first terms in the right-hand sides 
of eqs. (\ref{22down}) - (\ref{2-1down}) 
and (\ref{21up}) - (\ref{2-2up}). 
The latter $X(m,\sigma;\varsigma)$'s are obtained from 
the second terms in them. 
The negative sign of the former 
means that 
the existence probability of the pure d orbital of $m$ 
decreases owing to hybridization with the other d orbital 
in the presence of the spin--orbit interaction 
(see the gray areas in Fig. \ref{dos_sign}(b)). 
In contrast, 
the positive sign of the latter 
represents 
the addition of the existence probability of the other d orbital 
(see the black areas in Fig. \ref{dos_sign}(b)). 
Note that 
the spin of the other d orbital is opposite to 
that of the pure d orbital 
under the influence of $S_\pm$ in the spin-mixing term.

\begin{figure}[ht]
\begin{center}
\includegraphics[width=0.8\linewidth]{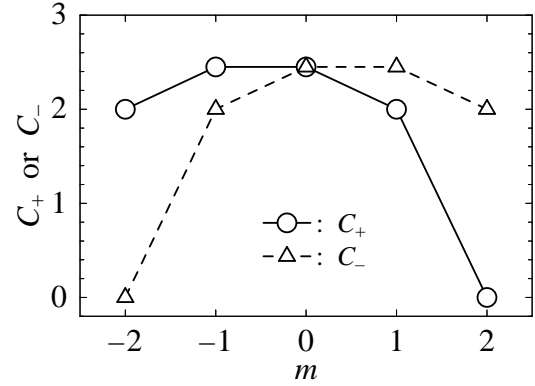} 
\caption{
$m$ dependence of $C_\pm=\sqrt{(L\mp m)(L\pm m +1)}$ 
with $L$=2 and $m=-2$, $-1$, 0, 1, 2. 
Here, we have 
$L_\pm \phi_{m,\sigma}({\mbox{\boldmath $r$}})$=
$C_\pm \phi_{m \pm 1,\sigma}({\mbox{\boldmath $r$}})$, 
where 
$\phi_{m,\sigma}({\mbox{\boldmath $r$}})$ 
is given by eq. (\ref{d_orbital}). 
}
\label{ls}
\end{center}
\end{figure}

Furthermore, 
we find a relation of 
$|X(0,\sigma;\varsigma)|>|X(\pm 2,\sigma;\varsigma)|$ 
for each set of $\sigma$ and $\varsigma$. 
The relation 
is attributed to the mixing effect of 
the d orbitals due to $L_\pm=L_x \pm {\rm i} L_y$ 
in the spin-mixing term. 
This effect is verified from the $m$ dependence of 
$C_\pm$ (=$\sqrt{(L\mp m)(L\pm m +1)}$) in Fig. \ref{ls}, 
where 
$L_\pm \phi_{m,\sigma}({\mbox{\boldmath $r$}})$=
$C_\pm \phi_{m \pm 1,\sigma}({\mbox{\boldmath $r$}})$ 
and $L$=2. 
The coefficient $C_\pm$ at $m=0$ becomes larger than that at $m=\pm 2$; 
that is, 
the mixing effect at $m=0$ is larger than that at $m=\pm 2$.

Using such $X(m,\sigma;\varsigma)$'s, 
we can obtain $Z_{\sigma;\varsigma}$ of eq. (\ref{AAA}) 
as shown in Table \ref{tab_wf_diff}. 
In addition, we find 
the following relation between 
the sign of $Z_{\sigma;\varsigma}$ 
and 
the s--d scattering process $s \sigma \to d \varsigma$: 
$Z_{\downarrow;\downarrow} < 0$ for $s\downarrow \to d \downarrow$, 
$Z_{\uparrow;\downarrow} > 0$ for $s\uparrow \to d \downarrow$, 
$Z_{\uparrow;\uparrow} < 0$ for $s \uparrow \to d \uparrow$, and 
$Z_{\downarrow;\uparrow} > 0$ for $s \downarrow \to d \uparrow$ 
(see Table \ref{tab_wf_diff}). 
Here, $s \sigma \to d \varsigma$ 
indicates that 
the conduction electron of the $\sigma$ spin 
is scattered into $\phi_{m,\sigma}({\mbox{\boldmath $r$}})$ 
in $\Phi_{M,\varsigma}^{(d)}({\mbox{\boldmath $r$}})$ of $M=-2$ - 2. 
The $\sigma$ spin is conserved in the scattering process. 
The spins $\sigma$ and $\varsigma$ of 
$s \sigma \to d \varsigma$ are extracted from 
$X(m,\sigma;\varsigma)$. 
Roughly speaking, 
the negative sign of $Z_{\downarrow;\downarrow}$ and $Z_{\uparrow;\uparrow}$ 
originates from the decrease of 
the existence probability of the pure d orbital, 
while 
the positive sign of $Z_{\uparrow;\downarrow}$ and $Z_{\downarrow;\uparrow}$
is due to 
the addition of the existence probability of the other d orbital 
(see Fig. \ref{dos_sign}(b)).

Since $Z_{\sigma;\varsigma}$ may correspond approximately to 
$\rho_{\parallel}-\rho_\perp$ of the AMR ratio, 
we can roughly determine the relation between 
the sign of the AMR ratio and 
the s--d scattering process. 
Namely, when the dominant s--d scattering process 
is $s\downarrow \to d \downarrow$ or $s\uparrow \to d \uparrow$, 
the AMR ratio tends to become negative. 
In contrast, 
when the dominant s--d scattering process 
is $s\uparrow \to d \downarrow$ or $s\downarrow \to d \uparrow$, 
the AMR ratio tends to be positive. 
Such a relation 
agrees with a trend for real materials, 
as will be shown in \S \ref{AMR and scattering}.

\begin{figure}[ht]
\begin{center}
\includegraphics[width=0.92\linewidth]{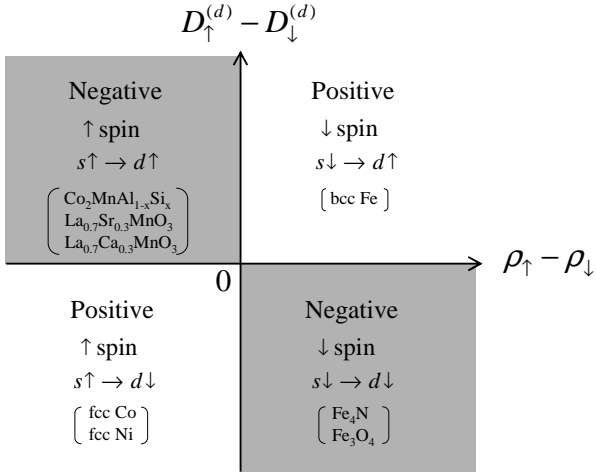} 
\caption{
Sign of the AMR ratio $\Delta \rho/\rho$ 
and the dominant s--d scattering process $s \sigma \to d \varsigma$ 
in a simple model with 
$n_\uparrow=n_\downarrow$, 
$m_\uparrow^*=m_\downarrow^*$, 
$V_{s\uparrow \to d \uparrow}=V_{s\downarrow \to d \downarrow}$, 
and $\rho_{\uparrow \downarrow}=\rho_{\downarrow \uparrow}$=0. 
They are 
shown in the $(\rho_{\uparrow} - \rho_{\downarrow})$-$(D_{\uparrow}^{(d)}- D_{\downarrow}^{(d)})$ plane, 
where 
$\rho_\sigma=\rho_{s\sigma} + \rho_{s \sigma \to d \sigma}$. 
In each quadrant, 
the first, second, and third lines from the top 
denote the sign of the AMR ratio, 
the spin of the conduction electrons 
contributing dominantly to the transport, 
and $s \sigma \to d \varsigma$, respectively. 
Here, the sign of the AMR ratio can be judged from eq. (\ref{scatt_sign3}). 
In addition, 
$s \sigma \to d \varsigma$ 
is extracted from
$\rho_{s \sigma \to d \varsigma}$, 
which contributes dominantly to the sign of the AMR ratio. 
Namely, this $\rho_{s \sigma \to d \varsigma}$ 
corresponds to 
the greater of 
$\rho_{s\downarrow \to d \uparrow}$ and $\rho_{s\downarrow \to d \downarrow}$ 
in the case of $\rho_\uparrow > \rho_\downarrow$ 
and the greater of 
$\rho_{s\uparrow \to d \downarrow}$ and $\rho_{s\uparrow \to d \uparrow}$ 
in the case of $\rho_\uparrow < \rho_\downarrow$. 
Furthermore, 
materials in Table \ref{tab1} 
are assigned to 
the respective quadrants 
on the basis of results of 
(i) - (v) of \S \ref{appl_sign}. 
}
\label{fig_region}
\end{center}
\end{figure}

\subsection{Sign of the AMR ratio and s--d scattering of real material}
\label{AMR and scattering}

Within a unified framework, 
we find 
the sign of the AMR ratio and the dominant scattering process 
of each material in Table \ref{tab1}.
We here utilize 
$\rho_{s\downarrow}/\rho_{s\uparrow}$ and 
$D_{\uparrow}^{(d)}/D_{\downarrow}^{(d)}$ 
from Table \ref{tab1}.

\subsubsection{A simple model}
Toward the unified framework, 
we present a simple model with 
$n_\uparrow=n_\downarrow$ ($\ne 0$), 
$m_\uparrow^*=m_\downarrow^*$, 
$V_{s\uparrow \to d \uparrow}=V_{s\downarrow \to d \downarrow}$, 
and $\rho_{\uparrow \downarrow}=\rho_{\downarrow \uparrow}$=0. 
This model has a relation of 
$\rho_{\parallel,\uparrow}+\rho_{\parallel,\downarrow}=\rho_{\perp,\uparrow}+\rho_{\perp,\downarrow}$ from eqs. (\ref{rho_j_s}) - (\ref{rho_j1_s}). 
The AMR ratio of eq. (\ref{ratio}) is then expressed as
\begin{eqnarray}
\label{scatt_sign}
\frac{\Delta \rho}{\rho} = 
\frac{ \rho_{\parallel \uparrow} \rho_{\parallel \downarrow} 
- \rho_{\perp \uparrow} \rho_{\perp \downarrow}}
{\rho_{\perp \uparrow}\rho_{\perp \downarrow}}.
\end{eqnarray}
Using eqs. (\ref{rho_j_s}) - (\ref{tau_sd}), 
eq. (\ref{scatt_sign}) is rewritten as
\begin{eqnarray}
\label{scatt_sign1}
&&
\hspace{-0.8cm}
\frac{\Delta \rho}{\rho}
= \gamma 
\left( \frac{\rho_{s\downarrow \to d \uparrow}-\rho_{s\downarrow \to d \downarrow}}{\rho_\downarrow} + \frac{\rho_{s\uparrow \to d \downarrow}-\rho_{s\uparrow \to d \uparrow}}{\rho_\uparrow} \right) \\
\label{scatt_sign2}
&&
\hspace{-0.2cm}
\propto \gamma 
\left( \frac{D_{\uparrow}^{(d)}-D_{\downarrow}^{(d)}}{\rho_\downarrow}+
\frac{D_{\downarrow}^{(d)}-D_{\uparrow}^{(d)}}{\rho_\uparrow} \right) \\
\label{scatt_sign3}
&&
\hspace{-0.2cm}
= \gamma 
\left( D_{\uparrow}^{(d)}-D_{\downarrow}^{(d)} \right)
\left( \frac{1}{\rho_\downarrow} - \frac{1}{\rho_\uparrow} \right), 
\end{eqnarray}
with 
\begin{eqnarray}
\label{r_sigma}
\rho_\sigma=\rho_{s\sigma} + \rho_{s \sigma \to d \sigma}, 
\end{eqnarray}
where $\rho_{s\sigma}$ is given by eq. (\ref{r_s_s}), 
and $\rho_{s \sigma \to d \sigma}$ is written by eq. (\ref{rho_sd}) 
with $\varsigma=\sigma$. 
This $\rho_\sigma$ corresponds approximately to the resistivity of 
the $\sigma$ spin for a system with no spin--orbit interaction, i.e., 
eqs. (\ref{rho_j_s}) - (\ref{rho_j1_s}) 
with $\lambda$=0. 
Note here that 
$D_\sigma^{(d)}$ in $\rho_{s \sigma \to d \sigma}$ in eq. (\ref{r_sigma}) 
actually contains the effect of the spin--orbit interaction, 
as found from eq. (\ref{D_Ms^d}).

From eqs. (\ref{scatt_sign1}) - (\ref{scatt_sign3}), 
we can find the relation 
between the sign of the AMR ratio and the dominant s--d scattering process. 
First, the sign of the AMR ratio is shown 
in each quadrant of the 
$(\rho_{\uparrow} - \rho_{\downarrow})$-$(D_{\uparrow}^{(d)} - D_{\downarrow}^{(d)})$ plane of Fig. \ref{fig_region}. 
The AMR ratio becomes positive 
in the case of $\rho_{\uparrow} > \rho_{\downarrow}$ and 
$D_{\uparrow}^{(d)} > D_{\downarrow}^{(d)}$ or 
in the case of 
$\rho_{\uparrow} < \rho_{\downarrow}$ and 
$D_{\uparrow}^{(d)} < D_{\downarrow}^{(d)}$. 
In contrast, 
the AMR ratio is negative 
in the case of $\rho_{\uparrow} > \rho_{\downarrow}$ and 
$D_{\uparrow}^{(d)} < D_{\downarrow}^{(d)}$ or 
in the case of 
$\rho_{\uparrow} < \rho_{\downarrow}$ and 
$D_{\uparrow}^{(d)} > D_{\downarrow}^{(d)}$. 
Here, 
the case of 
$\rho_\uparrow > \rho_\downarrow$ 
($\rho_\uparrow < \rho_\downarrow$) 
shows that 
the down spin electrons (the up spin electrons) 
contribute dominantly to the transport. 
Furthermore, 
the dominant s--d scattering process is indicated by 
$s \sigma \to d \varsigma$ 
in each quadrant of Fig. \ref{fig_region}. 
The process $s \sigma \to d \varsigma$ is 
extracted from
$\rho_{s \sigma \to d \varsigma}$, 
which contributes dominantly to the sign of the AMR ratio. 
Concretely speaking, 
this $\rho_{s \sigma \to d \varsigma}$ 
corresponds to 
the greater of 
$\rho_{s\downarrow \to d \uparrow}$ and $\rho_{s\downarrow \to d \downarrow}$ 
in the case of $\rho_\uparrow > \rho_\downarrow$ 
and the greater of 
$\rho_{s\uparrow \to d \downarrow}$ and $\rho_{s\uparrow \to d \uparrow}$ 
in the case of $\rho_\uparrow < \rho_\downarrow$. 
It is also noteworthy that the relation 
in Fig. \ref{fig_region} 
is consistent with the result 
in \S \ref{sign_Y} or Table \ref{tab_wf_diff}.

\subsubsection{Application to materials}
\label{appl_sign}
Applying 
$\rho_{s\downarrow}/\rho_{s\uparrow}$ and 
$D_\uparrow^{(d)}/D_\downarrow^{(d)}$ of Table \ref{tab1} 
to the results of Fig. \ref{fig_region}, 
we can roughly determine 
the dominant s--d scattering and the sign of the AMR ratio of 
each material. 
The determined signs 
agree with 
the experimental results of Table \ref{tab1}. 
The details are written as follows: 
\begin{itemize}
\item[(i)] bcc Fe \\
The dominant s--d scattering is 
$s\downarrow \to d \uparrow$ 
because of 
$D_\uparrow^{(d)}>D_\downarrow^{(d)}$ and 
$\rho_{\uparrow} > \rho_{\downarrow}$. 
The AMR ratio is thus positive. 
Here, $\rho_{\uparrow} > \rho_{\downarrow}$ 
originates from 
$\rho_{s\uparrow} > \rho_{s\downarrow}$ 
and $\rho_{s\uparrow \to d \uparrow} > \rho_{s\downarrow \to d \downarrow}$ 
due to $D_\uparrow^{(d)}>D_\downarrow^{(d)}$. 
\item[(ii)] fcc Co and fcc Ni \\
The dominant s--d scattering is 
$s\uparrow \to d\downarrow$ 
because of 
$D_\uparrow^{(d)}<D_\downarrow^{(d)}$ and 
$\rho_{\uparrow} < \rho_{\downarrow}$. 
The AMR ratio is then positive. 
Here, $\rho_{\uparrow} < \rho_{\downarrow}$ 
is obtained from 
$\rho_{s\uparrow} < \rho_{s\downarrow}$ 
and $\rho_{s\uparrow \to d \uparrow} < \rho_{s\downarrow \to d \downarrow}$ 
due to $D_\uparrow^{(d)}<D_\downarrow^{(d)}$. 
\item[(iii)] Fe$_4$N\\
The dominant s--d scattering is $s\downarrow \to d \downarrow$ 
because of 
$D_\uparrow^{(d)}<D_\downarrow^{(d)}$ and 
$\rho_{\uparrow} > \rho_{\downarrow}$. 
The AMR ratio is thus negative. 
Here, $\rho_{\uparrow}>\rho_{\downarrow}$ 
mainly results from 
$\rho_{s\uparrow}/\rho_{s\downarrow} = (1.6 \times 10^{-3})^{-1}$ 
(see Table \ref{tab1}). 
The relation $\rho_{s\uparrow \to d \uparrow}$=0 
is assumed by considering that 
$D_\uparrow^{(d)}$ 
is considerably smaller than $D_\downarrow^{(d)}$, where 
it is reported that this model has $n_\sigma \ne 0$. 
In addition, 
we assume that 
$0.01 \lesssim \rho_{s\downarrow \to d \downarrow}/\rho_{s\uparrow}\lesssim 0.5$, which will be estimated in \S \ref{Strong FM}. 
\item[(iv)] Co$_2$MnAl$_{1-x}$Si$_x$, 
La$_{0.7}$Sr$_{0.3}$MnO$_3$, and La$_{0.7}$Ca$_{0.3}$MnO$_3$  \\
The dominant s--d scattering is $s\uparrow \to d \uparrow$ 
because of 
$D_\uparrow^{(d)}>D_\downarrow^{(d)}$ and 
$\rho_{\uparrow} < \rho_{\downarrow}$. 
The AMR ratio is thus negative. 
Here, $\rho_{\uparrow}<\rho_{\downarrow}$ 
mainly originates from 
$\rho_{s\downarrow}/\rho_{s\uparrow} \gtrsim 10^6$ 
(see (i) of \S \ref{suggest} or \S \ref{Evaluation of AMR ratio}). 
The relation 
$\rho_{s\downarrow \to d \downarrow}$=0 
is roughly set 
on the basis of $D_\downarrow^{(d)}\sim 0$, 
where $n_\sigma \ne 0$. 
In addition, 
we assume that 
$\rho_{s\uparrow \to d \uparrow} \sim \rho_{s\uparrow}$, which 
will be estimated in \S \ref{Evaluation of AMR ratio}. 
\item[(v)] Fe$_3$O$_4$\\
The dominant s--d scattering is $s\downarrow \to d \downarrow$ 
because of 
$D_\uparrow^{(d)}<D_\downarrow^{(d)}$ and 
$\rho_{\uparrow} > \rho_{\downarrow}$. 
The AMR ratio is then negative. 
Here, $\rho_{\uparrow}>\rho_{\downarrow}$ 
mainly stems from 
$\rho_{s\uparrow}/\rho_{s\downarrow} \gtrsim 10^6$ 
(see (i) of \S \ref{suggest} or \S \ref{Evaluation of AMR ratio}). 
The relation 
$\rho_{s\uparrow \to d \uparrow}$=0 
is roughly set 
on the basis of $D_\uparrow^{(d)}\sim 0$, 
where $n_\sigma \ne 0$. 
In addition, we assume that 
$\rho_{s\downarrow \to d \downarrow} \sim \rho_{s\downarrow}$, 
which 
will be estimated in \S \ref{Evaluation of AMR ratio}. 
Note that, 
in this system, 
the direction of each spin in (iv) 
has been reversed 
by taking into account the DOS 
of Fig. \ref{fig_dos}(e). 
\end{itemize}

\section{Application 1: Weak or Strong Ferromagnet}
\label{Applications}
On the basis of the theory of \S \ref{Theory}, 
we obtain the expressions of the AMR ratios of 
``bcc Fe of the weak ferromagnet'' and 
``fcc Co, fcc Ni, and Fe$_4$N of the strong ferromagnet.'' 
Using the expressions, we analyze their AMR ratios.

\subsection{AMR ratio}
From eq. (\ref{general AMR}), 
we first derive an expression of the AMR ratio 
of the weak or strong ferromagnet. 
The weak or strong ferromagnet has 
the sp band DOS of the up and down spins at $E_{\mbox{\tiny F}}$ 
(see Figs. \ref{fig_dos}(a), \ref{fig_dos}(b), and \ref{fig_dos}(c)). 
We thus use 
the conventional approximation 
in order to reduce parameters. 
Namely, 
we set 
$n_\uparrow=n_\downarrow$, $m_\uparrow^*=m_\downarrow^*$, 
$V_{s\uparrow \to d \uparrow}=V_{s\downarrow \to d \downarrow}$, 
and $\tau_{\uparrow \downarrow}=\tau_{\downarrow \uparrow}$. 
Meanwhile, 
the $\sigma$ dependence of 
$D_\sigma^{(s)}$ and 
the $\varsigma$ dependence of 
$D_\varsigma^{(d)}$ are taken into account 
(see eqs. (\ref{r_s_s}), (\ref{1/tau_s_s}), 
(\ref{rho_sd}), and (\ref{tau_sd})). 
The AMR ratio of eq. (\ref{general AMR}) is then given simply by
\begin{eqnarray}
\label{del_rho/rho0}
&&\hspace*{-1.cm}\frac{\Delta \rho }{\rho}= \nonumber \\
&&\hspace*{-1.cm}
\frac{ \gamma (\rho_{s \to d\uparrow} - 
\rho_{s \to d\downarrow})
(\rho_{s\uparrow} - \rho_{s\downarrow} 
+ \rho_{s \to d\uparrow} - 
\rho_{s \to d\downarrow})}
{(\rho_{s\uparrow}+\rho_{s \to d\uparrow})
(\rho_{s\downarrow}+\rho_{s \to d\downarrow})
+ \rho_{\uparrow \downarrow} 
(\rho_{s\uparrow} + \rho_{s\downarrow}
+\rho_{s \to d\uparrow}
+\rho_{s \to d\downarrow})}, \nonumber \\
\end{eqnarray}
where 
\begin{eqnarray}
\label{rho_ss usual}
&&\rho_{s\sigma}=\frac{m^*}{n e^2 \tau_{s\sigma}}, \\
\label{rho_sd usual}
&&\rho_{s \to d \varsigma}=\frac{m^*}{n e^2 \tau_{s \to d\varsigma}}. 
\end{eqnarray}
Here, we have 
$m_\sigma^* \equiv m^*$, $n_\sigma \equiv n$, 
and $\tau_{s\sigma \to d\varsigma} \equiv \tau_{s \to d\varsigma}$, 
where 
$1/\tau_{s \sigma} \propto D_\sigma^{(s)}$ and 
$1/\tau_{s \to d\varsigma} \propto D_\varsigma^{(d)}$. 
In addition, 
$\rho_{\sigma \sigma'}$ of eq. (\ref{rho_sf}) is rewritten by 
$\rho_{\sigma \sigma'}=m^*/(n e^2 \tau_{\sigma \sigma'})$. 
It is noteworthy that 
$\rho_{\uparrow \downarrow}$ has no influence on the sign of the AMR ratio 
of eq. (\ref{del_rho/rho0}). 
Also, 
eq. (\ref{del_rho/rho0}) with $\rho_{\uparrow \downarrow}$=0 
corresponds to 
an expression of the AMR ratio obtained by Malozemoff.\cite{Malozemoff1}

\subsection{Weak ferromagnet: Fe}
\label{Weak ferromagnet}
Using eq. (\ref{del_rho/rho0}), 
we analyze the AMR ratio of bcc Fe of the weak ferromagnet. 
Here, 
$\rho_{s \to d \uparrow}/\rho_{s \to d \downarrow}$ 
(=$D_\uparrow^{(d)}/D_\downarrow^{(d)}$) 
is assumed to be 
$\rho_{s \to d \uparrow}/\rho_{s \to d \downarrow}$=2.0 
on the basis of 
$D_\uparrow^{(d)}/D_\downarrow^{(d)}$=2.0 of Table \ref{tab1}.\cite{Wang} 
The constant 
$\gamma$ is chosen to be $\gamma$=0.01 as a typical value. 
Meanwhile, 
we ignore $\rho_{\uparrow \downarrow}$ 
which does not change the sign of the AMR ratio. 
It is noteworthy that 
the spin-dependent disorder\cite{FeNi,CoPd}, 
which gives rise to the spin-flip scattering, 
may be 
weak for the present ferromagnets with nonmagnetic impurities.

\begin{figure}[ht]
\begin{center}
\includegraphics[width=0.8\linewidth]{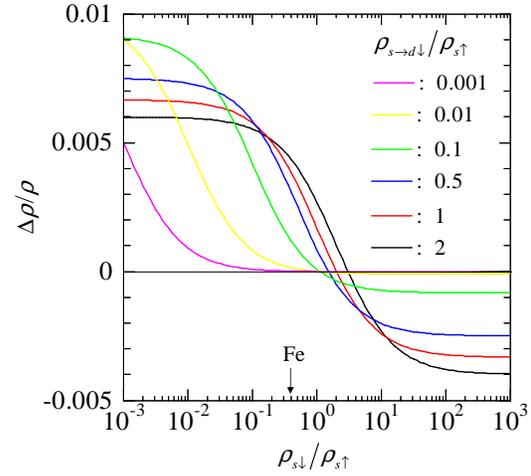} 
\caption{
(Color) 
Quantity $\rho_{s\downarrow}/\rho_{s\uparrow}$ dependence of 
the AMR ratio $\Delta \rho/\rho$ of bcc Fe 
for any $\rho_{s \to d\downarrow}/\rho_{s\uparrow}$. 
The expression of the AMR ratio is given by eq. (\ref{del_rho/rho0}). 
Here, $\gamma$=0.01, 
$\rho_{s\to d\uparrow}/\rho_{s\to d\downarrow}$=2.0, 
and $\rho_{\uparrow \downarrow}$=0 are set. 
In addition, 
an arrow indicates 
the theoretical value of 
$\rho_{s\downarrow}/\rho_{s\uparrow}$ (=3.8 $\times$ 10$^{-1}$) 
(see Table \ref{tab1}). 
}
\label{fig_fe}
\end{center}
\end{figure}

In Fig. \ref{fig_fe}, we show 
the $\rho_{s\downarrow}/\rho_{s\uparrow}$ dependence of 
the AMR ratio for any $\rho_{s\to d \downarrow}/\rho_{s\uparrow}$. 
The AMR ratio 
behaves as 
a smooth step-like function. 
In addition, 
the AMR ratio tends to be positive for 
$\rho_{s\downarrow}/\rho_{s\uparrow} \lesssim 1$ 
or negative for 
$\rho_{s\downarrow}/\rho_{s\uparrow} \gtrsim 1$. 
In the case of 
$\rho_{s\downarrow}/\rho_{s\uparrow}$=3.8$\times 10^{-1}$ of Table \ref{tab1}, 
the AMR ratio 
becomes positive 
irrespective of $\rho_{s \to d \downarrow}/\rho_{s\uparrow}$. 
In particular, 
when $\rho_{s\to d \downarrow}/\rho_{s\uparrow}$=0.5, 
the AMR ratio agrees fairly well with the experimental value, 
i.e., 0.003.

\begin{figure}[ht]
\begin{center}
\includegraphics[width=0.8\linewidth]{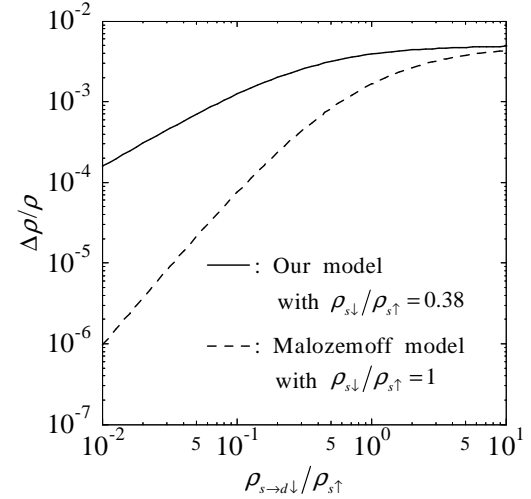} 
\caption{
Quantity $\rho_{s \to d \downarrow}/\rho_{s\uparrow}$ 
dependence of 
the AMR ratio $\Delta \rho/\rho$ of bcc Fe. 
The solid curve represents 
our model, i.e., eq (\ref{del_rho/rho0}) 
with $\rho_{s\downarrow}/\rho_{s\uparrow}$=3.8 $\times$ 10$^{-1}$ 
(see Table \ref{tab1}) 
and $\rho_{\uparrow \downarrow}$=0. 
The dashed curve is 
the Malozemoff model 
with $\rho_{s\downarrow}/\rho_{s\uparrow}$=1,\cite{Malozemoff1} 
i.e., eq. (\ref{Malozemoff_AMR}). 
Here, $\gamma$=0.01 
and $\rho_{s\to d\uparrow}/\rho_{s\to d\downarrow}$=2.0 
are set. 
}
\label{fig_fe_sd}
\end{center}
\end{figure}

Figure \ref{fig_fe_sd} shows 
the $\rho_{s\to d \downarrow}/\rho_{s\uparrow}$ dependence of 
the AMR ratio. 
Our model with $\rho_{s\downarrow}/\rho_{s\uparrow}$=3.8$\times 10^{-1}$ 
is compared with 
the Malozemoff model 
with $\rho_{s\downarrow}/\rho_{s\uparrow}$=1,\cite{Malozemoff1} 
i.e., eq. (\ref{Malozemoff_AMR}). 
The difference of the AMR ratio between them becomes prominent 
for $\rho_{s \to d \downarrow}/\rho_{s\uparrow}\lesssim 1$. 
For example, 
in the case of 
the above-mentioned $\rho_{s \to d \downarrow}/\rho_{s\uparrow}$=0.5, 
the AMR ratio of our model 
is about four times as large as that of the Malozemoff model.

\subsection{Strong ferromagnet: Co, Ni, and Fe$_4$N}
\label{Strong FM}
\begin{figure}[ht]
\begin{center}
\includegraphics[width=0.8\linewidth]{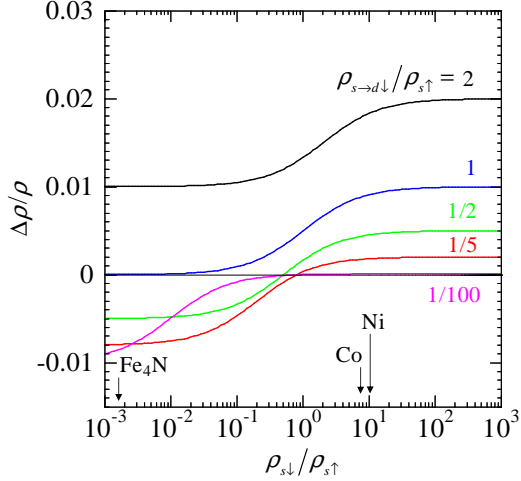} 
\caption{
(Color) 
Quantity $\rho_{s\downarrow}/\rho_{s\uparrow}$ dependence of 
the AMR ratio $\Delta \rho/\rho$ 
of the strong ferromagnet 
for any $\rho_{s \to d\downarrow}/\rho_{s\uparrow}$. 
The expression of 
the AMR ratio 
is given by eq. (\ref{AMR1}). 
Here, $\gamma$=0.01 and $\rho_{\uparrow \downarrow}$=0 are set. 
In addition, 
arrows indicate theoretical values of 
$\rho_{s\downarrow}/\rho_{s\uparrow}$ 
of the respective materials, i.e., 
7.3 for Co, 
1.0 $\times$ 10 for Ni, 
and 1.6 $\times 10^{-3}$ for Fe$_4$N (see Table \ref{tab1}). 
}
\label{AMR_fig1}
\end{center}
\end{figure}

\begin{figure}[ht]
\begin{center}
\hspace*{-0.5cm}
\includegraphics[width=0.8\linewidth]{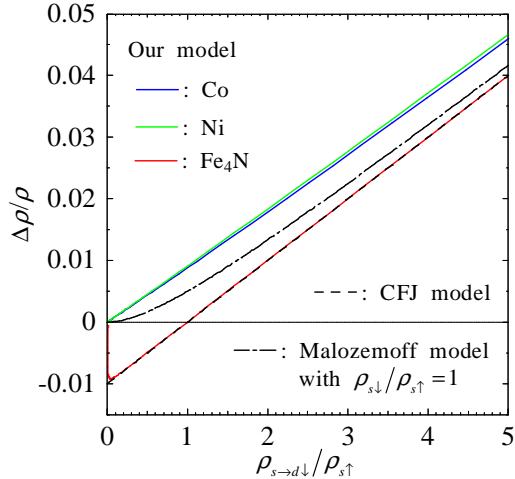} 
\caption{
(Color) 
Quantity $\rho_{s \to d \downarrow}/\rho_{s\uparrow}$ 
dependence of the AMR ratio $\Delta \rho/\rho$ 
of Co, Fe, Ni, and Fe$_4$N. 
The AMR ratio of our model 
is given by 
eq. (\ref{AMR1}) 
with $\rho_{\uparrow \downarrow}$=0, 
where 
$\rho_{s\downarrow}/\rho_{s\uparrow}$ 
is set to be 
7.3 for Co, 
1.0 $\times$ 10 for Ni, 
and 1.6 $\times$ 10$^{-3}$ for Fe$_4$N (see Table \ref{tab1}). 
The dashed curve represents the CFJ model of eq. (\ref{CFJ}), 
where 
$\alpha$ 
is given by 
$\alpha \approx \rho_{s \to d\downarrow}/\rho_{s\uparrow}$. 
The dot-dashed curve is 
the Malozemoff model with $\rho_{s\downarrow}/\rho_{s\uparrow}$=1, 
i.e., eq. (\ref{Malozemoff_AMR}), 
where $\rho_{s \to d\uparrow}$=0 is adopted. 
Here, $\gamma$=0.01 is set. 
}
\label{rho_sd1}
\end{center}
\end{figure}

Utilizing eq. (\ref{del_rho/rho0}), we investigate 
the AMR ratios of fcc Co, fcc Ni, and Fe$_4$N of the strong ferromagnet. 
The DOS of this system 
is schematically illustrated in Figs. \ref{fig_dos}(b) and \ref{fig_dos}(c). 
The fcc Co\cite{Matar} and fcc Ni\cite{Vargas,papa} 
have little d band DOS of the up spin 
at $E_{\mbox{\tiny F}}$. 
As to Fe$_4$N,\cite{Sakuma} 
the d band DOS of the up spin is considerably smaller 
than that of the down spin at $E_{\mbox{\tiny F}}$. 
We thus assume $D_{\uparrow}^{(d)}$=0 
and then have $\rho_{s \to d\uparrow}$=0. 
Substituting $\rho_{s \to d\uparrow}$=0 into eq. (\ref{del_rho/rho0}), 
we obtain the AMR ratio as
\begin{eqnarray}
\label{AMR1}
\frac{\Delta \rho }{\rho } 
= 
\frac{ \gamma \rho_{s \to d\downarrow} \left(
-\rho_{s\uparrow} + \rho_{s\downarrow} + 
\rho_{s \to d\downarrow}\right)}
{\rho_{s\uparrow} 
\left( \rho_{s\downarrow} + \rho_{s \to d\downarrow} \right) 
+ \rho_{\uparrow \downarrow} 
\left( \rho_{s\uparrow} + \rho_{s\downarrow}
+\rho_{s \to d\downarrow}\right)
}.
\end{eqnarray}
Here, when $\rho_{s\downarrow}/\rho_{s\uparrow}$ is 
sufficiently small or sufficiently large, 
eq. (\ref{AMR1}) 
with $\rho_{\uparrow \downarrow}$=0 
is approximated as 
\begin{eqnarray}
\frac{\Delta \rho}{\rho} \approx 
\left\{
\begin{array}{l}
\label{000}
\gamma \left( 
\displaystyle{\frac{\rho_{s \to d\downarrow}}{\rho_{s\uparrow}}}- 1 \right),\hspace{0.8cm}
{\rm for}\hspace{0.3cm}
\displaystyle{\frac{\rho_{s\downarrow}}{\rho_{s\uparrow}}} 
\ll 1, \displaystyle{\frac{\rho_{s \to d\downarrow}}{\rho_{s\uparrow}}}, \\
\gamma 
\displaystyle{\frac{\rho_{s \to d\downarrow}}{\rho_{s\uparrow}}},
\hspace{0.8cm}
{\rm for}\hspace{0.3cm}
\displaystyle{\frac{\rho_{s\downarrow}}{\rho_{s\uparrow}}} 
\gg 1, \displaystyle{\frac{\rho_{s \to d\downarrow}}{\rho_{s\uparrow}}},
\end{array}
\right.
\end{eqnarray}
where 
$\rho_{s \to d\downarrow}/\rho_{s\uparrow}$ 
is set to be 
$0 \le \rho_{s \to d\downarrow}/\rho_{s\uparrow} \le 5$ 
in the present calculation. 
The respective expressions of eq. (\ref{000}) increase 
with increasing $\rho_{s \to d\downarrow}/\rho_{s\uparrow}$ and $\gamma$, 
while the magnitude of 
the difference between the two expressions 
is given by $\gamma$. 
We also mention that 
$\gamma ( \rho_{s \to d\downarrow}/\rho_{s\uparrow}- 1)$ 
corresponds approximately to 
the CFJ model\cite{Campbell} of eq. (\ref{CFJ}), 
which is applicable to the strong ferromagnet. 
Here, $\alpha$ in eq. (\ref{CFJ}) is originally defined by 
$\alpha=\rho_{\perp,\downarrow}/\rho_{\perp,\uparrow}$ 
(see eqs. (\ref{rho_j3}) and (\ref{rho_j1_s})). 
This $\alpha$ can be rewritten as 
$\alpha \approx \rho_{s \to d\downarrow}/\rho_{s\uparrow}$ 
under the following conditions: 
One is the condition of the CFJ model, 
i.e., 
$\rho_{s\sigma \to d \uparrow}=0$, 
$\rho_{s\downarrow}/\rho_{s\downarrow \to d \downarrow} \to 0$, 
$\gamma \ll 1$, and 
$\rho_{s\sigma \to d \downarrow} \equiv \rho_{s \to d \downarrow}$. 
The other is the condition of 
$\gamma \rho_{s \uparrow \to d\downarrow}/\rho_{s\uparrow} \ll 1$. 
The latter reflects that 
$\gamma$=0.01 and 
$\rho_{s \uparrow \to d\downarrow}/\rho_{s\uparrow} < 10$ 
are set in the present study (see Figs. \ref{AMR_fig1} and \ref{rho_sd1}).

In Fig. \ref{AMR_fig1}, 
we show 
the $\rho_{s\downarrow}/\rho_{s\uparrow}$ dependence of 
the AMR ratio of eq. (\ref{AMR1}) 
with $\rho_{\uparrow \downarrow}$=0. 
The quantity $\gamma$ is chosen to be $\gamma$=0.01 
as a typical value. 
We find that the AMR ratio 
behaves as a smooth step-like function 
with the limiting values of eq. (\ref{000}). 
In particular, 
the AMR ratio 
is positive for $\rho_{s\downarrow}/\rho_{s\uparrow} > 1$, 
while 
it can be negative for 
$\rho_{s\downarrow}/\rho_{s\uparrow} \ll 1$ 
and $\rho_{s \to d \downarrow}/\rho_{s\uparrow} \lesssim 1$. 
Note that 
the system of $\rho_{s\downarrow}/\rho_{s\uparrow} > 1$ corresponds to 
Co and Ni, 
while that of $\rho_{s\downarrow}/\rho_{s\uparrow} \ll 1$ 
corresponds to Fe$_4$N.

When $\rho_{s\downarrow}/\rho_{s\uparrow}$'s of 
Co, Ni, and Fe$_4$N are respectively set to be 
7.3, 1.0$\times 10$, and 1.6$\times 10^{-3}$ of Table \ref{tab1}, 
we obtain 
the $\rho_{s \to d \downarrow}/\rho_{s\uparrow}$ dependence of 
the AMR ratios as shown in Fig. \ref{rho_sd1}. 
The main results are as follows: 
\begin{itemize}
\item[(i)] 
The fcc Co and fcc Ni exhibit a positive AMR ratio irrespective of 
$\rho_{s \to d \downarrow}/\rho_{s\uparrow}$, 
while Fe$_4$N 
can take the negative AMR ratio 
depending on $\rho_{s \to d \downarrow}/\rho_{s\uparrow}$. 
Such tendencies roughly correspond to the experimental results 
(see Table \ref{tab1}). 
On the basis of the experimental values of the AMR ratios, 
$\rho_{s \to d \downarrow}/\rho_{s\uparrow}$'s 
of Co, Ni, and Fe$_4$N 
are evaluated to be 
$\rho_{s \to d \downarrow}/\rho_{s\uparrow}\sim2.2$, 
$\rho_{s \to d \downarrow}/\rho_{s\uparrow}\sim2.5$, 
and 
$0.01 \lesssim \rho_{s \to d \downarrow}/\rho_{s\uparrow} 
\lesssim 0.5$, respectively. 
It is noted here that 
the large AMR ratio of Fe$_4$N (e.g., $-$0.07) 
cannot be obtained in the present theory. 
Eventually, a theoretical model 
that takes into account a realistic band structure 
may be necessary for a quantitative analysis.\cite{TB_model} 
\item[(ii)] 
The AMR ratios calculated for fcc Co and fcc Ni 
are clearly different from 
the CFJ model of eq. (\ref{CFJ}) 
because $\rho_{s\downarrow}/\rho_{s\uparrow}$'s of Co and Ni 
are largely different from 
that in the CFJ model 
(i.e., $\rho_{s\downarrow}/\rho_{s\uparrow} \to 0$). 
In contrast, the AMR ratio calculated for Fe$_4$N 
agrees well with the CFJ model, 
because $\rho_{s\downarrow}/\rho_{s\uparrow}$ (=1.6 $\times$ 10$^{-3}$) 
of Fe$_4$N is much smaller than 1. 
\item[(iii)] 
The AMR ratios calculated for fcc Co, fcc Ni, and Fe$_4$N 
deviate from 
the Malozemoff model 
with $\rho_{s\downarrow}/\rho_{s\uparrow}$=1, 
i.e., eq. (\ref{Malozemoff_AMR}). 
The reason is that 
their $\rho_{s\downarrow}/\rho_{s\uparrow}$'s 
are different from 1. 
\end{itemize}

\section{Application 2: Half-Metallic Ferromagnet}
\label{Half-metallic ferromagnet}
On the basis of the theory of \S \ref{Theory}, 
we derive an expression of the AMR ratio of 
the half-metallic ferromagnet. 
Using the expression, 
we obtain an accurate condition for the negative or positive AMR ratio 
and further analyze the AMR ratio.

\subsection{AMR ratio}
\label{suggest}
We first report the feature of the half-metallic ferromagnet 
of Table \ref{tab1}. 
The DOS of Co$_2$MnAl$_{1-x}$Si$_x$,\cite{Galanaki} 
La$_{0.7}$Sr$_{0.3}$MnO$_3$,\cite{Park,Teresa} 
or La$_{0.7}$Ca$_{0.3}$MnO$_3$\cite{Pickett} 
is schematically illustrated in Fig. \ref{fig_dos}(d). 
The conductive and localized d band DOS's of the up spin are present 
at $E_{\mbox{\tiny F}}$, 
while 
there is little DOS of the down spin. 
In real systems, 
however, 
there may be a slight DOS of the down spin 
in the presence of disorders or defects. 
According to previous studies, 
such a feature of the DOS of Co$_2$MnAl$_{1-x}$Si$_x$ 
originates from 
atomic disorders,\cite{Miura} 
while 
that of La$_{0.7}$Sr$_{0.3}$MnO$_3$\cite{LSMO,LSMO1} 
or La$_{0.7}$Ca$_{0.3}$MnO$_3$ 
may be due to oxygen vacancies.\cite{Shirai} 
It is also noted that, 
by reversing the direction of each spin, 
we can treat the opposite case 
(i.e., Fe$_3$O$_4$\cite{Camphausen,Zhang} of Fig. \ref{fig_dos}(e)), 
in which the DOS of the down spin is present at $E_{\mbox{\tiny F}}$, 
while there is little DOS of the up spin.

Focusing on 
the half-metallic ferromagnet with the DOS 
of Fig. \ref{fig_dos}(d), 
we now 
obtain an expression of the AMR ratio as accurately as possible. 
We here utilize the AMR ratio of eq. (\ref{general AMR}) 
because 
$n_\sigma$ and $m_\sigma^*$ are considered to 
have the significant $\sigma$ dependence. 
Meanwhile, 
$\rho_{\uparrow \downarrow}$ and $\rho_{\downarrow \uparrow}$ 
are ignored in the same manner as in \S \ref{Weak ferromagnet}. 
The AMR ratio of eq. (\ref{general AMR}) 
with $\rho_{\uparrow \downarrow}=\rho_{\downarrow \uparrow}=0$ 
is rewritten as 
\begin{eqnarray}
\label{AMR_HM0}
\frac{\Delta \rho}{\rho} = 
-\gamma 
\left( \frac{u-t}{u+1} \right) 
\left[ \frac{r - 
\displaystyle{\frac{v-w}{u-t}} \left( \displaystyle{\frac{u+1}{w+1}} 
\right)^2 }
{r + \displaystyle{\frac{u+1}{w+1}} } \right], 
\end{eqnarray}
with
\begin{eqnarray}
\label{rrr}
&&\hspace*{-1.cm}
r=\frac{\rho_{s \downarrow}}{\rho_{s\uparrow}}
=\left( \frac{m_\downarrow^*}{m_\uparrow^*} \right)^4 
\left( \frac{D_\uparrow^{(s)}}{D_\downarrow^{(s)}} \right)^2, \\
\label{ttt}
&&\hspace*{-1.cm}
t=\frac{\rho_{s \uparrow \to d \downarrow}}{\rho_{s\uparrow}}
=\frac{\tau_{s \uparrow \to d \downarrow}^{-1}}{\tau_{s\uparrow}^{-1}} 
=\beta_\uparrow \frac{D_\downarrow^{(d)}}{D_\uparrow^{(s)}}, \\
\label{uuu}
&&\hspace*{-1.cm}
u=\frac{\rho_{s \uparrow \to d \uparrow}}{\rho_{s\uparrow}}
=\frac{\tau_{s \uparrow \to d \uparrow}^{-1}}{\tau_{s\uparrow}^{-1}}
=\beta_\uparrow \frac{D_\uparrow^{(d)}}{D_\uparrow^{(s)}}, \\
\label{vvv}
&&\hspace*{-1.cm}
v=\frac{\rho_{s \downarrow \to d \uparrow}}{\rho_{s\downarrow}}=
\frac{\tau_{s \downarrow \to d \uparrow}^{-1}}{\tau_{s\downarrow}^{-1}}
=\beta_\downarrow \frac{D_\uparrow^{(d)}}{D_\downarrow^{(s)}}, \\
\label{www}
&&\hspace*{-1.cm}
w=\frac{\rho_{s \downarrow \to d \downarrow}}{\rho_{s\downarrow}}=
\frac{\tau_{s \downarrow \to d \downarrow}^{-1}}{\tau_{s\downarrow}^{-1}}
=\beta_\downarrow \frac{D_\downarrow^{(d)}}{D_\downarrow^{(s)}}, \\
\label{beta}
&&\hspace*{-1.cm}
\beta_\sigma=N_{\rm n}\frac{|V_{s\sigma \to d\sigma}|^2}{|V_{s}|^2}, 
\end{eqnarray}
where 
eq. (\ref{rrr}) has been derived in the Appendix \ref{parameters} 
and  
eqs. (\ref{ttt}) - (\ref{www}) 
have been obtained by using 
eqs. (\ref{r_s_s}), (\ref{1/tau_s_s}), (\ref{rho_sd}), and (\ref{tau_sd}). 
We also have assumed 
$D_\downarrow^{(d)} \ne 0$ and $D_\downarrow^{(s)} \ne 0$ 
on the basis of 
the above-mentioned feature of the DOS of the down spin. 
Here, the conduction state (named as $s$ in $D_\sigma^{(s)}$) 
may correspond to the conductive d state 
in the case of the present half-metallic ferromagnet 
(see Figs. \ref{fig_dos}(d) and \ref{fig_dos}(e)). 
From eqs. (\ref{ttt}) - (\ref{www}), 
we find the following relation: 
\begin{eqnarray}
\frac{t}{u}=\frac{w}{v}. 
\end{eqnarray}
Using this relation, 
we express eq. (\ref{AMR_HM0}) as
\begin{eqnarray}
\label{AMR_HM2}
\frac{\Delta \rho}{\rho} = 
\frac{ - \gamma}{u^{-1} + 1} 
\left( 1 - \frac{w}{v}  \right) 
\left[\frac{ r - \displaystyle{\frac{v}{u}} 
\left( \displaystyle{\frac{u+1}{w+1}} \right)^2}{ r + \displaystyle{ \frac{u+1}{w+1}} }\right]. 
\end{eqnarray}
Here, 
parameters in eq. (\ref{AMR_HM2}), 
$r$, $u$, $v$, and $w$, 
are suggested 
as follows: 
\begin{enumerate}
\item[(i)] 
The parameter $r$ of eq. (\ref{rrr}) may become extremely large 
owing to 
$\rho_{s \downarrow} \gg \rho_{s\uparrow}$. 
This relation 
is based on the fact that 
the resistivity of semiconductors is more than 10$^4$ times larger than 
that of metals.\cite{Kittel} 
As a typical system, we consider $r$ to be $r \gtrsim 10^6$ 
on the assumption of 
$D_\uparrow^{(s)}/D_\downarrow^{(s)}\gtrsim 10^5$ and 
$m_\downarrow^*/m_\uparrow^* \sim 0.1$. 
Here, $m_\downarrow^*/m_\uparrow^*$ has been roughly estimated on the basis of 
the effective mass of the carrier of the semiconductor 
divided by the electron mass.\cite{Kittel} 
\item[(ii)]
The parameter $u$ of eq. (\ref{uuu}) 
takes a finite value, 
where  
$D_{\uparrow}^{(d)} \ne 0$ 
and $D_{\uparrow}^{(s)} \ne 0$. 
In the present calculation, 
$u$ is treated as a variable number of $0.01 \le u \le 50$. 
\item[(iii)]
The parameter $v$ of eq. (\ref{vvv}) 
may be sufficiently large 
because of $D_{\uparrow}^{(d)} \gg D_{\downarrow}^{(s)}$. 
In the case of the $D_\uparrow^{(s)}/D_\downarrow^{(s)}\gtrsim 10^5$ 
reported above, 
we find the relation of 
$v/u=(\beta_\downarrow/\beta_\uparrow)D_\uparrow^{(s)}/D_\downarrow^{(s)} 
\gtrsim  10^5$, 
where $\beta_\uparrow \sim \beta_\downarrow$ has been assumed.  
\item[(iv)]
The parameter $w$ of eq. (\ref{www}) may take a finite value, 
although both $D_{\downarrow}^{(d)}$ 
and $D_{\downarrow}^{(s)}$ are extremely small. 
In addition, the relation of 
$w/v=D_\downarrow^{(d)}/D_\uparrow^{(d)} \ll 1$ 
is realized. 
\end{enumerate}

On the basis of 
eqs. (\ref{rrr}) - (\ref{www}) and the above suggestions, 
we next obtain an approximate expression of eq. (\ref{AMR_HM2}). 
We here assume 
$\beta_\uparrow \sim \beta_\downarrow$ 
and $u \sim w$ 
and also take into account 
$w/v \ll 1$ in (iv), 
$r \gg 1$, and 
$r \gg v/u \sim D_\uparrow^{(s)}/D_\downarrow^{(s)}$, 
where 
$D_\uparrow^{(s)}/D_\downarrow^{(s)}\gtrsim 10^5$ 
and 
$m_\downarrow^*/m_\uparrow^* \sim 0.1$ in (i) 
have been adopted. 
Equation (\ref{AMR_HM2}) has thus been written as
\begin{eqnarray}
\label{AMR_HM3}
\frac{\Delta \rho}{\rho} =\frac{ - \gamma}{u^{-1} + 1}. 
\end{eqnarray}
The AMR ratio of eq. (\ref{AMR_HM3}) always takes a negative value.

\subsection{Sign of AMR ratio}
\label{half_sign_AMR}
From eq. (\ref{AMR_HM2}), 
we can find the condition for the negative or positive AMR ratio 
of the half-metallic ferromagnet. 
This condition is more accurate than 
the result in the unified framework of \S \ref{AMR and scattering}. 
Because of $w/v \ll 1$ in (iv), 
we focus on the numerator in $[~~]$ of eq. (\ref{AMR_HM2}). 
The numerator is written by 
$rf(u)$ with 
\begin{eqnarray}
\label{f(u)}
&&\hspace*{-1cm}f(u) = - \frac{(u+1)^2}{ \xi u} +1, \\
\label{xi_def}
&&\hspace*{-1cm}\xi=\frac{r(w+1)^2}{v}, 
\end{eqnarray}
where $\xi>0$ and $u>0$. 
Here, $f(u)>0$ and $f(u)<0$ correspond to 
the negative and positive AMR ratios, respectively. 
From eq. (\ref{f(u)}), 
we first find that the AMR ratio becomes positive when $\xi < 4$. 
Second, in the case of $\xi \ge 4$, 
the AMR ratio is negative for
\begin{eqnarray}
\label{condition}
\mu_- < u < \mu_+,
\end{eqnarray}
while it is positive for
\begin{eqnarray}
0 < u < \mu_-~{\rm and}~\mu_+ < u, 
\end{eqnarray}
with 
$\mu_-=( \xi- 2 - \sqrt{\xi^2 -4 \xi} )/2$ and 
$\mu_+=(\xi - 2 + \sqrt{\xi^2 -4 \xi})/2$.
Note that the AMR ratio becomes 0 at $u=\mu_\pm$.

\begin{figure}[ht]
\begin{center}
\includegraphics[width=0.8\linewidth]{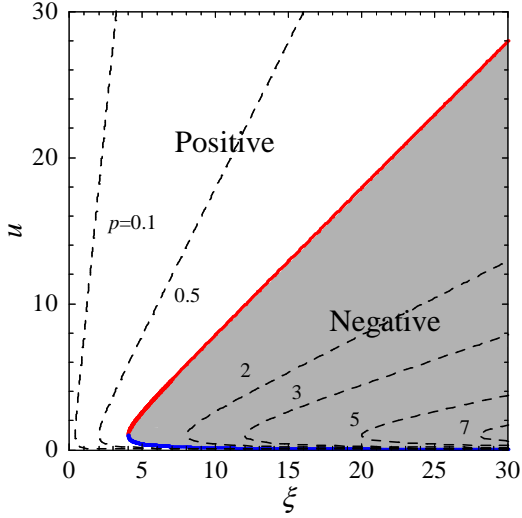} 
\caption{
(Color) 
Sign of the AMR ratio $\Delta \rho/\rho$ of the half-metallic ferromagnet 
in the $\xi$-$u$ plane. 
The negative and positive AMR ratios are shown 
by the dark and white regions, respectively. 
The AMR ratio becomes zero at $u=\mu_\pm$. 
Here, $u=\mu_-$ and $u=\mu_+$ are shown 
by the solid curves with $u \ge 1$ and $u < 1$, respectively. 
The relation between $\xi$ and $u$ of a half-metallic ferromagnet, 
eq. (\ref{xi}), 
is shown by the dashed curves, 
where $p$=0.1, 0.5, 2, 3, 5, and 7. 
In addition, eq. (\ref{xi}) of $p$=1 corresponds to 
$\mu_-$ and $\mu_+$. 
}
\label{half_sign}
\end{center}
\end{figure}

Figure \ref{half_sign} shows 
the sign of the AMR ratio 
in the $\xi$-$u$ plane based on 
the above results. 
From this figure, 
we can find signs of the AMR ratios of various systems. 
We here focus on 
a simple system with $\beta_\uparrow=\beta_\downarrow$ and 
$D_\uparrow^{(d)}/D_\uparrow^{(s)}=D_\downarrow^{(d)}/D_\downarrow^{(s)}$ 
(i.e., $u=w$). 
For this system, 
we first 
determine the specific sets of $\xi$ and $u$. 
The relation between $\xi$ and $u$ has been obtained as
\begin{eqnarray}
\label{xi}
\xi=p \left( u + \frac{1}{u} +2 \right), 
\end{eqnarray}
with 
$p=(m_\downarrow^*/m_\uparrow^*)^4 D_\uparrow^{(s)}/D_\downarrow^{(s)}$ (see eq. (\ref{xi_ex})). 
In Fig. \ref{half_sign}, 
we show eq. (\ref{xi}) with $p$=0.1, 0.5, 2, 3, 5, and 7 
by the dashed curves, 
where eq. (\ref{xi}) with $p$=1 corresponds to 
$\mu_-$ and $\mu_+$. 
It is found that eq. (\ref{xi}) with $p>$1 exists 
in the region of the negative AMR ratio. 
For example, 
the case of 
$D_\uparrow^{(s)}/D_\downarrow^{(s)}\gtrsim 10^5$ 
and 
$m_\downarrow^*/m_\uparrow^* \sim 0.1$ in (i) leads to $p \gtrsim 10$. 
This case thus can take the negative AMR ratio. 
Negative AMR ratios been experimentally observed, 
as shown in Table \ref{tab1}.

\subsection{Evaluation of AMR ratio}
\label{Evaluation of AMR ratio}
\begin{figure}[ht]
\begin{center}
\includegraphics[width=0.8\linewidth]{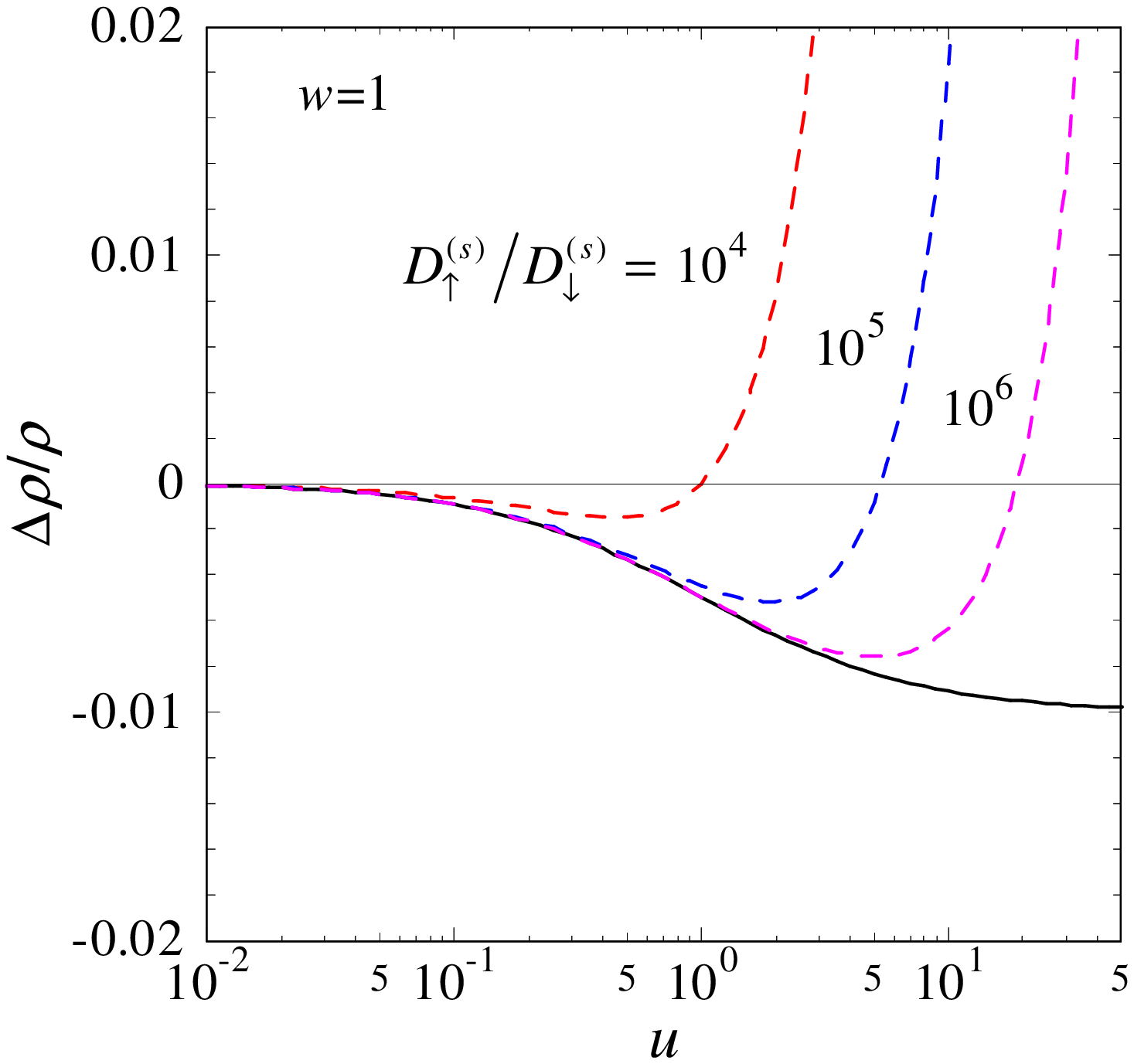}
\includegraphics[width=0.8\linewidth]{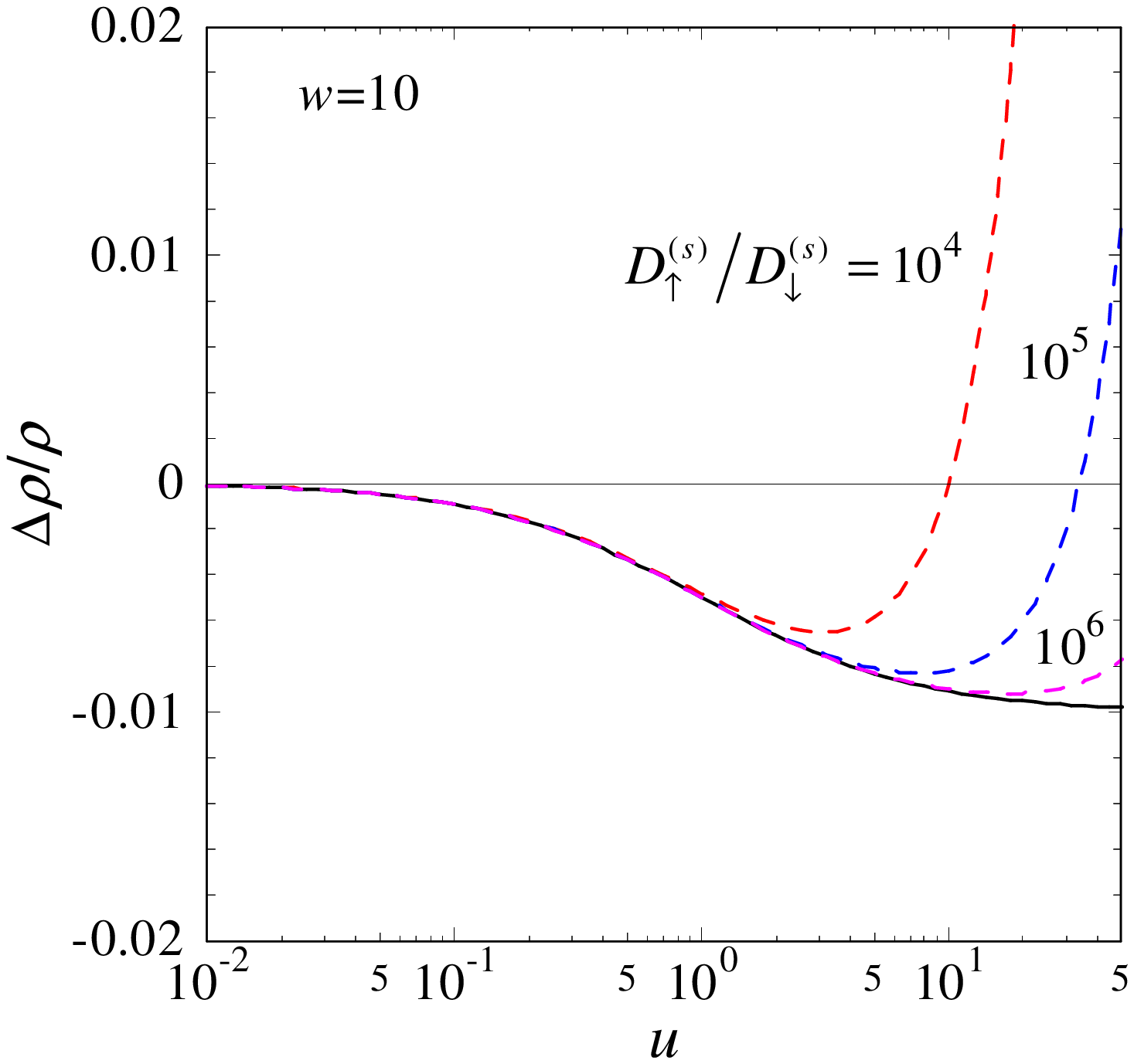}
\caption{
(Color) 
Quantity $u$ dependence of 
the AMR ratio $\Delta \rho/\rho$ 
of the half-metallic ferromagnet. 
Upper panel: $w$=1. 
Lower panel: $w$=10. 
In each panel, 
the dashed curves show 
the AMR ratios of eq. (\ref{AMR_HM2}) with 
$D_\uparrow^{(s)}/D_\downarrow^{(s)}$=10$^4$, 10$^5$, and 10$^6$. 
In addition, the solid curve is the AMR ratio of eq. (\ref{AMR_HM3}). 
Here, $\gamma$=0.01, $m_\downarrow^*/m_\uparrow^*$=0.1, 
and $\beta_\uparrow=\beta_\downarrow$ are set. 
}
\label{AMR_HMFM}
\end{center}
\end{figure}

Using the results of \S \ref{suggest} and \S \ref{half_sign_AMR}, 
we evaluate the AMR ratio. 
The $u$ dependence of 
the AMR ratio is shown in Fig. \ref{AMR_HMFM}. 
The dashed curves represent eq. (\ref{AMR_HM2}) 
with 
the parameters of 
$\gamma$=0.01, 
$0 \le u \le 50$, 
$v=(D_\uparrow^{(s)}/D_\downarrow^{(s)})u$, 
$r=(0.1)^4 (D_\uparrow^{(s)}/D_\downarrow^{(s)})^2$, 
$w$=1, 10, 
and $D_\uparrow^{(s)}/D_\downarrow^{(s)}$=10$^4$, 10$^5$, 10$^6$, 
where $m_\downarrow^*/m_\uparrow^*$=0.1 
and $\beta_\uparrow=\beta_\downarrow$. 
The parameters 
have been chosen on the basis of 
(i) - (iv) in \S \ref{suggest}. 
We observe that each AMR ratio exhibits a convex downward curve with 
a negative minimum value. 
The AMR ratio approaches 0 with decreasing $u$, 
while it changes from negative to positive 
with increasing $u$. 
In addition, the AMR ratio comes close to 
eq. (\ref{AMR_HM3}) with $\gamma$=0.01 (the solid curve) 
with increasing $D_\uparrow^{(s)}/D_\downarrow^{(s)}$. 
It is noted that eq. (\ref{AMR_HM3}) is obtained 
from eq. (\ref{AMR_HM2}) under the condition of 
$r \gg (v/u)[(u+1)/(w+1)]^2$, $r \gg (u+1)/(w+1)$, and $w/v \ll 1$ in (iv). 
Also, in the case of $D_\uparrow^{(s)}/D_\downarrow^{(s)} \gtrsim 10^5$, 
the AMR ratio becomes about $-$0.004 at $u=w=1$ 
(see the upper panel of Fig. \ref{AMR_HMFM}), 
where 
the system of $u=w$ corresponds to 
the simple system in \S \ref{half_sign_AMR}. 
This AMR ratio agrees well with 
the experimental results of Table \ref{tab1}.

\subsection{Sign change of the AMR ratio in Fe$_3$O$_4$}
Utilizing eq. (\ref{AMR_HM2}), we analyze 
an experimental result of Fe$_3$O$_4$, 
in which 
the sign of the AMR ratio 
changes from negative to positive 
as the temperature increases.\cite{Ziese,Ziese2} 
Here, Fe$_3$O$_4$ 
has been theoretically predicted to have a half-metallic property 
at the ground state in the absence of the spin--orbit interaction.\cite{Zhang} 
The DOS of Fe$_3$O$_4$ 
is schematically illustrated 
in Fig. \ref{fig_dos}(e):\cite{Camphausen,Zhang} 
the DOS of the down spin is present at $E_{\mbox{\tiny F}}$, 
while there is little DOS of the up spin. 

\begin{figure}[ht]
\begin{center}
\includegraphics[width=0.8\linewidth]{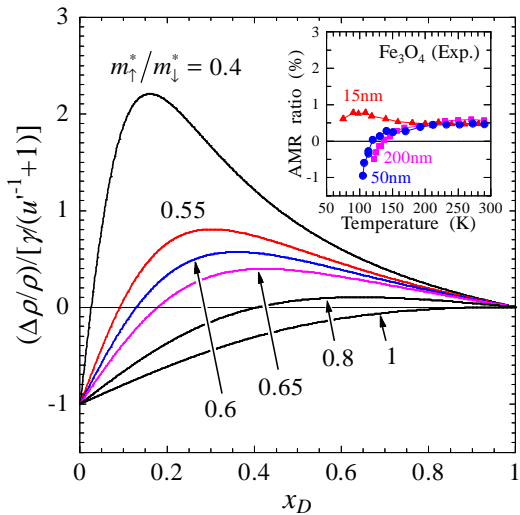}
\caption{
(Color) 
Quantity $x_D$ 
(=$D_\uparrow^{(s)}/D_\downarrow^{(s)}=D_\uparrow^{(d)}/D_\downarrow^{(d)}$) 
dependence of the AMR ratio $\Delta \rho/\rho$ of eq. (\ref{AMR_x}) 
for any $m_\uparrow^*/m_\downarrow^*$. 
The inset shows an experimental result of 
the temperature dependence of the AMR ratio of 
Fe$_3$O$_4$ films on MgO 
obtained by Ziese.\cite{Ziese} 
The respective film thicknesses are 15 nm, 50 nm, and 200 nm. 
Note also that the DOS of Fe$_3$O$_4$ is schematically illustrated 
in Fig. \ref{fig_dos}(e). 
}
\label{Fe3O4}
\end{center}
\end{figure}

Recently, Ziese has experimentally observed that 
the Fe$_3$O$_4$ film on MgO 
with film thickness of 50 nm or 200 nm 
changed the sign of the AMR ratio 
from negative to positive 
with increasing temperature 
(see the inset of Fig. \ref{Fe3O4}).\cite{Ziese,Ziese2} 
This Fe$_3$O$_4$ eventually exhibited 
positive AMR ratios of about 0.005 
at temperatures higher than 200 K. 
As a cause of this phenomenon, he considered that 
the majority spin band (i.e., $e_{g\uparrow}$ band) 
came close to $E_{\mbox{\tiny F}}$ with increasing temperature, 
and, furthermore, 
this band was present at $E_{\mbox{\tiny F}}$ 
in the high temperature region (e.g., the region higher than 200 K). 
On the basis of such an idea, 
he proposed a two-band model composed of 
$t_{2g\downarrow}$ and $e_{g\uparrow}$ bands; 
$t_{2g\downarrow}$ and $e_{g\uparrow}$ bands have been shown in 
Fig. \ref{fig_dos}(e). 
Using the model, he primarily found that 
the AMR ratio became 0.005 for the specific values of 
the minority-to-majority resistivity ratio 
and the reduced spin-flip scattering resistivity. 
Meanwhile, he also showed that the sign of the AMR ratio changed 
from negative to positive with increasing 
$\rho_{s \to d \downarrow}/\rho_{s \to d \uparrow}$.\cite{Ziese_comment} 
Here, $\rho_{s \to d \downarrow}/\rho_{s \to d \uparrow}$ 
is reduced to 
$D_\downarrow^{(d)}/D_\uparrow^{(d)}$ in our formulation 
(see eq. (\ref{rho_sd usual})). 
From the standpoint of the AMR ratio versus 
$D_\downarrow^{(d)}/D_\uparrow^{(d)}$, however, 
we see a problem; that is, 
the sign change of this model appears to be contrary to 
the experimental trend of the inset of Fig. \ref{Fe3O4} 
or the above idea. 
In fact, 
with decreasing $D_\downarrow^{(d)}/D_\uparrow^{(d)}$, 
the sign may change from negative to positive. 
In addition, we notice that 
this model consists of only 
the resistivities due to the s--d scattering 
but neglects 
the resistivity of the conductive d states, 
$\rho_{s\sigma}$, due to 
the scattering process between the conductive d states.\cite{Ziese_comment1} 
For this situation, 
we believe that there is a need to reexamine 
the sign change of the AMR ratio 
by using a model 
that takes into account both resistivities.

We, therefore, 
demonstrate 
the sign change of the AMR ratio 
using our model with both resistivities. 
On the basis of the behavior of the $e_{g\uparrow}$ band reported above, 
we assume that 
the DOS of the up spin at $E_{\mbox{\tiny F}}$ 
increases with increasing temperature. 
Our concern, thus, is with how the DOS 
of the up spin influences the AMR ratio. 
To clearly show the influence, 
we consider a simple case of 
$D_\uparrow^{(s)}/D_\downarrow^{(s)}=D_\uparrow^{(d)}/D_\downarrow^{(d)}$ 
(or 
$D_\downarrow^{(d)}/D_\downarrow^{(s)}=D_\uparrow^{(d)}/D_\uparrow^{(s)}$) 
and $\beta_\uparrow=\beta_\downarrow$. 
By paying attention to the DOS of Fig. \ref{fig_dos}(e), 
i.e., 
the reversion of 
the direction of each spin of eq. (\ref{AMR_HM2}), 
eq. (\ref{AMR_HM2}) is then rewritten as
\begin{eqnarray}
\label{AMR_x}
\frac{\Delta \rho}{\rho} =
\frac{ - \gamma}{u'^{-1} + 1} (1-x_D) 
\left[ \frac{ \left(\displaystyle{m_\uparrow^*/m_\downarrow^*}\right)^4 -x_D}{ \left(\displaystyle{m_\uparrow^*/m_\downarrow^*}\right)^4 +x_D^2} \right], 
\end{eqnarray}
with $x_D=D_\uparrow^{(s)}/D_\downarrow^{(s)}=D_\uparrow^{(d)}/D_\downarrow^{(d)}$ and 
$u'=\rho_{s \downarrow \to d \downarrow}/\rho_{s\downarrow}$
=$\beta_\downarrow D_\downarrow^{(d)}/D_\downarrow^{(s)}$. 
Figure \ref{Fe3O4} shows 
the $x_D$ dependence of the AMR ratio 
of eq. (\ref{AMR_x}) 
for $m_{\uparrow}^*/m_{\downarrow}^*$=0.4, 0.55, 0.6, 0.65, 0.8, and 1. 
The AMR ratios of $m_{\uparrow}^*/m_{\downarrow}^*$=0.4, 0.55, 0.6, 0.65, 
and 0.8 
change from negative to positive with increasing $x_D$, 
although that of $m_{\uparrow}^*/m_{\downarrow}^*$=1 is always negative. 
The sign change appears to 
originate from the feature in which 
the s--d scatterings of $s \downarrow \to d \uparrow$ 
and $s \uparrow \to d \downarrow$ 
increase with increasing $D_\uparrow^{(s)}$ and $D_\uparrow^{(d)}$. 
Here, it is noteworthy that 
these s--d scatterings tend to lead to the positive AMR ratio 
(see \S \ref{Origin of AMR} and \S \ref{AMR and scattering}). 
In addition, roughly speaking, 
the $x_D$ dependence of the AMR ratio 
appears to be qualitatively similar to 
the experimental trend of the inset of Fig. \ref{Fe3O4}. 
In particular, 
the AMR ratios of $m_{\uparrow}^*/m_{\downarrow}^*$=0.6 and 0.65 
may correspond well to 
the experimental results 
for film thicknesses of 50 nm and 200 nm, respectively. 
In addition, 
the AMR ratio of $m_{\uparrow}^*/m_{\downarrow}^*$=0.55 
may partially correspond to 
the experimental result 
for film thicknesses of 15 nm.

\section{Conclusion}
\label{Conclusion}

We systematically analyzed the AMR effects of 
bcc Fe of the weak ferromagnet, 
fcc Co, fcc Ni, and Fe$_4$N of the strong ferromagnet, 
and the half-metallic ferromagnet. 
We here used the two-current model for a system 
consisting of a spin-polarized conduction state and 
localized d states with spin--orbit interaction.

From such a model, 
we first derived 
general expressions of resistivities 
composed of 
$\rho_{s\sigma}$ and $\rho_{s \sigma \to d\varsigma}$. 
The resistivity 
$\rho_{s \sigma}$ arose from the s--s scattering, 
in which 
the conduction electron of the $\sigma$ spin 
was scattered into the conduction state of the $\sigma$ spin 
by nonmagnetic impurities. 
The resistivity $\rho_{s \sigma \to d\varsigma}$ was 
due to the s--d scattering, in which 
the conduction electron of the $\sigma$ spin was scattered into 
the $\sigma$ spin state 
in the localized d states of the $\varsigma$ spin by the impurities, 
where 
the $\varsigma$ spin represented 
the spin of the dominant state in the d states (i.e., the spin-mixed states).

Using the resistivities, 
we next obtained a general expression of the AMR ratio. 
On the basis of 
the AMR ratio and the resistivities, 
we showed that 
the AMR effect reflected 
the difference of ``changes of the d orbitals 
due to the spin--orbit interaction'' 
between different $m$'s, 
where $m$ was the magnetic quantum number of the d orbital. 
In addition, 
we roughly determined 
a relation between 
the sign of the AMR ratio and the scattering process. 
In brief, when the dominant s--d scattering process 
was $s\uparrow \to d \downarrow$ or $s\downarrow \to d \uparrow$, 
the AMR ratio tended to become positive. 
In contrast, 
when the dominant s--d scattering process 
was $s\uparrow \to d \uparrow$ or $s\downarrow \to d \downarrow$, 
the AMR ratio tended to be negative.

Finally, 
from the general expression of the AMR ratio, 
we obtained expressions of AMR ratios 
appropriate to 
the respective materials. 
Using the expressions, we analyzed their AMR ratios. 
The results for the respective materials 
were written as follows: 
\begin{enumerate}
\item[(i)] bcc Fe of weak ferromagnet\\
Using 
the AMR ratio of eq. (\ref{del_rho/rho0}) with 
$\rho_{s\downarrow}/\rho_{s\uparrow}=3.8 \times 10^{-1}$ in 
Table \ref{tab1} and $\rho_{\uparrow \downarrow}$=0, 
we found that the AMR ratio became positive 
irrespective of $\rho_{s \to d \downarrow}/\rho_{s\uparrow}$, 
where 
$\rho_{s \sigma \to d \varsigma}=\rho_{s \to d \varsigma}$ 
has been set. 
In particular, 
when $\rho_{s\to d \downarrow}/\rho_{s\uparrow}$=0.5, 
the AMR ratio  
agreed fairly well with the experimental value in Table \ref{tab1}, 
i.e., 0.003. 
Here, the positive AMR ratio originated from 
the dominant s--d scattering process of 
$s \downarrow \to d \uparrow$. 
Regarding 
the $\rho_{s \to d \downarrow}/\rho_{s\uparrow}$ dependence of 
the AMR ratio, 
the difference of the AMR ratio between 
our model with $\rho_{s\downarrow}/\rho_{s\uparrow}$=3.8$\times 10^{-1}$ 
and the Malozemoff model with $\rho_{s\downarrow}/\rho_{s\uparrow}$=1 
was clearly observed 
for $\rho_{s \to d \downarrow}/\rho_{s\uparrow}\lesssim 1$. 
\item[(ii)] fcc Co, fcc Ni, and Fe$_4$N of strong ferromagnet\\
Using 
the AMR ratio of eq. (\ref{AMR1}) with $\rho_{\uparrow \downarrow}$=0 
and 
$\rho_{s\downarrow}/\rho_{s\uparrow}$'s in Table \ref{tab1}, i.e., 
7.3 for fcc Co, 
1.0$\times 10$ for fcc Ni, 
and 1.6$\times 10^{-3}$ for Fe$_4$N, 
we found that 
fcc Co and fcc Ni exhibited a positive AMR ratio irrespective of 
$\rho_{s \to d \downarrow}/\rho_{s\uparrow}$, 
while Fe$_4$N 
could take the negative AMR ratio 
depending on $\rho_{s \to d \downarrow}/\rho_{s\uparrow}$. 
In particular, 
when $\rho_{s \to d \downarrow}/\rho_{s\uparrow}$'s of 
fcc Co, fcc Ni, and Fe$_4$N 
were, respectively, chosen to be 
$\rho_{s \to d \downarrow}/\rho_{s\uparrow}\sim$2.2, 
$\rho_{s \to d \downarrow}/\rho_{s\uparrow}\sim$2.5, and 
$0.01 \lesssim \rho_{s \to d \downarrow}/\rho_{s\uparrow} \lesssim 0.5$, 
their AMR ratios corresponded well to 
the respective experimental values in Table \ref{tab1}, i.e., 
0.020 for fcc Co, 0.022 for fcc Ni, and $-$0.01 - $-$0.005 for Fe$_4$N. 
It is noted, however, that 
the large AMR ratio of Fe$_4$N (e.g., $-$0.07 - $-$0.02) 
could not be obtained in the present theory. 
The positive AMR ratios of fcc Co and fcc Ni 
originated from 
the dominant s--d scattering process of 
$s \uparrow \to d \downarrow$. 
In contrast, 
the negative AMR ratio of Fe$_4$N was due to 
the dominant s--d scattering process of 
$s \downarrow \to d \downarrow$. 
As for the $\rho_{s \to d \downarrow}/\rho_{s\uparrow}$ dependence of 
the AMR ratios, 
the calculation result of fcc Co and fcc Ni by our model 
was obviously different from 
those by the CFJ model 
and the Malozemoff model. 
The reason was that 
$\rho_{s\downarrow}/\rho_{s\uparrow}$ ($>1$) of fcc Co or fcc Ni 
was largely different from 
$\rho_{s\downarrow}/\rho_{s\uparrow}$ ($\ll$1) of 
the CFJ model and 
$\rho_{s\downarrow}/\rho_{s\uparrow}$ (=1) of 
the Malozemoff model. 
In the case of 
Fe$_4$N, 
the result by our model 
agreed well with that by the CFJ model 
because 
$\rho_{s\downarrow}/\rho_{s\uparrow}$ (=1.6 $\times$ 10$^{-3}$) of Fe$_4$N 
corresponded well to 
$\rho_{s\downarrow}/\rho_{s\uparrow}$ ($\ll$1) of the CFJ model. 
\item[(iii)] half-metallic ferromagnet \\
Using the AMR ratio of eq. (\ref{AMR_HM2}), 
which took into account the spin dependence of 
the effective mass and the number density 
of electrons in the conduction band, 
we showed that 
the AMR ratio 
could become negative 
for a typical system with $D_\uparrow^{(s)}/D_\downarrow^{(s)}\gtrsim 10^5$ 
and $m_\downarrow^*/m_\uparrow^* \sim 0.1$. 
In particular, when 
$\rho_{s \uparrow \to d \uparrow}/\rho_{s\uparrow}=\rho_{s \downarrow \to d \downarrow}/\rho_{s\downarrow}=1$, 
the AMR ratio was evaluated to be about $-$0.004, 
which was close to the experimental values. 
Here, the negative AMR ratio 
of Co$_2$MnAl$_{1-x}$Si$_x$, La$_{0.7}$Sr$_{0.3}$MnO$_3$, 
and La$_{0.7}$Ca$_{0.3}$MnO$_3$ 
originated from the dominant s--d scattering process of 
$s \uparrow \to d \uparrow$, 
while 
the negative AMR ratio of Fe$_3$O$_4$ 
was due to the dominant s--d scattering process of 
$s \downarrow \to d \downarrow$. 
We also analyzed 
the experimental result of the AMR effect of Fe$_3$O$_4$, 
in which 
the sign of the AMR ratio 
changed from negative to positive 
as the temperature increased. 
Such a sign change occurred with increasing 
the DOS of the majority spin at $E_{\mbox{\tiny F}}$, 
$D_\uparrow^{(s)}$ and $D_\uparrow^{(d)}$. 
The increase of $D_\uparrow^{(s)}$ and $D_\uparrow^{(d)}$ 
appeared to enhance 
the s--d scatterings of $s \uparrow \to d \downarrow$ 
and $s \downarrow \to d \uparrow$, 
which tended to lead to the positive AMR ratio. 
\end{enumerate}

\begin{acknowledgments}

We acknowledge the stimulated discussion 
in the meeting of the Cooperative Research Project 
of the Research Institute of Electrical Communication, Tohoku University. 
This work has been supported by 
a Grant-in-Aid for Young Scientists (B) 
(No. 20710076) 
and 
a Grant-in-Aid for Scientific Research (B) 
(No. 23360130) 
from the Japan Society for the Promotion of Science. 

\end{acknowledgments}

\appendix 
\section{Localized d States}
\label{WF}
Applying the perturbation theory to 
${\cal H}$ of eq. (\ref{single_H}), 
we obtain the wave function of the localized d state 
(i.e., the spin-mixed state), 
$\Phi_{M,\varsigma}^{(d)} ({\mbox{\boldmath $r$}})$, 
with $M=-$2, $-$1, 0, 1, 2, 
and $\varsigma=\uparrow$ or $\downarrow$. 
Here, 
${\mbox{\boldmath $r$}}$ is the position vector, 
while $M$ and $\varsigma$ are, respectively, 
the magnetic quantum number and the spin 
of the dominant state in the spin-mixed state.

Within the second-order perturbation, 
$\Phi_{M,\uparrow}^{(d)} ({\mbox{\boldmath $r$}})$ 
is obtained as
\begin{eqnarray}
\label{22down}
&&\hspace*{-1.5cm}\Phi_{2,\downarrow}^{(d)}({\mbox{\boldmath $r$}})
=\left( 1 - \frac{1}{2} \epsilon^2 \right) 
\phi_{2,\downarrow}({\mbox{\boldmath $r$}})
+ \left( \epsilon + \frac{3}{2} \epsilon^2 \right) 
\phi_{1,\uparrow}({\mbox{\boldmath $r$}}), \\
&&\hspace*{-1.5cm}
\Phi_{1,\downarrow}^{(d)}({\mbox{\boldmath $r$}}) = \left( 1 - \frac{3}{4} \epsilon^2 \right) 
\phi_{1,\downarrow}({\mbox{\boldmath $r$}}) + \left( \frac{\sqrt{6}}{2}
\epsilon + \frac{\sqrt{6}}{4} \epsilon^2 \right) 
\phi_{0,\uparrow}({\mbox{\boldmath $r$}}), \\
&&\hspace*{-1.5cm}
\Phi_{0,\downarrow}^{(d)} ({\mbox{\boldmath $r$}})= \left( 1 - \frac{3}{4} \epsilon^2 \right) 
\phi_{0,\downarrow} ({\mbox{\boldmath $r$}})+ 
\left( \frac{\sqrt{6}}{2} \epsilon - \frac{\sqrt{6}}{4} \epsilon^2 \right) 
\phi_{-1,\uparrow}({\mbox{\boldmath $r$}}), \\
\label{2-1down}
&&\hspace*{-1.5cm}
\Phi_{-1,\downarrow}^{(d)}({\mbox{\boldmath $r$}}) = \left( 1 - \frac{1}{2} \epsilon^2 \right) 
\phi_{-1,\downarrow}({\mbox{\boldmath $r$}}) 
+ \left( \epsilon - \frac{3}{2} \epsilon^2 \right) 
\phi_{-2,\uparrow}({\mbox{\boldmath $r$}}), \\
\label{2-2down}
&&\hspace*{-1.5cm}
\Phi_{-2,\downarrow}^{(d)}({\mbox{\boldmath $r$}}) = 
\phi_{-2,\downarrow}({\mbox{\boldmath $r$}}), 
\end{eqnarray}
while 
$\Phi_{M,\downarrow}^{(d)} ({\mbox{\boldmath $r$}})$ is 
\begin{eqnarray}
\label{22up}
&&\hspace*{-1.5cm}
\Phi_{2,\uparrow}^{(d)}({\mbox{\boldmath $r$}}) = 
\phi_{2,\uparrow}({\mbox{\boldmath $r$}}), \\
\label{21up}
&&\hspace*{-1.5cm}
\Phi_{1,\uparrow}^{(d)}({\mbox{\boldmath $r$}}) 
= \left( 1 - \frac{1}{2} \epsilon^2 \right) 
\phi_{1,\uparrow}({\mbox{\boldmath $r$}}) 
- \left( \epsilon + \frac{3}{2} \epsilon^2 \right) 
\phi_{2,\downarrow}({\mbox{\boldmath $r$}}), \\
&&\hspace*{-1.5cm}
\Phi_{0,\uparrow}^{(d)}({\mbox{\boldmath $r$}}) 
= \left( 1 - \frac{3}{4} \epsilon^2 \right) 
\phi_{0,\uparrow}({\mbox{\boldmath $r$}}) - \left( \frac{\sqrt{6}}{2}
\epsilon + \frac{\sqrt{6}}{4} \epsilon^2 \right) 
\phi_{1,\downarrow}({\mbox{\boldmath $r$}}), \\
&&\hspace*{-1.5cm}
\Phi_{-1,\uparrow}^{(d)} ({\mbox{\boldmath $r$}})
= \left( 1 - \frac{3}{4} \epsilon^2 \right) 
\phi_{-1,\uparrow}({\mbox{\boldmath $r$}}) - 
\left( \frac{\sqrt{6}}{2} \epsilon - \frac{\sqrt{6}}{4} \epsilon^2 \right) 
\phi_{0,\downarrow}({\mbox{\boldmath $r$}}), \nonumber \\\\
\label{2-2up}
&&\hspace*{-1.5cm}
\Phi_{-2,\uparrow}^{(d)}({\mbox{\boldmath $r$}}) 
= \left( 1 - \frac{1}{2} \epsilon^2 \right) 
\phi_{-2,\uparrow}({\mbox{\boldmath $r$}}) - 
\left( \epsilon - \frac{3}{2} \epsilon^2 \right) 
\phi_{-1,\downarrow}({\mbox{\boldmath $r$}}), 
\end{eqnarray}
with $\epsilon=\lambda/H_{\rm ex}$. 
Here, $\phi_{m,\sigma}({\mbox{\boldmath $r$}})$ 
represents the d orbital 
of the magnetic quantum number $m$ and the spin $\sigma$, 
defined by
\begin{eqnarray}
\label{d_orbital}
\phi_{m,\sigma}({\mbox{\boldmath $r$}})=u_{m}({\mbox{\boldmath $r$}})\chi_\sigma, 
\end{eqnarray}
with 
$u_{\pm 2}({\mbox{\boldmath $r$}})=R(r)( x \pm {\rm i} y)^2/(2 \sqrt{2})$, 
$u_{\pm 1}({\mbox{\boldmath $r$}})=\mp R(r) z( x \pm {\rm i} y)/\sqrt{2}$, 
$u_{0}({\mbox{\boldmath $r$}})=R(r) ( 3 z^2 - r^2)/(2 \sqrt{3})$, 
$r=|{\mbox{\boldmath $r$}}|$, $x=\sin \theta \cos \phi$, 
$y=\sin \theta \sin \phi$, 
and $z=\cos \theta$, 
where $R(r)$ is the radial part of the d orbital 
and $\chi_\sigma$ ($\sigma=\uparrow$ or $\downarrow$) 
is the spin state.

Here, we mention 
the right-hand sides of eqs. (\ref{22down}) - (\ref{2-1down}) 
and (\ref{21up}) - (\ref{2-2up}). 
The coefficient $\left( 1 - \frac{3}{4} \epsilon^2 \right)$ or 
$\left( 1 - \frac{1}{2} \epsilon^2 \right)$ 
means that 
the probability amplitude of the pure orbital 
decreases from 1 owing to 
hybridization with the other orbital. 
In contrast, 
$\left( \epsilon \pm  \frac{3}{2} \epsilon^2 \right)$ or 
$\left( \frac{\sqrt{6}}{2}
\epsilon \pm  \frac{\sqrt{6}}{4} \epsilon^2 \right)$ 
corresponds to the probability amplitude of the other orbital. 
Here, 
$-\frac{3}{4} \epsilon^2$ and 
$- \frac{1}{2} \epsilon^2$ in the former 
and 
$\epsilon$ and $\frac{\sqrt{6}}{2} \epsilon$ in the latter
arise from 
the Smit\cite{Smit} spin-mixing mechanism\cite{Jaoul,Malozemoff2} 
with $(\lambda/2)(L_+ S_- + L_- S_+)$. 
On the other hand, 
$\pm  \frac{3}{2} \epsilon^2$ and 
$\pm \frac{\sqrt{6}}{4} \epsilon^2$ in the latter 
stem from a combination of 
the $\lambda L_zS_z$ operator and 
the Smit\cite{Smit} spin-mixing mechanism. 
In deriving the resistivities 
of eqs. (\ref{rho_j_s}) - (\ref{rho_j1_s}), however, 
the terms related to the $\lambda L_zS_z$ operator are eliminated 
by ignoring terms higher than the second order of $\epsilon$.

\section{s--d Scattering Rate}
\label{s-d}
We derive an expression of 
the s--d scattering rate for the case of the $\ell$ configuration 
($\ell=\parallel$ or $\perp$), 
$1/\tau_{s \sigma \to d M \varsigma}^{(\ell)}$ (see eq. (\ref{tau_inv}). 
This scattering means that 
the conduction electron is scattered into 
the localized d states by nonmagnetic impurities. 
Here, we consider a system in which 
some atoms of the host lattice are substituted by the impurity atoms. 
In addition, the conduction state is represented by a plane wave, 
while the localized d states are described by a tight-binding model.

The scattering rate 
$1/\tau_{s \sigma \to d M \varsigma}^{(\ell)}$ is written as
\begin{eqnarray}
\label{W_sd_ap}
&&\hspace*{-0.4cm}
\frac{1}{\tau_{s \sigma \to d M \varsigma}^{(\ell)}} =
\frac{2 \pi}{\hbar} 
\sum_{\mbox{\boldmath $k$}'}
\left\langle \left| \left\langle 
\Psi_{\mbox{\boldmath $k$}',M,\varsigma}^{(d)} 
\Big|V_{\rm imp}({\mbox{\boldmath $r$}})\Big|
\Psi_{\mbox{\boldmath $k$}_{\mbox{\tiny F},\sigma}^{(\ell)},\sigma}^{(s)} 
\right\rangle \right|^2 \right\rangle_{\rm imp} \nonumber \\
&&\hspace*{1.5cm}\times
\delta \left(
E_{\mbox{\tiny F}} 
- E_{\mbox{\boldmath $k$}',M,\varsigma}^{(d)} \right), 
\end{eqnarray}
with 
\begin{eqnarray}
\label{plane_wave}
&&\hspace*{-1cm}\Psi_{{\mbox{\boldmath $k$}}_{\mbox{\tiny F},\sigma}^{(\ell)},\sigma}^{(s)} ({\mbox{\boldmath $r$}}) 
= \frac{1}{\sqrt{\Omega}} \exp\left({\rm i} {\mbox{\boldmath $k$}}_{\mbox{\tiny F},\sigma}^{(\ell)}\cdot
{\mbox{\boldmath $r$}}\right) \chi_\sigma, \\
\label{Psi}
&&\hspace*{-1cm}\Psi_{\mbox{\boldmath $k$}',M,\varsigma}^{(d)} ({\mbox{\boldmath $r$}}) 
=
\frac{1}{\sqrt{N}} \sum_j 
\exp \left({\rm i} {\mbox{\boldmath $k$}}'\cdot
{\mbox{\boldmath $R$}}_j \right)
\Phi_{M,\varsigma}^{(d)} 
({\mbox{\boldmath $r$}} - {\mbox{\boldmath $R$}}_j), \\
&&\hspace*{-1cm}
\Phi_{M,\varsigma}^{(d)} 
({\mbox{\boldmath $r$}} - {\mbox{\boldmath $R$}}_j)
=
\sum_{m,\sigma} c_{m,\sigma,M,\varsigma} \phi_{m,\sigma}
({\mbox{\boldmath $r$}} - {\mbox{\boldmath $R$}}_j), \\
\label{V(r)}
&&\hspace*{-1cm}
V_{\rm imp}({\mbox{\boldmath $r$}}) = \sum_i 
v_{\rm imp} ({\mbox{\boldmath $r$}}- {\mbox{\boldmath $R$}}_i), \\
\label{v_imp(r)}
&&\hspace*{-1cm}
v_{\rm imp} ({\mbox{\boldmath $r$}}- {\mbox{\boldmath $R$}}_i)
=\frac{\Delta Z e^2}{4\pi \epsilon_0|{\mbox{\boldmath $r$}}- {\mbox{\boldmath $R$}}_i|} 
\exp \left(-q |{\mbox{\boldmath $r$}}- {\mbox{\boldmath $R$}}_i| \right). 
\end{eqnarray}
The function 
$\Psi_{{\mbox{\boldmath $k$}}_{\mbox{\tiny F},\sigma}^{(\ell)}}^{(s)} ({\mbox{\boldmath $r$}})$ 
is the plane wave, 
where 
${\mbox{\boldmath $r$}}$ is the position vector, 
${\mbox{\boldmath $k$}}_{\mbox{\tiny F},\sigma}^{(\ell)}$ 
is 
the Fermi wavevector of the $\sigma$ spin 
in the current direction for the case of the $\ell$ configuration, 
$\Omega$ is the volume of the system, 
and $\chi_\sigma$ 
is the spin state.\cite{Malozemoff2} 
The eigenenergy of 
$\Psi_{{\mbox{\boldmath $k$}}_{\mbox{\tiny F},\sigma}^{(\ell)},\sigma}^{(s)} ({\mbox{\boldmath $r$}})$ 
is set to be $E_{\mbox{\tiny F}}$. 
The function 
$\Psi_{\mbox{\boldmath $k$}',M,\varsigma}^{(d)} ({\mbox{\boldmath $r$}})$ is 
the wave function of 
the tight-binding model.\cite{TB_model} 
Here, $\mbox{\boldmath $k$}'$ is the wavevector, 
$N$ is the number of unit cells, and 
$\Phi_{M,\varsigma}^{(d)} 
({\mbox{\boldmath $r$}} - {\mbox{\boldmath $R$}}_j)$ 
is the spin-mixed state in the atom located at ${\mbox{\boldmath $R$}}_j$, 
where $c_{m,\sigma,M,\varsigma}$ 
is the coefficient of 
$\phi_{m,\sigma}({\mbox{\boldmath $r$}} - {\mbox{\boldmath $R$}}_j)$ 
(see Appendix \ref{WF}). 
The eigenenergy of 
$\Psi_{\mbox{\boldmath $k$}',M,\varsigma}^{(d)} ({\mbox{\boldmath $r$}})$ is 
given by $E_{\mbox{\boldmath $k$}',M,\varsigma}^{(d)}$. 
The function $V_{\rm imp}({\mbox{\boldmath $r$}})$ is 
the scattering potential created by nonmagnetic impurities 
located randomly,\cite{Mahan} 
where 
$v_{\rm imp} ({\mbox{\boldmath $r$}}- {\mbox{\boldmath $R$}}_i)$ 
is a spherically symmetric scattering potential 
due to the impurity at ${\mbox{\boldmath $R$}}_i$.\cite{Potter} 
The quantity $\Delta Ze$ is the difference 
of the effective nuclear charge between the impurity and the host lattice, 
$q$ is the screening length, 
and $\epsilon_0$ is the dielectric constant. 
In addition, 
$\langle X \rangle_{\rm imp}$ represents 
the average of 
$X$ 
over the random distribution of the impurities, 
defined by $\langle X \rangle_{\rm imp}$
= $\sum_l X(\left\{{\mbox{\boldmath $R$}}\right\}_l)/
(\sum_l 1)$, 
where 
$\left\{{\mbox{\boldmath $R$}}\right\}_l$ 
(=$\left\{{\mbox{\boldmath $R$}}_1, {\mbox{\boldmath $R$}}_2, 
{\mbox{\boldmath $R$}}_3, \cdot \cdot \cdot \right\}_l$) is 
the $l$th set of the random distribution of the impurities.

\begin{figure}[ht]
\begin{center}
\includegraphics[width=0.65\linewidth]{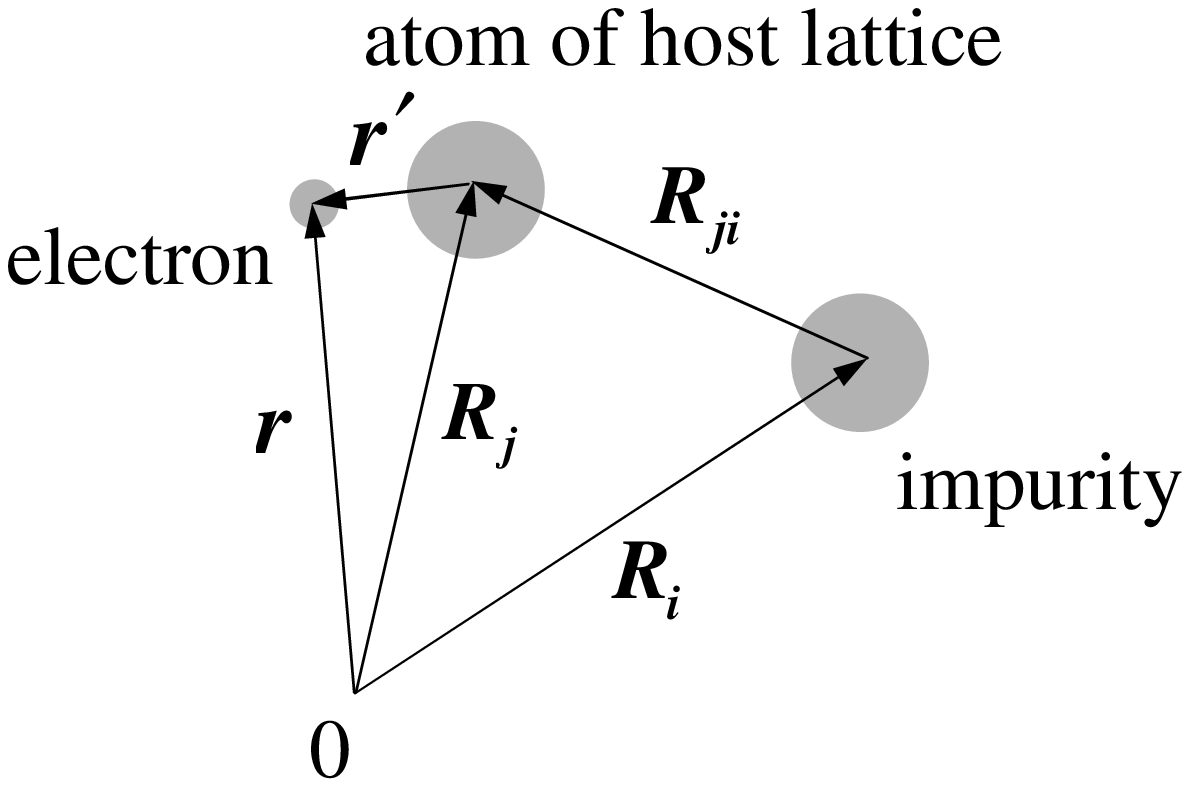} 
\caption{
Vectors 
${\mbox{\boldmath $r$}}$, ${\mbox{\boldmath $r$}}'$, 
${\mbox{\boldmath $R$}}_i$, ${\mbox{\boldmath $R$}}_j$, 
and ${\mbox{\boldmath $R$}}_{ji}$. 
Here, ${\mbox{\boldmath $r$}}$, 
${\mbox{\boldmath $R$}}_j$, 
and ${\mbox{\boldmath $R$}}_i$ 
are, respectively, the position vectors of the electron, 
the $j$th atom of the host lattice, 
and the $i$th impurity measured from the origin 0. 
In addition, 
${\mbox{\boldmath $r$}}'$ is the position vector of 
the electron measured from the $j$th atom of the host lattice, 
while 
${\mbox{\boldmath $R$}}_{ji}$ is 
the position vector of the $j$th atom of the host lattice 
measured from the $i$th impurity. 
}
\label{fig_vector}
\end{center}
\end{figure}

To rewrite eq. (\ref{W_sd_ap}) as a more specific expression, 
we consider 
\begin{eqnarray}
\label{matrix0}
&&\hspace*{-1cm}
\left\langle 
\left| \left\langle 
\Psi_{\mbox{\boldmath $k$}',M,\varsigma}^{(d)} 
\Big| V_{\rm imp}({\mbox{\boldmath $r$}}) \Big|
\Psi_{{\mbox{\boldmath $k$}}_{\mbox{\tiny F},\sigma}^{(\ell)},\sigma}^{(s)} \right\rangle 
\right|^2 \right\rangle_{\rm imp} \nonumber \\
&&= \frac{1}{N \Omega}
\left\langle \left| \sum_{m} c_{m,\sigma,M,\varsigma}^* \Lambda_{m\sigma} 
\right|^2 \right\rangle_{\rm imp}, \\
&&\hspace*{-1cm}
\Lambda_{m\sigma} = \sum_{i,j} \int 
\exp \left(-{\rm i} {\mbox{\boldmath $k$}}'
\cdot {\mbox{\boldmath $R$}}_j \right) 
\phi_{m,\sigma}^* ({\mbox{\boldmath $r$}}-{\mbox{\boldmath $R$}}_j) 
\nonumber \\
&& \times v_{\rm imp} ({\mbox{\boldmath $r$}}-{\mbox{\boldmath $R$}}_i) 
\exp \left( {\rm i} {\mbox{\boldmath $k$}}_{\mbox{\tiny F},\sigma}^{(\ell)}
\cdot {\mbox{\boldmath $r$}} \right) 
{\rm d} {\mbox{\boldmath $r$}}, 
\end{eqnarray}
where 
the inner product between $\chi_\sigma$ and 
the spin state of $\phi_{m,\sigma}$ has been taken 
in eq. (\ref{matrix0}). 
Note here that 
the case of $i=j$ corresponds to 
the scattering from the conduction state to the d states of the impurity atom. 
Such a case may be suitable for 
a system containing transition-metal impurities. 
In the present study, however, 
the impurity is considered to be 
a light element, such as carbon, 
in which 2s and 2p orbitals contribute to the transport. 
We, therefore, treat the case of $i \ne j$. 
Using 
${\mbox{\boldmath $R$}}_{ji}$ 
(=${\mbox{\boldmath $R$}}_{j}-{\mbox{\boldmath $R$}}_{i}$), 
we 
represent $\Lambda_{m\sigma}$ as 
\begin{eqnarray}
&&\hspace*{-1cm}\Lambda_{m\sigma} =
\sum_i \sum_{j} \displaystyle{\int} 
\exp \left( -{\rm i} {\mbox{\boldmath $k$}}'\cdot ({\mbox{\boldmath $R$}}_i + {\mbox{\boldmath $R$}}_{ji}) \right)
\phi_{m\sigma}^* \left({\mbox{\boldmath $r$}} - ({\mbox{\boldmath $R$}}_i + {\mbox{\boldmath $R$}}_{ji}) \right) \nonumber \\
&& \times v_{\rm imp} ({\mbox{\boldmath $r$}} - {\mbox{\boldmath $R$}}_i)  
\exp \left({\rm i} {\mbox{\boldmath $k$}}_{\mbox{\tiny F},\sigma}^{(\ell)} \cdot {\mbox{\boldmath $r$}} \right)
{\rm d} {\mbox{\boldmath $r$}}.
\end{eqnarray}
By replacing 
${\mbox{\boldmath $r$}} - ({\mbox{\boldmath $R$}}_i + {\mbox{\boldmath $R$}}_{ji})$ by ${\mbox{\boldmath $r$}}'$ (see Fig. \ref{fig_vector}), 
$\Lambda_{m\sigma}$ becomes 
\begin{eqnarray}
\label{Lambda1}
&&\hspace*{-2cm}\Lambda_{m\sigma}
=\sum_i \sum_{j} 
\exp \left( {\rm i} ({\mbox{\boldmath $k$}}_{\mbox{\tiny F},\sigma}^{(\ell)} - {\mbox{\boldmath $k$}}') \cdot ({\mbox{\boldmath $R$}}_i + {\mbox{\boldmath $R$}}_{ji} ) \right)  \nonumber \\
&&\hspace*{-1.cm}\times \displaystyle{\int} 
\phi_{m\sigma}^* ({\mbox{\boldmath $r$}}') v_{\rm imp} ({\mbox{\boldmath $r$}}' + {\mbox{\boldmath $R$}}_{ji})  
\exp \left({\rm i} {\mbox{\boldmath $k$}}_{\mbox{\tiny F},\sigma}^{(\ell)} \cdot {\mbox{\boldmath $r$}}' \right)
{\rm d} {\mbox{\boldmath $r$}}'.
\end{eqnarray}
We now assume that 
$v_{\rm imp} ({\mbox{\boldmath $r$}}' + {\mbox{\boldmath $R$}}_{ji})$ 
acts between the impurity and its nearest-neighbor atoms. 
We then have 
$v_{\rm imp} ({\mbox{\boldmath $r$}}' + {\mbox{\boldmath $R$}}_{ji})=v_{\rm imp} ({\mbox{\boldmath $r$}}' + {\mbox{\boldmath $R$}}_{j1})$, 
indicating that 
$v_{\rm imp} ({\mbox{\boldmath $r$}}' + {\mbox{\boldmath $R$}}_{ji})$ 
is independent of $i$. 
In addition, 
since $R_{j1}$ is larger than 
the orbital radius of the 3d electron $r'$, 
$|{\mbox{\boldmath $r$}}' + {\mbox{\boldmath $R$}}_{j1}|$ 
is roughly replaced by the dominant component $R_{j1}$. 
Namely, we have 
$|{\mbox{\boldmath $r$}}' + {\mbox{\boldmath $R$}}_{j1}|=(R_{j1}^2 
+r'^2 + 2{\mbox{\boldmath $r$}}' \cdot {\mbox{\boldmath $R$}}_{j1})^{1/2} \approx R_{j1}$ owing to 
$R_{j1}^2 > r'^2$, 
$2|{\mbox{\boldmath $r$}}' \cdot {\mbox{\boldmath $R$}}_{j1}|$. 
As a result, 
$v_{\rm imp} ({\mbox{\boldmath $r$}}'+{\mbox{\boldmath $R$}}_{j1})$ 
is approximated as follows: 
\begin{eqnarray}
\label{V_imp_app}
&&\hspace*{-1cm}
v_{\rm imp} ({\mbox{\boldmath $r$}}' + {\mbox{\boldmath $R$}}_{j1}) 
= \frac{\Delta Z e^2}{4\pi \epsilon_0 \left| {\mbox{\boldmath $r$}}' + {\mbox{\boldmath $R$}}_{j1} \right|} \exp \left( -q \left| {\mbox{\boldmath $r$}}'+ {\mbox{\boldmath $R$}}_{j1} \right| \right) \nonumber \\
&&\hspace*{0.9cm}
\approx 
\frac{\Delta Z e^2}{4\pi \epsilon_0 R_{j1}} \exp \left(-q R_{j1} \right) \nonumber \\
&&\hspace*{0.9cm}
\equiv v_{\rm imp} (R_{j1}). 
\end{eqnarray}
The distance 
$R_{j1}$ is here set to be constant independently of $j$; that is, 
$R_{j1}$ is written as $R_{j1} \equiv R_{\rm n}$, 
where $R_{\rm n}$ is constant. 
By substituting eq. (\ref{V_imp_app}) with $R_{j1}=R_{\rm n}$ 
into eq. (\ref{Lambda1}), 
$\Lambda_{m\sigma}$ becomes
\begin{eqnarray}
\label{Lambda}
&&\hspace*{-0.8cm}\Lambda_{m\sigma}
=\sum_i \exp \left( {\rm i} ({\mbox{\boldmath $k$}}_{\mbox{\tiny F},\sigma}^{(\ell)}- {\mbox{\boldmath $k$}}') \cdot {\mbox{\boldmath $R$}}_i \right)
\sum_{j~({\rm n.n.})} \exp \left( {\rm i} ({\mbox{\boldmath $k$}}_{\mbox{\tiny F},\sigma}^{(\ell)}- {\mbox{\boldmath $k$}}') \cdot {\mbox{\boldmath $R$}}_{j1} \right)  \nonumber \\
&& \hspace*{0.2cm} \times 
v_{\rm imp} (R_{\rm n}) 
\displaystyle{\int} 
\phi_{m\sigma}^* ({\mbox{\boldmath $r$}}') 
\exp \left({\rm i} {\mbox{\boldmath $k$}}_{\mbox{\tiny F},\sigma}^{(\ell)}\cdot {\mbox{\boldmath $r$}}' \right)
{\rm d} {\mbox{\boldmath $r$}}',
\end{eqnarray}
where 
$\sum_{j}$ of eq. (\ref{Lambda1}) 
has been replaced by $\sum_{j~({\rm n.n.})}$, 
i.e., 
the summation over the nearest-neighbor atoms around the impurity. 
Next, we consider 
$\left\langle \left| \sum_i \exp \left({\rm i} 
( {\mbox{\boldmath $k$}}_{\mbox{\tiny F},\sigma}^{(\ell)} - {\mbox{\boldmath $k$}}' )
\cdot {\mbox{\boldmath $R$}}_i \right) \right|^2 \right\rangle_{\rm imp}$, 
which 
is contained in eq. (\ref{matrix0}) (in addition, see eq. (\ref{Lambda})). 
This part is expressed as follows: 
\begin{eqnarray}
\label{N_imp}
&&\hspace*{-1cm}\left\langle \left| \sum_i \exp \left({\rm i} 
( {\mbox{\boldmath $k$}}_{\mbox{\tiny F},\sigma}^{(\ell)} - {\mbox{\boldmath $k$}}')
\cdot {\mbox{\boldmath $R$}}_i \right)  \right|^2 \right\rangle_{\rm imp} \nonumber \\
&&= 
\left\langle \sum_{i,i'} \exp \left({\rm i} 
( {\mbox{\boldmath $k$}}_{\mbox{\tiny F},\sigma}^{(\ell)} - {\mbox{\boldmath $k$}}')
\cdot \left({\mbox{\boldmath $R$}}_i - {\mbox{\boldmath $R$}}_{i'} \right) \right) 
\right\rangle_{\rm imp}
\nonumber \\
&&=
\left\langle 
\sum_{i,i'} \delta_{i,i'} + 
\sum_{i \ne i'} \exp \left({\rm i} 
( {\mbox{\boldmath $k$}}_{\mbox{\tiny F},\sigma}^{(\ell)} - {\mbox{\boldmath $k$}}')
\cdot ({\mbox{\boldmath $R$}}_i - {\mbox{\boldmath $R$}}_{i'} ) \right) 
\right\rangle_{\rm imp}
\nonumber \\
&&\approx
N_{\rm imp} + N_{\rm imp} \left( N_{\rm imp} -1 \right) 
\delta_{{\mbox{\boldmath $k$}}_{\mbox{\tiny F},\sigma}^{(\ell)},{\mbox{\boldmath $k$}}'}, 
\end{eqnarray}
where $N_{\rm imp}$ is the number of impurities in the volume of $\Omega$. 
In the calculation process of eq. (\ref{N_imp}), 
we have taken the summation about 
random points on a unit circle in a complex plane 
and the average 
over the impurity distributions.\cite{Mahan} 
In a similar manner, 
we deal with 
$\left| \sum_{j~({\rm n.n.})} \exp \left({\rm i} 
( {\mbox{\boldmath $k$}}_{\mbox{\tiny F},\sigma}^{(\ell)} 
- {\mbox{\boldmath $k$}}' )
\cdot {\mbox{\boldmath $R$}}_{j1} \right)  \right|^2 $ in eq. (\ref{matrix0}) 
to obtain a simple expression. 
Note, however, that 
$\langle ~~~ \rangle_{\rm imp}$ is in fact not contained in this expression 
and the number of $j$ (i.e., $\sum_{j~({\rm n.n.})}1$) 
is also much smaller than $N_{\rm imp}$. 
Though this treatment may be crude, we have
\begin{eqnarray}
\label{N_nn}
&&\hspace*{-0.5cm} \left| \sum_{j~({\rm n.n.})} \exp \left({\rm i} 
( {\mbox{\boldmath $k$}}_{\mbox{\tiny F},\sigma}^{(\ell)} - {\mbox{\boldmath $k$}}')
\cdot {\mbox{\boldmath $R$}}_{j1} \right)  \right|^2 \nonumber \\
&&= 
\sum_{j,j'~({\rm n.n.})} \delta_{j,j'} + 
\sum_{j \ne j'~({\rm n.n.})} \exp \left({\rm i} 
( {\mbox{\boldmath $k$}}_{\mbox{\tiny F},\sigma}^{(\ell)} - {\mbox{\boldmath $k$}}')
\cdot ({\mbox{\boldmath $R$}}_{j1} - {\mbox{\boldmath $R$}}_{j'1} ) \right) \nonumber \\
&&\approx
N_{\rm n} + N_{\rm n} \left( N_{\rm n} -1 \right) 
\delta_{{\mbox{\boldmath $k$}}_{\mbox{\tiny F},\sigma}^{(\ell)},{\mbox{\boldmath $k$}}'}, 
\end{eqnarray}
where $N_{\rm n}$ is the number of nearest-neighbor atoms around the impurity.

Using eqs. (\ref{W_sd_ap}), (\ref{matrix0}), (\ref{Lambda}), 
(\ref{N_imp}), and (\ref{N_nn}), we obtain 
\begin{eqnarray}
\label{rate}
&&\hspace*{-1.5cm}\frac{1}{\tau_{s \sigma \to d M\varsigma}^{(\ell)}} =
\frac{2\pi}{\hbar} \sum_{\mbox{\boldmath $k$}'}
\frac{N_{\rm imp}}{\Omega} 
\left[ 1 + (N_{\rm imp}-1) 
\delta_{{\mbox{\boldmath $k$}}_{\mbox{\tiny F},\sigma}^{(\ell)},{\mbox{\boldmath $k$}}'} \right]  \nonumber \\
&& \hspace*{0.3cm}\times N_{\rm n} \left[ 1 + (N_{\rm n} -1) \delta_{{\mbox{\boldmath $k$}}_{\mbox{\tiny F},\sigma}^{(\ell)},{\mbox{\boldmath $k$}}'} \right] \nonumber \\
&& \hspace*{0.3cm}\times \left|v_{M,\varsigma}({\mbox{\boldmath $k$}}_{\mbox{\tiny F},\sigma}^{(\ell)})\right|^2 
\frac{1}{N}\delta 
\left(E_{\mbox{\tiny F}} 
- E_{\mbox{\boldmath $k$}',M,\varsigma}^{(d)}\right), \\
\label{v_Ms_k}
&&\hspace*{-1.5cm}v_{M,\varsigma}({\mbox{\boldmath $k$}}_{\mbox{\tiny F},\sigma}^{(\ell)})=
v_{\rm imp} (R_{\rm n}) \nonumber \\
&& \hspace*{0.3cm}\times \sum_{m} c_{m,\sigma,M,\varsigma}^* 
\int \phi_{m,\sigma}^* ({\mbox{\boldmath $r$}}) 
\exp \left({\rm i} {\mbox{\boldmath $k$}}_{\mbox{\tiny F},\sigma}^{(\ell)} \cdot {\mbox{\boldmath $r$}} \right)  
{\rm d}{\mbox{\boldmath $r$}}. \nonumber \\
\end{eqnarray}
We consider a case in which 
$\sum_{{\mbox{\boldmath $k$}}'} \delta 
\left(E_{\mbox{\tiny F}} 
- E_{\mbox{\boldmath $k$}',M,\varsigma}^{(d)}\right)$ 
is much larger than 
$(N_{\rm imp}-1)(N_{\rm n}-1) \delta 
\left(E_{\mbox{\tiny F}} 
- E_{{\mbox{\boldmath $k$}}_{\mbox{\tiny F},\sigma}^{(\ell)},M,\varsigma}^{(d)}\right)$. 
Equation (\ref{rate}) may then be given 
by the following approximate expression: 
\begin{eqnarray}
\label{tau_sd^-1}
&&\hspace*{-1cm} \frac{1}{\tau_{s \sigma \to d M\varsigma}^{(\ell)}} =
\frac{2\pi}{\hbar} n_{\rm imp} 
N_{\rm n} \left| v_{M,\varsigma}({\mbox{\boldmath $k$}}_{\mbox{\tiny F},\sigma}^{(\ell)}) \right|^2 
D_{M,\varsigma}^{(d)}, \\
\label{D_Ms^d}
&&\hspace*{-1cm} 
D_{M,\varsigma}^{(d)}=\frac{1}{N}
\sum_{\mbox{\boldmath $k$}'}
\delta 
\left(E_{\mbox{\tiny F}} 
- E_{\mbox{\boldmath $k$}',M,\varsigma}^{(d)}\right), 
\end{eqnarray}
with 
$n_{\rm imp}=N_{\rm imp}/\Omega$. 
It is noted that 
the unit of $D_{M,\varsigma}^{(d)}$ of eq. (\ref{D_Ms^d}) is J$^{-1}$, 
while 
that of 
$D_\sigma^{(s)}$ of eq. (\ref{D_s^s}) is J$^{-1}$m$^{-3}$. 
The unit of $| v_{M,\varsigma}({\mbox{\boldmath $k$}}_{\mbox{\tiny F},\sigma}^{(\ell)}) |^2$ in eq. (\ref{tau_sd^-1}) is J$^2$m$^3$, 
while that of $|V_s|^2$ in eq. (\ref{1/tau_ssigma}) is J$^2$m$^6$. 
As to the calculation of $D_{\varsigma}^{(d)}/D_\sigma^{(s)}$ 
and $\beta_\sigma$ in eqs. (\ref{ttt}) - (\ref{beta}), 
$D_{\varsigma}^{(d)}$ and $| V_{s\sigma \to d \sigma}|^2$ 
should be replaced by 
$D_{\varsigma}^{(d)}/\Omega_{\rm unit}$ 
and $|V_{s\sigma \to d \sigma}|^2 \Omega_{\rm unit}$, respectively, 
where $\Omega_{\rm unit}$ is the unit cell volume. 

\section{s--s Scattering Rate}
\label{s-s}
We 
derive 
an expression of the s--s scattering rate $1/\tau_{s\sigma}$ 
of eq. (\ref{1/tau_s_s}).

The scattering rate $1/\tau_{s\sigma}$ 
is originally written as\cite{Localization,Inoue} 
\begin{eqnarray}
\label{1/tau_ss appen}
&&\hspace*{-1.1cm}
\frac{1}{\tau_{s\sigma}}=
\frac{2 \pi}{\hbar} \sum_{\mbox{\boldmath $k$}'_\sigma}
\left\langle \left| \left\langle 
\Psi_{{\mbox{\boldmath $k$}'_\sigma},\sigma}^{(s)} 
\Big|V_{\rm imp}({\mbox{\boldmath $r$}})\Big|
\Psi_{\mbox{\boldmath $k$}_{\mbox{\tiny F},\sigma},\sigma}^{(s)} 
\right\rangle \right|^2 \right\rangle_{\rm imp} 
\nonumber \\
&&\times 
\delta \left( E_{\mbox{\tiny F}}  - E_{{\mbox{\boldmath $k$}}'_\sigma} \right) 
\left( 1 - \cos \theta_{{\mbox{\boldmath $k$}_{\mbox{\tiny F},\sigma}},{\mbox{\boldmath $k$}}'_\sigma} \right), 
\end{eqnarray}
where 
$\Psi_{\mbox{\boldmath $k$}_{\mbox{\tiny F},\sigma},\sigma}^{(s)}$ 
and $V_{\rm imp}({\mbox{\boldmath $r$}})$ 
are given by eqs. (\ref{plane_wave}) and (\ref{V(r)}), respectively. 
Here, ${\mbox{\boldmath $k$}_{\mbox{\tiny F},\sigma}}$ 
is the wavevector of the incident electron of the $\sigma$ spin 
(i.e., the Fermi wavevector of the $\sigma$ spin in the current direction), 
${\mbox{\boldmath $k$}}'_\sigma$ is 
the wavevector of the scattered electron of the $\sigma$ spin, 
and 
$\theta_{{{\mbox{\boldmath $k$}}_{\mbox{\tiny F},\sigma}-{\mbox{\boldmath $k$}}'}_\sigma}$ is the relative angle 
between ${\mbox{\boldmath $k$}}_{\mbox{\tiny F},\sigma}$ and ${\mbox{\boldmath $k$}}'_\sigma$. 
In addition, 
$E_{\mbox{\tiny F}}$ ($E_{{\mbox{\boldmath $k$}}'_\sigma}$) 
is the energy of the incident electron (the energy of the scattered electron). 
Equation (\ref{1/tau_ss appen}) is also rewritten as\cite{Localization}
\begin{eqnarray}
\label{1/tau_ss appen1}
&&\hspace*{-1.1cm}
\frac{1}{\tau_{s\sigma}}= \frac{2 \pi}{\hbar} \frac{n_{\rm imp}}{\Omega} 
\sum_{\mbox{\boldmath $k$}'_\sigma}
\left| v_{{\mbox{\boldmath $k$}}_{\mbox{\tiny F},\sigma}
-{\mbox{\boldmath $k$}}'_\sigma} \right|^2 
\delta \left( E_{\mbox{\tiny F}}  - E_{{\mbox{\boldmath $k$}}'_\sigma} \right) \nonumber \\
&&\times \left( 1 - \cos \theta_{{\mbox{\boldmath $k$}_{\mbox{\tiny F},\sigma}},{\mbox{\boldmath $k$}}'_\sigma} \right), 
\end{eqnarray}
where 
$v_{{\mbox{\boldmath $k$}}_{\mbox{\tiny F},\sigma}-{\mbox{\boldmath $k$}}'_\sigma}$ is 
given by
\begin{eqnarray}
v_{{\mbox{\boldmath $k$}}_{\mbox{\tiny F},\sigma}-{\mbox{\boldmath $k$}}'_\sigma}= 
\int v_{\rm imp}({\mbox{\boldmath $r$}}) 
\exp \left( {\rm i}({\mbox{\boldmath $k$}}_{\mbox{\tiny F},\sigma} - {\mbox{\boldmath $k$}}'_\sigma )
\cdot {\mbox{\boldmath $r$}} \right) {\rm d}{\mbox{\boldmath $r$}}, 
\end{eqnarray}
where 
$v_{\rm imp}({\mbox{\boldmath $r$}})$ is 
a short-range potential due to the impurity, 
i.e., eq. (\ref{v_imp(r)}). 
In the case of the s--s scattering, 
$v_{\rm imp}({\mbox{\boldmath $r$}})$ 
may be replaced by an approximate potential on the impurity site 
because such a potential contributes dominantly to 
$v_{{\mbox{\boldmath $k$}}_{\mbox{\tiny F},\sigma}-{\mbox{\boldmath $k$}}'_\sigma}$. 
In brief, 
$v_{\rm imp}({\mbox{\boldmath $r$}})$ is approximated as 
$v_{\rm imp}({\mbox{\boldmath $r$}})=V_{s} \delta({\mbox{\boldmath $r$}})$, 
where $V_s$ is constant. 
We thus obtain 
$v_{{\mbox{\boldmath $k$}}_{\mbox{\tiny F},\sigma}-{\mbox{\boldmath $k$}}'_\sigma}$
=$V_{s}$, 
which is independent of the $\sigma$ spin and the wavevectors. 
As a result, eq. (\ref{1/tau_ss appen1}) 
is expressed as\cite{Localization,Inoue} 
\begin{eqnarray}
\label{1/tau_ssigma}
&&\hspace*{-1cm} 
\frac{1}{\tau_{s\sigma}}= \frac{2 \pi}{\hbar} n_{\rm imp}
|V_s |^2 D_\sigma^{(s)}, \\
\label{D_s^s}
&&\hspace*{-1cm} 
D_\sigma^{(s)}= \frac{1}{\Omega}\sum_{\mbox{\boldmath $k$}'_\sigma}
\delta \left( E_{\mbox{\tiny F}}- E_{{\mbox{\boldmath $k$}}'_\sigma} \right). 
\end{eqnarray}
Here, 
$\sum_{\mbox{\boldmath $k$}'_\sigma} 
\delta \left( E_{\mbox{\tiny F}} - E_{{\mbox{\boldmath $k$}}'_\sigma} \right)
\cos \theta_{{\mbox{\boldmath $k$}_{\mbox{\tiny F},\sigma}},{\mbox{\boldmath $k$}}'_\sigma}$ 
disappears.

\section{Matrix Elements}
\label{matrix}
We consider the matrix element 
in eqs. (\ref{tau_sd^-1}) and (\ref{v_Ms_k}), 
$v_{\rm imp}(R_{\rm n})\int \phi_{m,\sigma}^* ({\mbox{\boldmath $r$}}) 
\exp \left({\rm i} {\mbox{\boldmath $k$}}_{\mbox{\tiny F},\sigma}^{(\ell)} \cdot {\mbox{\boldmath $r$}} \right){\rm d}{\mbox{\boldmath $r$}}$, 
with $m$=$-2$ - 2 and $\ell$=$\parallel$ or $\perp$. 

The matrix elements are written by
\begin{eqnarray}
\label{mat_0_para}
&&\hspace*{-1cm}
v_{\rm imp}(R_{\rm n})\int \phi_{0,\sigma}^* ({\mbox{\boldmath $r$}}) 
\exp \left({\rm i} {\mbox{\boldmath $k$}}_{\mbox{\tiny F},\sigma}^{(\parallel)} \cdot {\mbox{\boldmath $r$}} \right){\rm d}{\mbox{\boldmath $r$}} 
\nonumber \\
&&\hspace*{-0.5cm}
=\frac{1}{\sqrt{3}}v_{\rm imp}(R_{\rm n})\int R(r) (z^2 -x^2)\exp \left({\rm i}k_{\mbox{\tiny F},\sigma} z\right)
{\rm d}{\mbox{\boldmath $r$}}, \\
\label{mat_0_perp}
&&\hspace*{-1cm}
v_{\rm imp}(R_{\rm n})\int \phi_{0,\sigma}^* ({\mbox{\boldmath $r$}}) 
\exp \left({\rm i} {\mbox{\boldmath $k$}}_{\mbox{\tiny F},\sigma}^{(\perp)} \cdot {\mbox{\boldmath $r$}} \right){\rm d}{\mbox{\boldmath $r$}} 
\nonumber \\
&&\hspace*{-0.5cm}
= \frac{1}{2\sqrt{3}}v_{\rm imp}(R_{\rm n})\int R(r) (z^2 -x^2)\exp \left({\rm i}k_{\mbox{\tiny F},\sigma} x\right)
{\rm d}{\mbox{\boldmath $r$}}, \\
\label{mat_2_perp}
&&\hspace*{-1cm}
v_{\rm imp}(R_{\rm n})\int \phi_{\pm 2,\sigma}^* ({\mbox{\boldmath $r$}}) 
\exp \left({\rm i} {\mbox{\boldmath $k$}}_{\mbox{\tiny F},\sigma}^{(\perp)} \cdot {\mbox{\boldmath $r$}} \right){\rm d}{\mbox{\boldmath $r$}} 
\nonumber \\
&&\hspace*{-0.5cm}
= \frac{1}{2\sqrt{2}}v_{\rm imp}(R_{\rm n})\int R(r) (x^2 -z^2)\exp \left({\rm i}k_{\mbox{\tiny F},\sigma} x\right)
{\rm d}{\mbox{\boldmath $r$}}, 
\end{eqnarray}
with 
${\mbox{\boldmath $k$}}_{\mbox{\tiny F},\sigma}^{(\parallel)}$=$(0,0,k_{\mbox{\tiny F},\sigma})$ 
and 
${\mbox{\boldmath $k$}}_{\mbox{\tiny F},\sigma}^{(\perp)}$=$(k_{\mbox{\tiny F},\sigma},0,0)$, 
where 
$\phi_{m,\sigma}({\mbox{\boldmath $r$}})$ 
is eq. (\ref{d_orbital}). 
In addition, we note 
$v_{\rm imp}(R_{\rm n})\int \phi_{m,\sigma}^* ({\mbox{\boldmath $r$}}) 
\exp \left({\rm i} {\mbox{\boldmath $k$}}_{\mbox{\tiny F},\sigma}^{(\parallel)} \cdot {\mbox{\boldmath $r$}} \right){\rm d}{\mbox{\boldmath $r$}}$=0 
for $m$=$\pm 1$, $\pm 2$, 
and 
$v_{\rm imp}(R_{\rm n})\int \phi_{\pm 1,\sigma}^* ({\mbox{\boldmath $r$}}) 
\exp \left({\rm i} {\mbox{\boldmath $k$}}_{\mbox{\tiny F},\sigma}^{(\perp)} \cdot {\mbox{\boldmath $r$}} \right){\rm d}{\mbox{\boldmath $r$}}$=0. 
As for 
$\left| v_{\rm imp}(R_{\rm n})\int \phi_{m,\sigma}^* ({\mbox{\boldmath $r$}})
\exp \left({\rm i} {\mbox{\boldmath $k$}}_{\mbox{\tiny F},\sigma}^{(\ell)} \cdot {\mbox{\boldmath $r$}} \right){\rm d}{\mbox{\boldmath $r$}} \right|^2$, 
we have $|V_{s\sigma \to d \sigma}|^2$ for eq. (\ref{mat_0_para}), 
$\frac{1}{4} |V_{s\sigma \to d \sigma}|^2$ for eq. (\ref{mat_0_perp}), 
and $\frac{3}{8} |V_{s\sigma \to d \sigma}|^2$ for eq. (\ref{mat_2_perp}), 
where $|V_{s\sigma \to d \sigma}|^2$ is eq. (\ref{V_ss-ds^2}).

\section{Parameters}
\label{parameters}
We obtain concrete expressions of 
$\rho_{s\sigma}$ of eq. (\ref{r_s_s}), 
$r$ of eq. (\ref{rrr}), and $\xi$ of eq. (\ref{xi_def}).

The resistivity $\rho_{s\sigma}$ of eq. (\ref{r_s_s}) is first written as 
\begin{eqnarray}
\label{rho_s,sigma}
\rho_{s\sigma}=
\frac{6^{1/3}{m_\sigma^*}^2 n_{\rm imp} |V_{s}|^2}{n_\sigma^{2/3} e^2 \pi^{1/3} \hbar^3}. 
\end{eqnarray}
Here, $1/\tau_{s\sigma}$ 
of eq. (\ref{1/tau_s_s}) has been given by
\begin{eqnarray}
&&\hspace*{-1cm}\frac{1}{\tau_{s\sigma}}= \frac{2\pi}{\hbar}n_{\rm imp} |V_{s}|^2 D_{\sigma}^{(s)} \nonumber \\
&&\hspace*{-0.35cm}=\frac{6^{1/3} m_\sigma^* n_{\rm imp} |V_{s}|^2 n_\sigma^{1/3}}
{\pi^{1/3} \hbar^3},  
\end{eqnarray}
where 
\begin{eqnarray}
\label{DOS_n}
&&D_{\sigma}^{(s)} =
\frac{1}{4\pi^2} 
\left( \frac{2m_\sigma^*}{\hbar^2} \right)^{3/2} \sqrt{E_{\mbox{\tiny F}} + \Delta_\sigma }
\nonumber \\
&&\hspace*{0.62cm}=
\frac{1}{4\pi^2} \frac{2m_\sigma^*}{\hbar^3} (6 \pi^2 \hbar^3 n_\sigma)^{1/3}, 
\end{eqnarray}
with 
$E_{\mbox{\tiny F}}+\Delta_\sigma=(\hbar k_{\mbox{\tiny F},\sigma})^2/(2m_\sigma^*)=(6 \pi^2 \hbar^3 n_\sigma)^{2/3}/(2m_\sigma^*)$ and 
$k_{\mbox{\tiny F},\sigma}=(6 \pi^2 n_\sigma)^{1/3}$.\cite{Kittel1} 
The quantity $n_\sigma$ ($m_\sigma^*$) 
is the number density\cite{Ibach,Grosso} (the effective mass\cite{Mathon}) 
of the electrons in the conduction band of the $\sigma$ spin. 
In addition, 
$\Delta_\sigma$ is the exchange splitting energy 
of the conduction electron, 
where $\Delta_\uparrow=\Delta$ and $\Delta_\downarrow=-\Delta$.

Using eqs. (\ref{rho_s,sigma}) and (\ref{DOS_n}), 
$r$ of eq. (\ref{rrr}) is expressed as 
\begin{eqnarray}
\label{rrr_ex}
r=
\left(\frac{m_\downarrow^*}{m_\uparrow^*} \right)^4 
\left(\frac{D_\uparrow^{(s)}}{D_\downarrow^{(s)}} \right)^2. 
\end{eqnarray}

Using eqs. (\ref{rrr_ex}), (\ref{vvv}), and (\ref{www}), 
$\xi$ of eq. (\ref{xi_def}) is obtained as 
\begin{eqnarray}
\xi= 
\left(\frac{m_\downarrow^*}{m_\uparrow^*} \right)^4 
\frac{1}{\beta_\downarrow}
\frac{(D_\uparrow^{(s)})^2}{D_\uparrow^{(d)}D_\downarrow^{(s)}} 
\left( \beta_\downarrow\frac{D_\downarrow^{(d)}}{D_\downarrow^{(s)}} +1 \right)^2, 
\end{eqnarray}
where $\beta_\sigma$ is eq. (\ref{beta}). 
Furthermore, 
in the case of a simple system with 
$\beta_\uparrow=\beta_\downarrow$ 
and 
$D_\uparrow^{(d)}/D_\uparrow^{(s)}=D_\downarrow^{(d)}/D_\downarrow^{(s)}$, 
$\xi$ becomes 
\begin{eqnarray}
\label{xi_ex}
\xi= p \left( u + \frac{1}{u} + 2 \right), 
\end{eqnarray}
with $p=(m_\downarrow^*/m_\uparrow^*)^4
D_\uparrow^{(s)}/D_\downarrow^{(s)}$, 
where 
$u$ is eq. (\ref{uuu}).


\begin{thebibliography}{9}
\bibitem{Kokado_un1}
Unpublished data. 
We evaluated $\rho_{s \downarrow}/\rho_{s \uparrow}$ of fcc Ni 
to be $\rho_{s \downarrow}/\rho_{s \uparrow}$=1.0 $\times$ 10 
using a combination of 
the first principles calculation and the Kubo formula 
within the semiclassical approximation.\cite{Tsymbal} 
The conductivity of the semiclassical approximation 
corresponded to the Drude formula.\cite{Drude} 
Here, we used the tight-binding parameters in ref. \citen{papa}.
\bibitem{Kokado_un}
Unpublished data. 
We evaluated $\rho_{s \downarrow}/\rho_{s \uparrow}$ of Fe$_4$N 
to be $\rho_{s \downarrow}/\rho_{s \uparrow}$=1.6 $\times$ 10$^{-3}$ 
using a combination of 
the first principles calculation and the Kubo formula 
within the semiclassical approximation.\cite{Tsymbal} 
The conductivity of the semiclassical approximation 
corresponded to the Drude formula.\cite{Drude} 
Here, we used the tight-binding parameters obtained 
in the previous study.\cite{Kokado}
\bibitem{link}
The resistivity $\rho$ of eq. (\ref{rho_j_ini}) is obtained from 
${\mbox{\boldmath $j$}}$=$-{\mbox{\boldmath $v$}}_\uparrow n_\uparrow e 
-{\mbox{\boldmath $v$}}_\downarrow n_\downarrow e$=
${\mbox{\boldmath $E$}}/\rho$, 
$-e {\mbox{\boldmath $E$}} 
+ m_\uparrow^*( 0 - {\mbox{\boldmath $v$}}_\uparrow )/\tau_\uparrow 
+ (m_\downarrow^* {\mbox{\boldmath $v$}}_\downarrow
- m_\uparrow^* {\mbox{\boldmath $v$}}_\uparrow)/\tau_{\uparrow \downarrow}$=0, 
and 
$-e {\mbox{\boldmath $E$}} 
+ m_\downarrow^* ( 0 - {\mbox{\boldmath $v$}}_\downarrow )/\tau_\downarrow 
+ (m_\uparrow^* {\mbox{\boldmath $v$}}_\uparrow
- m_\downarrow^* {\mbox{\boldmath $v$}}_\downarrow)/\tau_{\downarrow \uparrow}$=0, 
where the suffix of the configuration $\ell$ has been omitted. 
Here, 
${\mbox{\boldmath $j$}}$ is the current density, 
${\mbox{\boldmath $v$}}_\sigma$ is the velocity of the $\sigma$ spin, 
and 
${\mbox{\boldmath $E$}}$ is the electric field. 
As to $\rho$ with 
$\tau_{\uparrow \downarrow}$=$\tau_{\downarrow \uparrow}$, 
see refs. \citen{Campbell}, \citen{Ren}, and \citen{Fert}.
\bibitem{Ziese_comment}
The resistivity $\rho_{s \to d \sigma}$ of eq. (\ref{rho_sd usual}) 
in this paper 
corresponds to $\rho_\sigma$ in ref. \citen{Ziese}, 
with $\sigma$=$\uparrow$ or $\downarrow$. 
This correspondence is confirmed from, for example, 
$\rho_\sigma$ in eqs. (5) and (6) in ref. \citen{Ziese}, 
where 
$\Delta \rho_\sigma=\rho_{\parallel,\sigma}-\rho_{\perp,\sigma}$. 
Incidentally, 
using eqs. (\ref{rho_j_s}) - (\ref{rho_j1_s}) in this paper, 
we obtain 
$\Delta \rho_\uparrow=\gamma \rho_{s \to d\downarrow} - \gamma \rho_{s \to d\uparrow}$ and 
$\Delta \rho_\downarrow=-\gamma \rho_{s \to d\downarrow} + \gamma \rho_{s \to d\uparrow}$, 
where $\rho_{s \sigma \to d\varsigma}=\rho_{s \to d\varsigma}$ has been set. 
For $\Delta \rho_\sigma$, 
also, see eqs. (3) and (4) in ref. \citen{Malozemoff2}.
\bibitem{Ziese_comment1}
As found from $\rho_\sigma$ in eqs. (2), (4), (5), and (6) 
in ref. \citen{Ziese}, 
$\rho_\sigma$ in eqs. (2) and (4) 
might be approximated as 
$\rho_\sigma \approx \rho_{s \to d \sigma}$, 
with $\sigma$=$\uparrow$ or $\downarrow$. 
Here, $\rho_\sigma$ in eqs. (2) and (4) is usually given by 
$\rho_\sigma=\rho_s + \rho_{s\to d \sigma}$ 
(for example, see $\rho_\sigma$ in eq. (6) in ref. \citen{Malozemoff2}).
\end{thebibliography}
\end{document}